\documentclass[twocolumn,twocolappendix]{aastex63}
\usepackage[T1]{fontenc}
\usepackage{natbib}
\usepackage{color}
\usepackage{mathptmx}
\usepackage{amsmath}
\usepackage{url}
\usepackage{multirow}
\usepackage{verbatim}

\usepackage{numprint}
\usepackage{graphicx}
\usepackage{refcount}
\graphicspath{{./}{figures/}}
\received{December XX, 2023}
\revised{December XX, 2023}
\accepted{January XX, 2024}

\shorttitle{{\it AGORA} Comparison. IV: Cosmological Zoom-in Simulations Down To $z\leq2$}
\shortauthors{{\it AGORA} Collaboration et al.}
\begin{document}
\title{The {\it AGORA} High-resolution Galaxy Simulations Comparison Project IV: Halo and Galaxy Mass Assembly in a Cosmological Zoom-in Simulation at $z\le2$}

\author[0000-0002-6299-152X]{Santi Roca-F\`{a}brega}
\altaffiliation{Code leaders}
\affil{Lund Observatory, Division of Astrophysics, Department of Physics, Lund University, SE-221 00 Lund, Sweden}
\affil{Departamento de F\'{i}sica de la Tierra y Astrof\'{i}sica, Facultad de Ciencias F\'{i}sicas, Plaza Ciencias, 1, 28040 Madrid, Spain; \rm{\href{mailto:santi.roca_fabrega@fysik.lu.se}{santi.roca\_fabrega@fysik.lu.se}}}

\author[0000-0003-4464-1160]{Ji-hoon Kim}
\altaffiliation{Code leaders}
\affiliation{Seoul National University Astronomy Research Center, Seoul 08826, Korea; \rm{\href{mailto:mornkr@snu.ac.kr}{mornkr@snu.ac.kr}}}
\affiliation{Center for Theoretical Physics, Department of Physics and Astronomy, Seoul National University, Seoul 08826, Korea}

\author[0000-0001-5091-5098]{Joel R. Primack}
\affil{Department of Physics, University of California at Santa Cruz, Santa Cruz, CA 95064, USA; \rm{\href{mailto:joel@ucsc.edu}{joel@ucsc.edu}}}

\author[0000-0002-9144-1383]{Minyong Jung}
\altaffiliation{Code leaders}
\affiliation{Center for Theoretical Physics, Department of Physics and Astronomy, Seoul National University, Seoul 08826, Korea; \rm{\href{mailto:wispedia@snu.ac.kr}{wispedia@snu.ac.kr}}}

%TIER 2: CODE LEADERS
\author{Anna Genina}
\altaffiliation{Code leaders}
\affil{Max-Planck-Institut f\"{u}r Astrophysik, Karl-Schwarzschild-Str. 1, D-85748, Garching, Germany}

\author[0000-0002-4687-4948]{Loic Hausammann}
\altaffiliation{Code leaders}
\affil{Institute of Physics, Laboratoire d'Astrophysique, \'{E}cole Polytechnique F\'{e}d\'{e}rale de Lausanne (EPFL), CH-1015 Lausanne, Switzerland}
\affil{TS High Performance Computing, Eidgen\"ossische Technische Hochschule Z\"urich (ETHZ), 8092 Z\"urich, Switzerland}

\author[0000-0002-7820-2281]{Hyeonyong Kim}
\altaffiliation{Code leaders}
\affiliation{Center for Theoretical Physics, Department of Physics and Astronomy, Seoul National University, Seoul 08826, Korea}
\affiliation{Department of Aerospace Engineering, Seoul National University, Seoul 08826, Korea}

\author{Alessandro Lupi}
\altaffiliation{Code leaders}
\affil{DiSAT, Universit\`a degli Studi dell'Insubria, via Valleggio 11, I-22100 Como, Italy}
\affil{Dipartimento di Fisica ``G. Occhialini'', Universit\`a degli Studi di Milano-Bicocca, I-20126 Milano, Italy}

\author[0000-0001-7457-8487]{Kentaro Nagamine}
\altaffiliation{Code leaders}
\affiliation{Department of Earth and Space Science, Graduate School of Science, Osaka University, Toyonaka, Osaka, 560-0043, Japan}
\affiliation{Kavli IPMU (WPI), University of Tokyo, 5-1-5 Kashiwanoha, Kashiwa, Chiba, 277-8583, Japan}
\affiliation{Department of Physics \& Astronomy, University of Nevada Las Vegas, Las Vegas, NV 89154, USA}

\author[0000-0002-3764-2395]{Johnny W. Powell}
\altaffiliation{Code leaders}
\affil{Department of Physics, Reed College, Portland, OR 97202, USA}

\author{Yves Revaz}
\altaffiliation{Code leaders}
\affil{Institute of Physics, Laboratoire d'Astrophysique, \'{E}cole Polytechnique F\'{e}d\'{e}rale de Lausanne (EPFL), CH-1015 Lausanne, Switzerland}

\author{Ikkoh Shimizu}
\altaffiliation{Code leaders}
\affil{Shikoku Gakuin University, 3-2-1 Bunkyocho, Zentsuji, Kagawa, 765-8505, Japan}

\author[0000-0001-9695-4017]{Clayton Strawn}
\altaffiliation{Code leaders}
\affil{Department of Physics, University of California at Santa Cruz, Santa Cruz, CA 95064, USA}

\author{H\'{e}ctor Vel\'{a}zquez}
\altaffiliation{Code leaders}
\affil{Instituto de Astronom\'{i}a, Universidad Nacional Aut\'{o}noma de M\'{e}xico, A.P. 70-264, 04510, Mexico, D.F., Mexico}

%TIER 3: Those who helped to run the codes and/or made an important contribution through the meetings since Paper III's release
\author[0000-0002-5969-1251]{Tom Abel}
\affil{Kavli Institute for Particle Astrophysics and Cosmology, Stanford University, Stanford, CA 94305, USA}
\affil{Department of Physics, Stanford University, Stanford, CA 94305, USA}
\affil{SLAC National Accelerator Laboratory, Menlo Park, CA 94025, USA}

\author{Daniel Ceverino}
\affil{Universidad Aut\'{o}noma de Madrid, Ciudad Universitaria de Cantoblanco, E-28049 Madrid, Spain}
\affil{CIAFF, Facultad de Ciencias, Universidad Aut\'{o}noma de Madrid, E-28049 Madrid, Spain}

\author{Bili Dong}
\affil{Department of Physics, Center for Astrophysics and Space Sciences, University of California at San Diego, La Jolla, CA 92093, USA}

\author[0000-0001-5510-2803]{Thomas R. Quinn}
\affil{Department of Astronomy, University of Washington, Seattle, WA 98195, USA}

\author[0000-0002-4639-5285]{Eun-jin Shin}
\affiliation{Center for Theoretical Physics, Department of Physics and Astronomy, Seoul National University, Seoul 08826, Korea}

\author[0000-0002-0415-3077]{Alvaro Segovia-Otero}
\affil{Lund Observatory, Division of Astrophysics, Department of Physics, Lund University, SE-221 00 Lund, Sweden}

%TIER 4: Those who helped this paper or the Collaboration by providing guidance and comments

\author[0000-0002-4287-1088]{Oscar Agertz}
\affil{Lund Observatory, Division of Astrophysics, Department of Physics, Lund University, SE-221 00 Lund, Sweden}

\author[0000-0002-8638-1697]{Kirk S.~S.~Barrow}
\affiliation{Department of Astronomy, University of Illinois at Urbana-Champaign, Urbana, IL 61801, USA}

\author[0000-0003-2285-0332]{Corentin~Cadiou}
\affiliation{Lund Observatory, Division of Astrophysics, Department of Physics, Lund University, SE-221 00 Lund, Sweden}

\author{Avishai Dekel}
\affil{Center for Astrophysics and Planetary Science, Racah Institute of Physics, The Hebrew University, Jerusalem 91904, Israel}

\author[0000-0002-3817-8133]{Cameron Hummels}
\affiliation{TAPIR, California Institute of Technology, Pasadena, CA 91125, USA}

\author[0000-0003-4597-6739]{Boon Kiat Oh}
\affiliation{Department of Physics, University of Connecticut, U-3046, Storrs, CT 06269, USA}
\affiliation{Center for Theoretical Physics, Department of Physics and Astronomy, Seoul National University, Seoul 08826, Korea}

\author{Romain Teyssier}
\affil{Department of Astrophysical Sciences, Princeton University, Princeton, NJ 08544, USA}

\author{the {\it AGORA} Collaboration}
\affiliation{\rm \url{http://www.AGORAsimulations.org}}
\affiliation{\rm The authors marked with * as code leaders contributed to the article by leading the effort within each code group to perform and analyze simulations.} 
%\footnote{The list of authors is provisional and, after the corresponding authors, it is just in alphabetical order. Before the beta release of this paper, the order will be discussed with all the coauthors and a final list will be included.}

\begin{abstract}

In this fourth paper from the {\it AGORA} Collaboration, we study the evolution down to redshift $z=2$ and below of a set of cosmological zoom-in simulations of a Milky Way mass galaxy by eight of the leading hydrodynamic simulation codes. We also compare this {\tt CosmoRun} suite of simulations with dark matter-only simulations by the same eight codes. We analyze general properties of the halo and galaxy at $z=4$ and 3, and before the last major merger, focusing on the formation of well-defined rotationally-supported disks, the mass-metallicity relation, the specific star formation rate, the gas metallicity gradients, and the non-axisymmetric structures in the stellar disks. Codes generally converge well to the stellar-to-halo mass ratios predicted by semi-analytic models at $z\sim$2. We see that almost all the hydro codes develop rotationally-supported structures at low redshifts. Most agree within 0.5 dex with the observed MZR at high and intermediate redshifts, and reproduce the gas metallicity gradients obtained from analytical models and low-redshift observations. We confirm that the inter-code differences in the halo assembly history reported in the first paper of the collaboration also exist in {\tt CosmoRun}, making the code-to-code comparison more difficult. We show that such differences are mainly due to variations in code-dependent parameters that control the time-stepping strategy of the gravity solver. We find that variations in the early stellar feedback can also result in differences in the timing of the low-redshift mergers. All the simulation data down to $z=2$ and the auxiliary data will be made publicly available.

\end{abstract}

\keywords{cosmology: theory -- galaxies: formation -- galaxies: evolution -- galaxies: kinematics and dynamics -- galaxies: structure -- galaxies: ISM -- galaxies: CGM -- methods: numerical -- hydrodynamics}


\section{Introduction}\label{sec:intro}

\vspace{1mm}

Leveraging collaborative efforts, the {\it AGORA}  ({\it Assembling Galaxies of Resolved Anatomy}) High-resolution Galaxy Simulations Comparison Project\footnote{See the Project website at \url{http://www.AGORAsimulations.org/} for more information about the {\it AGORA} Collaboration and its previous papers.} 
has been, since 2012, a platform to 
%JPinstill confidence in 
determine the reliability and predictive power of modern simulations of galaxy formation and evolution.\footnote{For publicly available datasets, visit \url{http://www.AGORAsimulations.org/} or \url{http://flathub.flatironinstitute.org/agora/}. \label{dataset_url}}
%Several other international research teams presented code-to-code comparisons \citep[e.g.,]{https://ui.adsabs.harvard.edu/abs/2020ApJ...903...32F/abstract} oriented to 
With the analysis of the inter-code differences in a set of dark matter-only (DMO) cosmological simulations \citep[the flagship paper of the Collaboration,][Paper I hereafter]{Kim2014} and in the  isolated Milky Way (MW)-size disks when using the same stellar feedback \citep[][Paper II hereafter]{Kim2016}, the {\it AGORA} Collaboration built a robust infrastructure for the production and analysis of simulations obtained with any of the contemporary codes widely-used in the community. 
Using this infrastructure, the {\it AGORA} Collaboration was finally able to undertake the herculean task of producing, analyzing, and comparing a set of {\it cosmological} ``zoom-in'' simulations of a MW-mass galaxy, %JP at $z=4$, 
which was presented in the 2021 {\it AGORA}  paper \citep[][Paper III hereafter]{RocaFabrega2021}. 
This work used seven of the most common numerical codes for high-resolution cosmological galaxy simulations, carried down to redshift $z=4$; we named this set of simulations the {\tt CosmoRun} suite. 
This was the first such complex and time-consuming comparison of zoom-in cosmological galaxy simulations.
Furthermore, in Paper III we presented a new calibration strategy designed to reduce the number of free parameters when comparing results by different code groups with different feedback strategies. 
This calibration strategy was designed so that it can be easily adopted by new code groups to produce new simulation datasets, and make a fair comparison with the ones presented in Paper III. 

\begin{table*}
\vspace*{1mm}
\footnotesize
\caption{Stellar feedback implementation, metal species tracked, and the diffusion scheme adopted by each participant code group}
\vspace*{-2mm}
\centering
\begin{tabular}{c || c | c | c }
\hline\hline
Code & Stellar feedback\tablenotemark{\textdagger} & Tracked metals & Diffusion scheme \\
\hline
{\sc Art-I}$^*$ & T+K, RP &$\alpha$, Fe & diffusion simulated self-consistently \\
{\sc Enzo} & T &$\alpha$ & diffusion simulated self-consistently \\
{\sc Ramses} & T, DC &$\alpha$ &  diffusion simulated self-consistently \\
{\sc Changa}$^{**}$ & T+S &$\alpha$, Fe & diffused explicitly using the scheme of \citet{Shen2010} \\
{\sc Gadget-3} & T+K, RP, DC &$\alpha$, Fe & smoothed by the SPH kernel of each gas particle, mimicking metal diffusion\\
{\sc Gear} & T, DC  & $\alpha$, Fe & smoothed by the SPH kernel of each gas particle, mimicking metal diffusion \\
%{\sc Arepo}$^{***}$ & T+K &  $\alpha$, Fe   & a Diffusion simulated self-consistently and occurs through mass fluxes between Voronoi cells \\
{\sc Arepo}$^{***}$ & T &  $\alpha$   & diffusion simulated self-consistently through mass fluxes between Voronoi cells \\
{\sc Gizmo} & T+K  & $\alpha$ & diffusion simulated self-consistently via a kernel-based scheme using a set of unstructured cells\\
\hline
\end{tabular}
\tablenotetext{$\textdagger$}{\scriptsize T = thermal feedback, K = kinetic feedback, RP = radiation pressure, DC = delayed cooling, S = ``superbubble'', $^*$ = same as in the original {\tt CosmoRun} model in Paper III,  but now injecting slightly less momentum of $p = 2.5\times10^6\,{\rm M}_\odot\,{\rm km\,s}^{-1}$/SN, $^{**}$ = same as in the original {\tt CosmoRun} model in Paper III, but now injecting slightly less thermal energy of $E_{\rm thermal}= 3.25 \times10^{51}$ ergs/SN,  $^{***}$ = injecting $E_{\rm thermal}= 2 \times10^{52}$ ergs/SN of energy 3 Myr after star formation (see also Table \ref{tab:arepogeneral} and Appendix \ref{sec:code-arepo} for more information on the {\sc Arepo} run).  For more information on the stellar feedback run time parameters, see Table 1 in Paper III.}
\label{tab:1}
\vspace*{1mm}
\end{table*}

The original {\tt CosmoRun} suite included seven simulations obtained using the three AMR (adaptive mesh refinement) codes, {\sc Art-I}, {\sc Enzo}, and {\sc Ramses}; three SPH (smoothed particle hydrodynamics) codes, {\sc Changa}, {\sc Gadget-3}, and {\sc Gear}; and a meshless Godunov code {\sc Gizmo}. 
All code groups started their simulations from a common initial condition (IC) generated with {\sc Music} \citep[][]{Hahn2011},\footnote{The website is \url{https://www-n.oca.eu/ohahn/MUSIC/}.} and 
%JP run by keeping 
for all these codes kept the same physics prescriptions (e.g., UV background, gas cooling and heating, star formation parameters), although some variations were made in each code (see Paper III for details on these differences). 
Only the decisions concerning the stellar feedback prescription and metal production to be used were left to each of the code groups, which were asked to use prescriptions close to the most widely-used practice in each code community. 
Spatial resolution was $\sim$80 physical pc at $z = 4$ to resolve the internal structure of a target halo, and to make our physics prescriptions less reliant on platform-specific models.  

Currently, the {\it AGORA} Collaboration is analyzing the original {\tt CosmoRun} suite of simulations while producing new ones both by adding new code groups and stellar feedback strategies, and by pushing the original models down to lower redshifts (this paper: {\it AGORA} Paper IV). For the same simulations we are also analyzing the satellite galaxies (Jung et al. 2024 accepted: Paper V hereafter), and the circumgalactic medium (CGM; Strawn et al. 2023 accepted: Paper VI hereafter), among other Paper III follow-up projects. 

In the present paper, we present the results from the analysis of the global properties of the {\tt CosmoRun} suite down to low redshifts ($z\leq2$), and show the comparison with new models obtained by using the {\sc Arepo} code, and by changing the original stellar feedback prescriptions in {\sc Art-I} and {\sc Changa}. 
We focus on the analysis of the mass assembly history of our MW-mass halo and galaxy, and on the inter-code differences in the major merger epochs.
As in all of the previous {\it AGORA} comparisons, we emphasize that we do not intend to identify correct or incorrect codes, but to focus on juxtaposing different codes for physical insights and to learn how much scatter is expected among modern galaxy simulations.  

This paper is organized as follows. 
In Section~\ref{sec:codes}, we briefly summarize the main differences in the feedback implementations by each code group and the number of models we already have in our {\tt CosmoRun} library.\footnote{We encourage new code groups, and/or the groups using codes already in the Collaboration, to run the new {\tt CosmoRun} models with different feedback implementations, thus enlarging the library of available simulations.} 
In Section~\ref{sec:MergerTree} we present the process to get the halo mass assembly history using the {\sc  Rockstar}  halo finder, and show the resulting halo and galaxy merger trees. 
In Section~\ref{sec:lowz_cosmorun} we summarize the galaxy and halo main properties from $z=4$ to $z=2$ and below, and analyze the presence of rotationally-supported disks, the galactic morphology, and the mass-metallicity relation.
In Section \ref{sec:conc} we conclude the article with a discussion of the similarities and differences among the simulations, and lay out the groundwork for future comparisons using the low-redshift {\tt CosmoRun} data. In Section \ref{sec:agora} we give an overview of the past, present and future of the {\it AGORA} Collaboration.
Last but not least, in the Appendix we present the details of the new {\sc Art-I}, {\sc Changa}, and {\sc Arepo} models included in this work.

\section{Simulations} \label{sec:codes}

In Paper III, we described the main properties of the {\tt CosmoRun} suite of simulations, its common physics, and the calibration strategy. 
We also detailed the inter-code differences in the stellar feedback implementation, and metal production and diffusion. 
In Table~\ref{tab:1}, we summarize such differences, including information on the new {\sc Arepo} model.  
The calibration process for the new {\sc Arepo} model is fully described in  Appendix~\ref{sec:code-arepo}. %\textcolor{red}{discussion on Courant condition/timestep?}  (??)
Notice also that here we use slightly different feedback models for {\sc Art-I} and {\sc Changa} than the ones used in Paper III. The only differences with the old models are weaker kinetic feedback, and weaker thermal feedback, respectively (see Appendix~\ref{sec:ArtChanga} for detailed discussion), and, in {\sc Art-I}, a change in the condition for the minimum timestep at high redshift. 
Changes in the {\sc Art-I} and {\sc Changa} supernovae (SNe) feedback are to reach still a better convergence in the calibration step 4 than the one reached before ({\tt Cal-4}; see Paper III for details on the {\tt CosmoRun} calibration steps). 
These changes do not affect the results presented in Paper III, and both the old and new models will be made publicly available together, thus enriching the library of {\it AGORA} simulations. 
The change in the condition for the minimum timestep in {\sc Art-I} has been adopted to achieve better convergence on the halo merger trees presented in this work.

\begin{figure*}
        \centering
        \vspace{2mm}
        \includegraphics[width=1.01\linewidth]{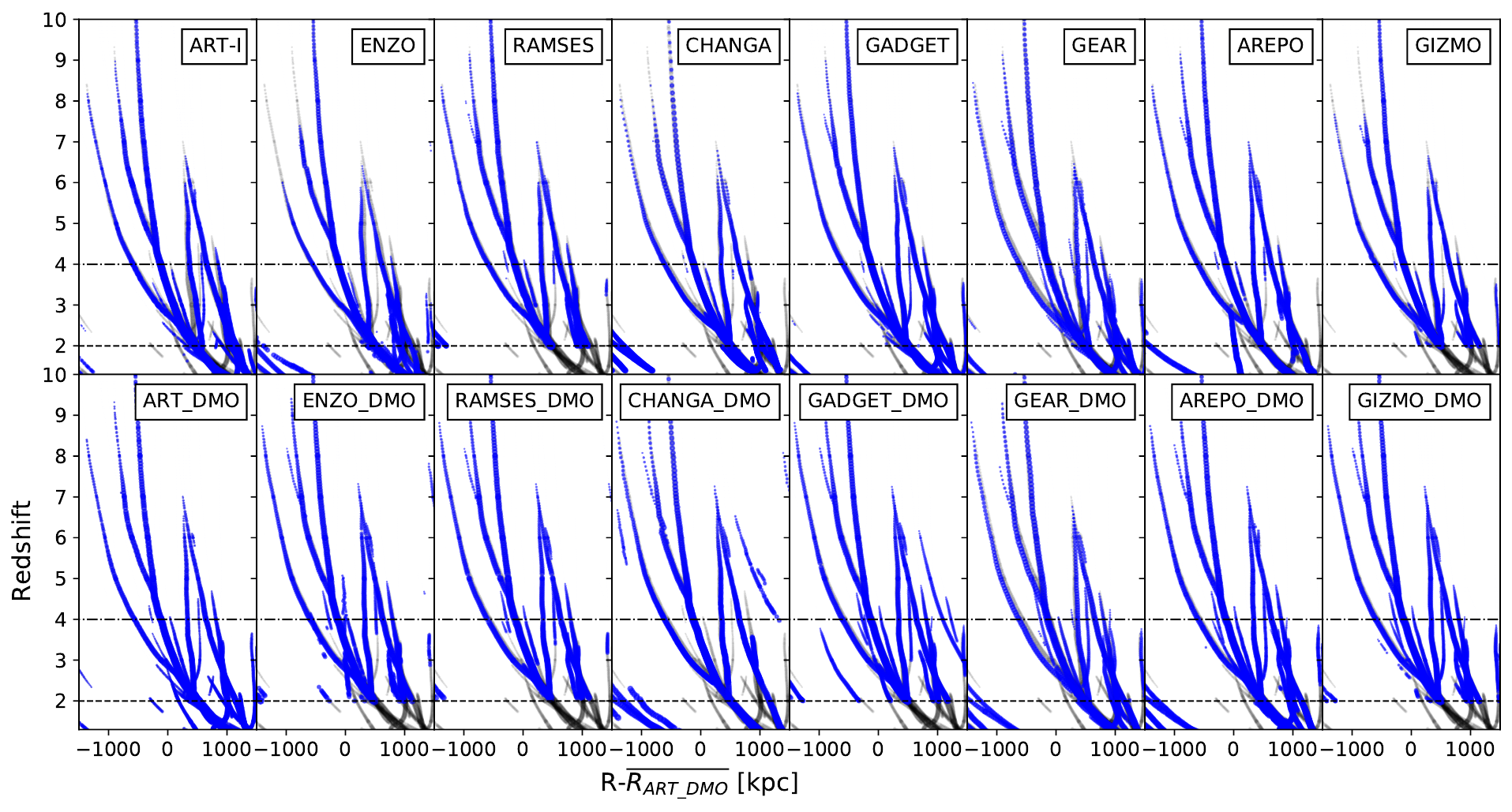}
        \vspace{-5mm}
        \caption{The merger history of the target halo and all subhalos with masses $>10^8 \, {\rm M}_{\odot}$  in the {\tt CosmoRun} simulations ({\it top row}) and the DMO runs ({\it bottom row}) between $z=10$ and 1.5. 
        In {\it gray circles}, we include the {\sc Art-I-dmo} merger tree to guide reader's eyes. 
        The diameter of the circle is proportional to the virial mass of each halo at each epoch (see the high-resolution version of this figure at the Project website \url{http://www.AGORAsimulations.org/} for better clarity on the circles and their size). 
        Horizontal {\it dot-dashed} and {\it dashed lines} indicate $z=4$ and $z=2$. 
%JP\textcolor{magenta}{Santi:      
       The $x$-axis shows the spherical radial position ($R$) of the target halo and the subhalos residing inside the zoom-in region in each run with respect to the median $R$ in the redshift interval from $z=15$ to 0 of the target halo in the {\sc Art-I-dmo}  (this is a reference point arbitrarily selected).  
        See Section \ref{sec:overview} for more information on {\tt CosmoRun} and this figure.
        	Simulations performed by:  Santi Roca-F\`{a}brega ({\sc Art-I}, {\sc Ramses}, and {\sc Art-I-dmo}), Ji-hoon Kim ({\sc Enzo}), Johnny Powell and H\'ector Vel\'azquez ({\sc Changa} and {\sc Changa-dmo}), Kentaro Nagamine and Ikkoh Shimizu ({\sc Gadget-3}), Loic Hausammann and Yves Revaz ({\sc Gear} and {\sc Gear-dmo}), Anna Genina ({\sc Arepo} and {\sc Arepo-dmo}), Alessandro Lupi and Bili Dong ({\sc Gizmo}), Hyeonyong Kim ({\sc Enzo-dmo}, {\sc Ramses-dmo}, {\sc Gadget-2-dmo}, and {\sc Gizmo-dmo}). 
	}
        \vspace{4mm}
        \label{fig:1}
\end{figure*}

All eight {\tt CosmoRun} simulations have been run at least down to $z=2$. 
The reader will notice that some of the codes were run down to lower redshifts; this was a decision made by each code group and is not indicative of differences in the computational costs or in the code performance. 
Throughout this paper, we show results at a lower redshift than $z=2$ for the codes that have this information available. 
In future papers on the Collaboration, we will present results at $z=0$ including only the code groups that decided to participate in the low-redshift comparison.\footnote{We warn the reader that our models do not include feedback from active galactic nuclei (AGN); therefore, some of the low-$z$ results will be somewhat unphysical, although useful for comparison with future more realistic models. For instance, while most observed massive galaxies ($M_{\rm vir}\gtrsim 10^{12}{\rm M}_{\odot}$ start to quench at $z\sim2$ because of the effects of AGN, in the {\tt CosmoRun} suite the quenching occur after starbursts following major mergers.}
In addition, in several analyses, we also present the results of the DMO simulations obtained by all 8 participating code groups. 
The ICs and parameters for these models are the same as the ones for the collisionless component in the {\tt CosmoRun} suite.

\section{Mass assembly history}\label{sec:MergerTree}

%JP\textcolor{magenta}{Santi: 
The current {\tt CosmoRun} repository contains more than 200 snapshots for each participant code from $z=15$ to $z=2$. Some of the models were kept running down to $z=0$; when these finish, the repository will contain a minimum of 336 snapshots per model equally spaced in the logarithm of the scale factor from $z=15$ to $z=0$.\footnote{Visit \url{http://physics.snu.ac.kr/cosmo/agora/output\_z\_cosmorun2.txt} for a full list of the output scale factors.} 
This thin spacing of snapshots provides us with good coverage of the mass assembly history of the main galactic system in the {\tt CosmoRun} models, allowing us to analyze its evolution in high fidelity. 
Using the {\sc Rockstar} halo finder interface  \citep{Behroozi2013} within the {\tt yt} analysis toolkit \citep{Turk2011}\footnote{The website is \url{http://yt-project.org/}.} we obtained the merger trees of the target halo for both the {\tt CosmoRun} simulations and the DMO simulations. 

\subsection{Overview}\label{sec:overview}

In Figure~\ref{fig:1}, in blue circles, we show the position of the target halo of our ``zoom-in'' simulation and its subhalos residing inside the zoom-in region of each simulation.  
The distance in the $x$-axis is with respect to the median position ($R$) of the target halo in the redshift interval from $z=15$ to 0 in the {\sc Art-I-dmo}  (this is a reference point arbitrarily selected). The figure shows the distances of the {\tt CosmoRun} simulations (top row) and the DMO simulations (bottom row), for each one of the participating code groups. From left to right, we show the grid-based (Eulerian) codes ({\sc Art-I}, {\sc Enzo} and {\sc Ramses}), the particle-based (Lagrangian) codes ({\sc Changa}, {\sc Gadget} and {\sc Gear}), followed by the moving-mesh code {\sc Arepo} and the meshless finite-volume code {\sc Gizmo}\footnote{Since in both {\sc Arepo} and {\sc Gizmo}, the resolution elements follow the fluid as in Lagrangian codes, while the Riemann problem is solved across interfaces of neighbouring cells as in Eulerian codes, we will refer to the two codes together as “moving-mesh” codes throughout this work, although we emphasize that there are fundamental differences in the way the simulation volume is partitioned  (Voronoi cells in {\sc Arepo} and a kernel-based partitioning in {\sc Gizmo})  see \citet{Hopkins2016}.}. 
In all panels, we include the {\sc Art-I-dmo} merger tree (gray circles), to guide reader's eyes. 

\begin{figure*}
        \centering
        \vspace{2mm}
        \includegraphics[width=1.01\linewidth]{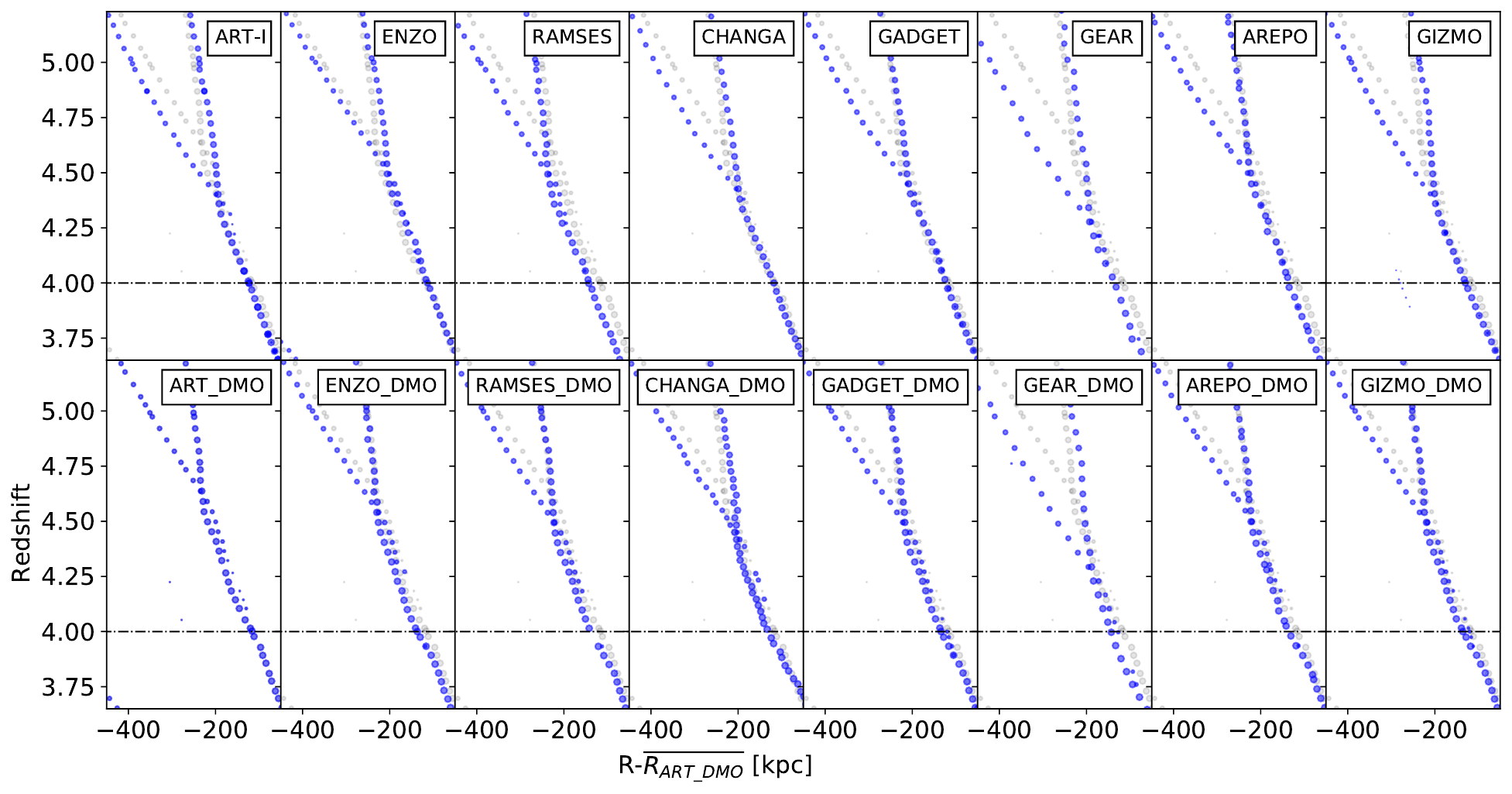}
        \vspace{-5mm}
        \caption{Zoom on the merger history of the target halo in all of the {\tt CosmoRun} simulations ({\it top row}) and the DMO runs ({\it bottom row}) around the $z\sim4.5$ major merger. 
        In {\it gray circles}, we include the {\sc Art-I-dmo} merger tree to guide reader's eyes.
        The sizes of the circles are proportional to the virial mass of each halo at each epoch. 
        Horizontal {\it dot-dashed lines} indicate $z=4$. 
      The $x$-axis shows the spherical radial position ($R$) of the target halo and the subhalos residing inside the zoom-in region in each run with respect to the median $R$ in the redshift interval from $z=15$ to 0 of the target halo in the {\sc Art-I-dmo}  (this is a reference point arbitrarily selected).
        See Section \ref{sec:highz} for more information on {\tt CosmoRun} and this figure.}
        \vspace{3mm}
        \label{fig:2}
\end{figure*}

Some differences observed in the number of branches and data points (e.g., {\sc Enzo} vs. {\sc Arepo} at $z> 6$) and in low-mass subhalos (e.g., small subhalos in the rightmost and leftmost ends of {\sc Changa-dmo} panel at $z\sim5-6$) are due to the process we follow to select the main branch of the merger tree from the  catalogs provided by {\sc Rockstar} rather than due to the intrinsic inter-code differences.\footnote{Members of the collaboration are now developing a new algorithm based on {\sc Rockstar} to improve the halo finding and the tracking of each halo and sub-halo. The results will be presented in a new paper of the collaboration and the code will be made public afterwards.}
Notice also that some code groups stopped their simulations at $z=2$ (i.e., {\sc Ramses} and {\sc Gizmo} in the top row, {\sc Enzo-dmo}, {\sc Ramses-dmo}, {\sc Gadget-3-dmo}, and {\sc Gizmo-dmo} in the bottom row), and for these simulations only the {\sc Art-I-dmo} merger tree is plotted below $z=2$.

In agreement with the results presented in Paper I of this Collaboration, the merger history and halo growth in DMO models are almost identical across all participating codes with a few exceptions (see the discussion in the following sections). 
This identity can be seen in the bottom panels where the guiding merger history of {\sc Art-I-dmo} (plotted in gray) is almost always overlapping with each merger history (plotted in blue). 
Interestingly, some changes appear when baryonic physics is included. 
In the top row, we show that, although the main merging processes are present in all codes, their times can change, sometimes dramatically. 
This difference in merger times is better presented in Figure~\ref{fig:2}, an enlarged version of part of Figure~\ref{fig:1}.
%JPand is especially evident when comparing e.g., {\sc Art-I} with {\sc Art-I-dmo}  (first column). Joel: I omitted this because ART is no longer an outlier
We find that the two major sources of these differences are the small timing mismatch in the numerical integration of the equations of motion, and the small variations in the angular momentum distribution after the first stars are created at high redshift. 
Later in this section and in Appendix~\ref{subsec:timdisc} we will discuss these two mechanisms for merger timing discrepancies in more detail.
To have a better picture of this so-called ``timing discrepancy'' of the major mergers among the codes, in Sections \ref{sec:highz} and \ref{sec:lowz} we focus on two of the last major mergers experienced by the target halo, at $z\sim 4.65$ and $z\sim 2.15$ (redshifts measured in the {\sc Art-I-dmo} run).

\begin{figure*}
        \centering
        \vspace{2mm}
        \includegraphics[width=1.01\linewidth]{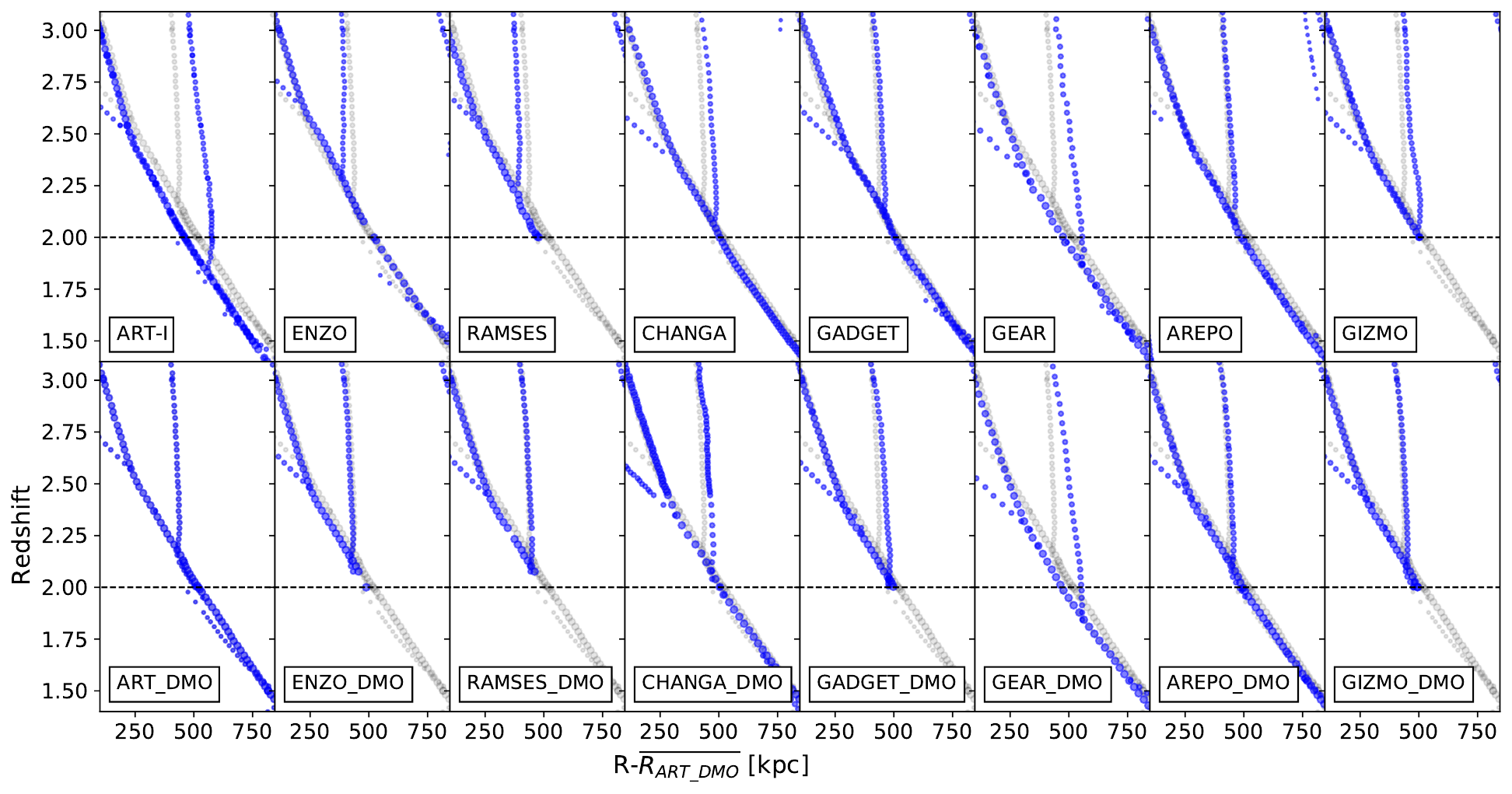}
        \vspace{-5mm}
        \caption{Zoom on the last merger in the merger history of the target halo in all of the {\tt CosmoRun} simulations ({\it top row}) and the DMO runs ({\it bottom row}) around the $z\sim2.2$ major merger. 
        In {\it gray circles}, we include the {\sc Art-I-dmo} merger tree to guide reader's eyes.
        The diameters of the circles are proportional to the virial mass of each halo at each epoch. 
        Horizontal {\it dashed lines} indicate $z=2$. 
      The $x$-axis shows the spherical radial position ($R$) of the target halo and the subhalos residing inside the zoom-in region in each run with respect to the median $R$ in the redshift interval from $z=15$ to 0 of the target halo in the {\sc Art-I-dmo}  (this is a reference point arbitrarily selected).
        See Section \ref{sec:lowz} for more information on {\tt CosmoRun} and this figure.}        
        \vspace{3mm}
        \label{fig:3}
\end{figure*}

\subsection{Major merger at $z\sim 4.5$}\label{sec:highz}

We now focus on the major merger that occurs at $z\sim4.65$ in the DMO simulations. 
In Figure~\ref{fig:2} we show the evolution of the two main branches of the target halo's merger tree between $z=3.8$ and 5.2. 
As in Figure~\ref{fig:1}, we show the merger tree of the labelled code, and as a guiding line the one corresponding to the {\sc Art-I-dmo} (bottom-leftmost panel) in gray.
From this figure, we conclude that just as in Figure~\ref{fig:1}, the merger trees differ slightly from one code to the other, even amongst the DMO simulations. 
The discrepancy is particularly evident in both the {\sc Gear} {\it and} the {\sc Gear-dmo} run, where the major merger occurs at $z\sim4.2$ instead of the $\sim4.5$ observed in most other codes. 
This difference between DMO simulations indicates that the origin of the timing discrepancies does not reside only in baryonic physics but also in the integration of the equations of motion. 
In Appendix~\ref{subsec:timdisc}, we show that the origin of these timing discrepancies in {\sc Gear}  and {\sc Gear-dmo} models is the choice of user parameters, not the code itself.
Another interesting feature in Figure~\ref{fig:2} is the coexistence of the two main halos for more than 0.5 redshift units, only in the DMO runs (AMR codes also show this coexistence but for a much shorter timespan). This coexistence is seen in the bottom row where the two halos are seen as independent structures after the first interaction (first crossing of the two blue paths in these diagrams) and for more than 0.5 redshift. 
After this time the secondary halo disappears and only a single blue dot path remains. 
The difference in the coalescence times between the DMO run and the {\tt CosmoRun} (DM+baryons) may be related to changes in the dynamical friction due to the presence of the collisional component \citep[e.g.][]{Zhang2016}, or to changes on the dark matter profile \citep[see, e.g.,][]{Duffy2010,Chua2019} that makes it more difficult for the {\sc Rockstar} halo finder to find merging halos in the {\tt CosmoRun} models. 
This topic will be studied further in a future {\it AGORA} paper that will focus on the CosmoRun mergers.

\subsection{Major merger at $z\sim2.2$}\label{sec:lowz}

In Figure~\ref{fig:3}, we analyze the target halo's last major merger. 
This merger occurs at $z\sim 2.15$ in the DMO models, but with some small variations among the codes (see the bottom row). 
The {\tt CosmoRun} simulations (i.e. including baryons) show larger variations (top row). These differences between the DMO and the {\tt CosmoRun} merger times suggest that the small timing discrepancies at low redshift are likely dictated by baryonic physics. 
An interesting exception is the {\sc Gear} code that shows an almost identical merger tree for both models with and without baryons, shifted towards a lower redshift concerning the other codes, a trend consistent with what was shown in Section~\ref{sec:highz}.

Also noticeable is that in the {\tt CosmoRun} simulations, the particle-based codes and the ``moving-mesh'' codes are shifted towards slightly lower redshift than the mesh-based codes in both major mergers (see Figures~\ref{fig:2} and \ref{fig:3}; except for {\sc Art-I} in Figure \ref{fig:3}). 
The AMR codes have intrinsically different resolution schemes from the particle-based codes or the ``moving-mesh'' codes (see Section 4 in Paper III for the choices of numerical resolution in the mesh-based and particle-based codes), which may explain the difference in merger timings. %, thus, this also points towards the baryonic physics being the cause of the timing discrepancies observed in most codes except for {\sc Gear} (described above) and for {\sc Art-I}.   (??)
The timing discrepancies in the {\sc Art-I} {\tt CosmoRun} at low redshift are particularly interesting. 
By including baryons, the timing of the major merger shifts from $z\sim2.1$ to $\sim1.8$, the largest shift among all the codes.
In the Appendix~\ref{subsec:timdisc}, we show that for {\sc Art-I} the timing discrepancies originated from the variations in the timesteps of the integration of the equations of motion at high redshift.  

\begin{figure*}
        \centering
        \vspace{1mm}
        \includegraphics[width=1.0\linewidth]{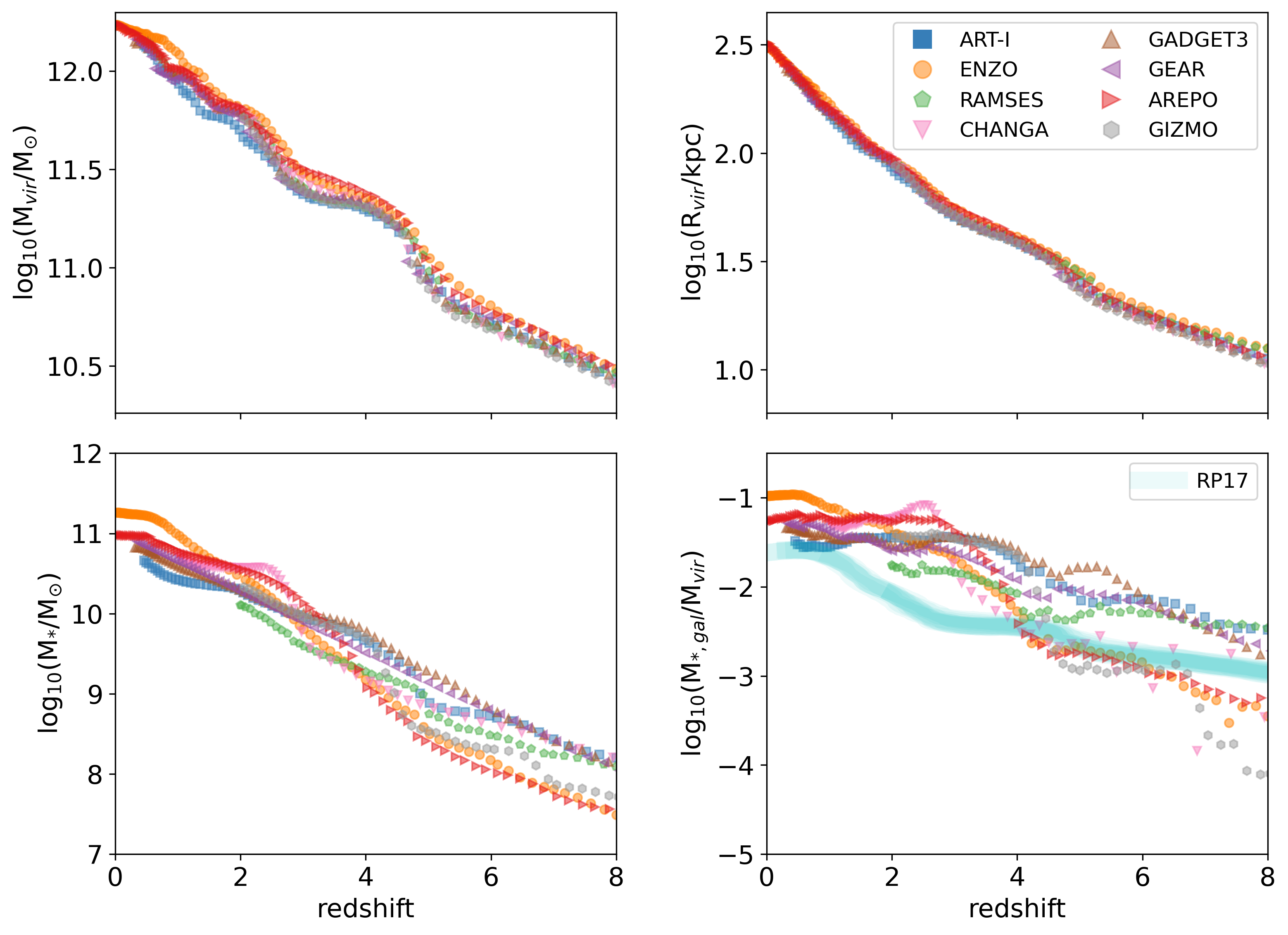}
        \vspace{-6mm}
        \caption{From the top-left to the bottom-right: the virial mass $M_{\rm vir}$, virial radius $R_{\rm vir}$, total stellar mass within $R_{\rm vir}$ $M_\star$, and the ratio of the galactic stellar mass within $R < 0.15 R_{\rm vir}$ with respect to the halo mass,  $M_{\star, \rm gal}/M_{\rm vir} $, as functions of redshift. 
        	The {\it cyan shadows} in the bottom right panel indicate the average stellar mass predicted by a semi-empirical model for the halo with the same total mass as the one in each simulation, based on an abundance matching technique by  \citet{RodriguezPuebla17}.
        See Section \ref{sec:massevolution} for more information on {\tt CosmoRun} and this figure.}	
        \label{fig:Mhz}
        \vspace{2mm}
\end{figure*}

\vspace{1mm}

\section{The {\it AGORA} \texorpdfstring{\tt C\MakeLowercase{osmo}R\MakeLowercase{un}}{CosmoRun} {\rm Simulations At Redshifts $z < 4$}}  \label{sec:lowz_cosmorun}

Following the presentation of the {\tt CosmoRun} simulations to redshift $z=4$ in Paper III, all code groups had the opportunity to run those simulations further down to lower redshifts. 
We currently have available data down to $z=2$ for all participating codes and, for some of them, also down to even lower redshifts.\footnote{The lowest redshift achieved by each code group does not reflect the performance of the code but instead the availability of manpower and CPU time at the computational facilities each group had  access to.} 
In this section, we show the general properties of the simulated galactic systems 
%JP at low redshift.  
down to the lowest redshifts presently available.
Companion papers from the {\it AGORA} Collaboration present further {\tt CosmoRun} results: the  distributions and properties of the satellite galaxies in Paper V (Jung et al. 2024 accepted) and the circumgalactic medium in Paper VI (Strawn et al. 2023 accepted).

\begin{table}
\vspace*{1mm}
\footnotesize
\caption{Redshifts we use in this paper that are computed for each code for the analyses in Section \ref{sec:lowz_cosmorun}, surrounding the time of the last major merger of the target halo.  See Section \ref{sec:lowz_cosmorun} for each redshift definition.}
\centering
\vspace*{2mm}
\begin{tabular}{c || c | c | c}
\hline\hline
Code & $z_{\rm blmm}$ & $z_{\rm lmm}$ & $z_{\rm almm}$ \\
\hline
{\sc Art-I} & 2.40  & 1.85 & 1.65\\
{\sc Enzo} &  2.80 & 2.35 & 2.10\\
{\sc Ramses} & 2.80 & 2.25 & 2.05\\
{\sc Changa} & 2.60  & 2.15 & 1.95\\
{\sc Gadget-3} & 2.70 & 2.20 & 2.00\\
{\sc Gear} &  2.40  & 1.90 & 1.70\\
{\sc Arepo} & 2.60 & 2.20 &  2.00\\
{\sc Gizmo} & 2.60 & 2.05 & 1.85\\
\hline
\end{tabular}
\label{tab:0}
%\vspace*{1mm}
\end{table}

\begin{table*}
\footnotesize
\vspace{2mm}
\caption{\footnotesize Global properties of the target galaxy progenitor in the {\it AGORA} {\tt CosmoRun} simulation suite at $z=4$, 3, $z_{\rm blmm}$\tablenotemark{\textdaggerdbl}, and 2.}
\centering
\npdecimalsign{.}
\nprounddigits{2}
\begin{tabular}{c c || cccccccc}
\hline\hline
Code & redshift $z$ & $R_{\rm vir}^{\,\,\,(a)}$\tablenotemark{\textdagger} & $M_{\rm vir}^{\,\,\,(b)}$ & $M_\star^{\,\,\,(c)}$ & $M_{\rm gas}^{\,\,\,(d)}$ & $M_{\rm gas, \,gal}^{\,\,\,(e)}$ & $M_{\rm gas, \,CGM}^{\,\,\,(f)}$ & log$_{10}(M_{\rm \star, \, gal} / M_{\rm vir})^{\,(g)}$ & log$_{10}(M_{\star, \rm hout}/M_*)^{\,\,\,(h)}$ \\
& & [kpc] & [10$^{11}\,$M$_{\odot}$] & [10$^{10}\,$M$_{\odot}$] & [10$^{10}\,$M$_{\odot}$] & [10$^{10}\,$M$_{\odot}$] & [10$^{10}\,$M$_{\odot}$] & & \\
\hline
{\sc Art-I} & 4  & 38.2   &  1.95  &  0.45  & 1.72   &  0.39  &  1.34   &  -1.75 &  -0.65 \\
&  3 & 50.9   &  2.37   &   0.90  &  2.05   &  0.69  &  2.06   &  -1.44  &  -1.26  \\
& $z_{\rm blmm}=2.4$  &  69.2  &   3.71  &  1.38   &  2.66   &   0.60  &  2.06   &   -1.48  &  -1.02  \\
& 2  & 85.0   &  4.94   &  1.94   &  3.11  &  0.73   &  2.38   &   -1.45   &  -1.06 \\
\hline
{\sc Enzo} & 4  & 41.1   &  2.27  &  0.15  & 2.34 & 0.54  &  1.79  & -2.27 & -0.70  \\
& 3  & 56.0 & 3.04  &  0.66   &  3.51   &  1.48  &   2.03  &    -1.68 &  -1.52 \\
&  $z_{\rm blmm}=2.8$ &  61.4  &  3.64   &   0.92  & 5.27   &  2.93   &   2.34  &   -1.62   &  -1.36 \\
& 2  & 93.2   &  6.54   &   2.88  &  7.30  &   4.87  &   2.53  &    -1.37  & -1.61  \\
\hline
{\sc Ramses}  & 4  & 39.4   &  2.02  & 0.19   & 1.29 & 0.39  & 0.90   &  -2.07 & -0.98  \\
&  3 & 53.2   &  2.57   &  0.40   &  1.84   &  0.48  &  1.36   &   -1.84  &  -1.30 \\
& $z_{\rm blmm}=2.8$  &  57.6  &  2.86   &  0.47   & 3.26   &  1.21   &    2.04 &   -1.81   &  -1.19 \\
& 2  & 92.5   &  6.14   &  1.28   &  5.74  &   2.24  &   3.50  &    -1.76  &  -0.81 \\
 \hline
{\sc Changa}  & 4  & 38.4   &  2.10  & 0.15  & 3.25 & 0.57  & 2.68   &  -2.38 & -1.18 \\
& 3  & 52.9   &  2.75   &   0.52  &  4.93   &  2.39  &  2.53   &   -1.80 &   -0.87 \\
& $z_{\rm blmm}=2.6$  &  62.4  &   3.40  &   2.87  & 3.87   & 2.19    & 1.68    &   -1.09  &  -1.50  \\
& 2  & 92.3   &   6.25  &  3.73   & 6.20   &  2.15   &  4.05   &   -1.24   &  -1.53 \\
\hline
{\sc Gadget-3}  & 4  & 38.9   &  2.11  &  0.59  & 1.98 & 0.75  &   1.23 & -1.59 & -1.07  \\
& 3  & 51.8   &  2.54   &  1.01   &  2.12   &  0.87  &   1.24  &    -1.44 &  -1.05 \\
& $z_{\rm blmm}=2.7$  &  58.4  &  2.93   &  1.17   &  3.78  &  1.43   &  2.34   &    -1.44  &  -0.99 \\
& 2  & 92.2   &  6.16  &  1.96   &  7.22  &  1.93  &   5.29  &   -1.54  & -1.06 \\
\hline
{\sc Gear}  &  4 & 38.5   & 2.05   &  0.32  & 2.74 &  1.36 &  1.38  &  -1.92 & -0.60 \\
&  3 & 50.9   &  2.41   &  0.78   &   3.21  &  2.51  &  0.70   &   -1.61 &  -0.61  \\
&  $z_{\rm blmm}=2.4$ &  70.2  &   3.97  &  1.32   & 5.82   &  2.88   & 2.94   &   -1.62 &   -0.58  \\
& 2  & 89.1   &  5.71   &  1.88  &  7.63  &  3.73   &   3.90  &   -1.60  &  -0.63  \\
 \hline
{\sc Arepo}  & 4  & 40.2   &  2.38 &  1.20 & 2.93 &  0.87 &     2.06 & -2.41 &  -1.78 \\
& 3  & 56.4   &  3.18  &   1.34  &  3.24   &  1.89  &  1.35  &   -1.38  &   -1.65 \\
&  $z_{\rm blmm}=2.6$ &  77.1  & 4.90    &   2.82 &  3.65  &  1.67   &   1.98  &   -1.24  &  -2.45  \\
& 2  & 93.9   &  6.49  &  3.58  &  4.41  &  1.85  &   2.56  &   -1.26  & -1.95  \\
\hline
{\sc Gizmo}  &  4 & 38.2   &  2.03  &  0.49  & 1.16 &  0.20 &  0.97  &  -1.65 & -1.09 \\
& 3  & 51.5   &  2.48   &   0.92  &  1.97   &  0.55  &  1.42   &   -1.46 &   -1.26 \\
&  $z_{\rm blmm}=2.6$ & 60.3   &  3.06   &   1.27  &  3.24  & 1.48   &   1.76  &    -1.40 &  -1.33  \\
&  2 & 91.3   &  6.02   &    2.25 &  5.66  &  2.47   & 3.19    &    -1.43 &  -1.65   \\
\hline
\end{tabular}
\tablenotetext{$\textdagger$}{\scriptsize Each column lists the following quantities at the corresponding redshift: $^{(a)}$virial radius, $^{(b)}$total halo mass, $^{(c)}$total stellar mass within $R_{\rm vir}$, $^{(e)}$gas mass inside the $R_{\rm vir}$, $^{(e)}$gas mass inside the main galaxy or the interstellar medium (ISM, which we define as regions with $R < 0.15 \,R_{\rm vir}$), $^{(f)}$gas mass in the CGM  (which we define as regions with $0.15 \,R_{\rm vir} < R < R_{\rm vir}$),  $^{(g)}$the ratio of stellar mass (in the main galaxy) to halo mass,  $^{(h)}$ the ratio of the outer stellar halo mass $M_{\star, \rm hout}$ (i.e. in $0.15 \,R_{\rm vir} < R < \,R_{\rm vir}$) and the total stellar mass $M_{\star}$.  See also footnote \ref{fnlabel}.}
\tablenotetext{$\textdaggerdbl$}{\scriptsize The $z_{\rm blmm}$ values for each of the codes in the {\tt CosmoRun} suite can be found in Table \ref{tab:0}.}
\label{tab:2}
\vspace*{2mm}
\end{table*}

In Section~\ref{sec:MergerTree} (and also in Appendix~\ref{subsec:timdisc}) we have shown the inter-code timing discrepancies in the {\tt CosmoRun}. 
To account for the merger-induced perturbations occurring at different epochs in different codes, we define three redshifts for the following discussion {\it (a)} $z_{\rm blmm}$ as a redshift at which the target halo remains isolated and stable before its last major merger event ({\it blmm}), i.e., when the centers of the two main ancestor halos are 3$R_{\rm vir}$ ($\sim250$~kpc) apart from each other (see Appendix~\ref{subsec:timdisc}). 
We also define {\it (b)} $z_{\rm lmm}$ as the redshift at which the major merger occurs (i.e., the centers of the two merging halos are closer than 5~kpc for the first time), and {\it (c)} $z_{\rm almm}$ as a redshift at which the coalescence is complete (i.e., $\sim$300~Myr after $z_{\rm lmm}$).  
These redshift values for each code are listed in Table~\ref{tab:0}. We analyze the galaxy properties at $z_{\rm blmm}$ as well as at several epochs from $z=4$ to 1.

\subsection{Evolution of the Total and Stellar Masses}\label{sec:massevolution}

In Figure~\ref{fig:Mhz} we show the evolution of the target halo's virial mass $M_{\rm vir}$ and virial radius $R_{\rm vir}$\footnote{Throughout this paper the virial radius, $R_{\rm vir}$, is defined as the radius of a sphere enclosing a mass of mean density $\bar\rho$ = $\Delta_{\rm vir}(z)\rho_c(z)$, where $\rho_c(z)$ is the critical density of the universe at 
%JP a given epoch
redshift $z$, and $\Delta_{\rm vir}$(z) is the density contrast defined in \citet{BryanNorman98}. The virial mass, $M_{\rm vir}$, is then defined as the mass inside $R_{\rm vir}$.  Notice that Table 2 in Paper III shows the $R_{200}$ values instead of $R_{\rm vir}$ used in Table~\ref{tab:2} here. \label{fnlabel}} in the top row, and the total stellar mass $M_\star$ and the ratio of the galactic stellar mass with respect to the halo mass,  $M_{\star, \rm gal}/M_{\rm vir} $, in the bottom row, as functions of redshift. 
In the top two panels, we illustrate that the evolution of the total halo mass and the virial radius show good convergence at high redshift for all codes ($z>6$), and also at low redshift when sufficiently far from the major merger events at $z\sim 4.5$ and $\sim 2.2$. 
%As expected, {\sc Art-I}, which shows a discrepancy in the major merger timings at $z\sim2.2$ (as described in Section~\ref{sec:lowz}), also exhibits the largest offset in the $M_{\rm vir}$ and $R_{\rm vir}$ evolution at $1<z<2$ with respect to the other codes. 
%\textcolor{red}{(Joel: OK?  Or should we not single out ART since
%its timing discrepancy is much reduced?)}
Our {\tt CosmoRun} results confirm the conclusion presented in Paper I on the overall convergence in the halo mass assembly history, but also highlight the effect of the timing discrepancies on the assembly of a simulated MW-mass halo.

\begin{figure*}
    \vspace{2mm}
    \centering
    \includegraphics[width = 1.01\linewidth]{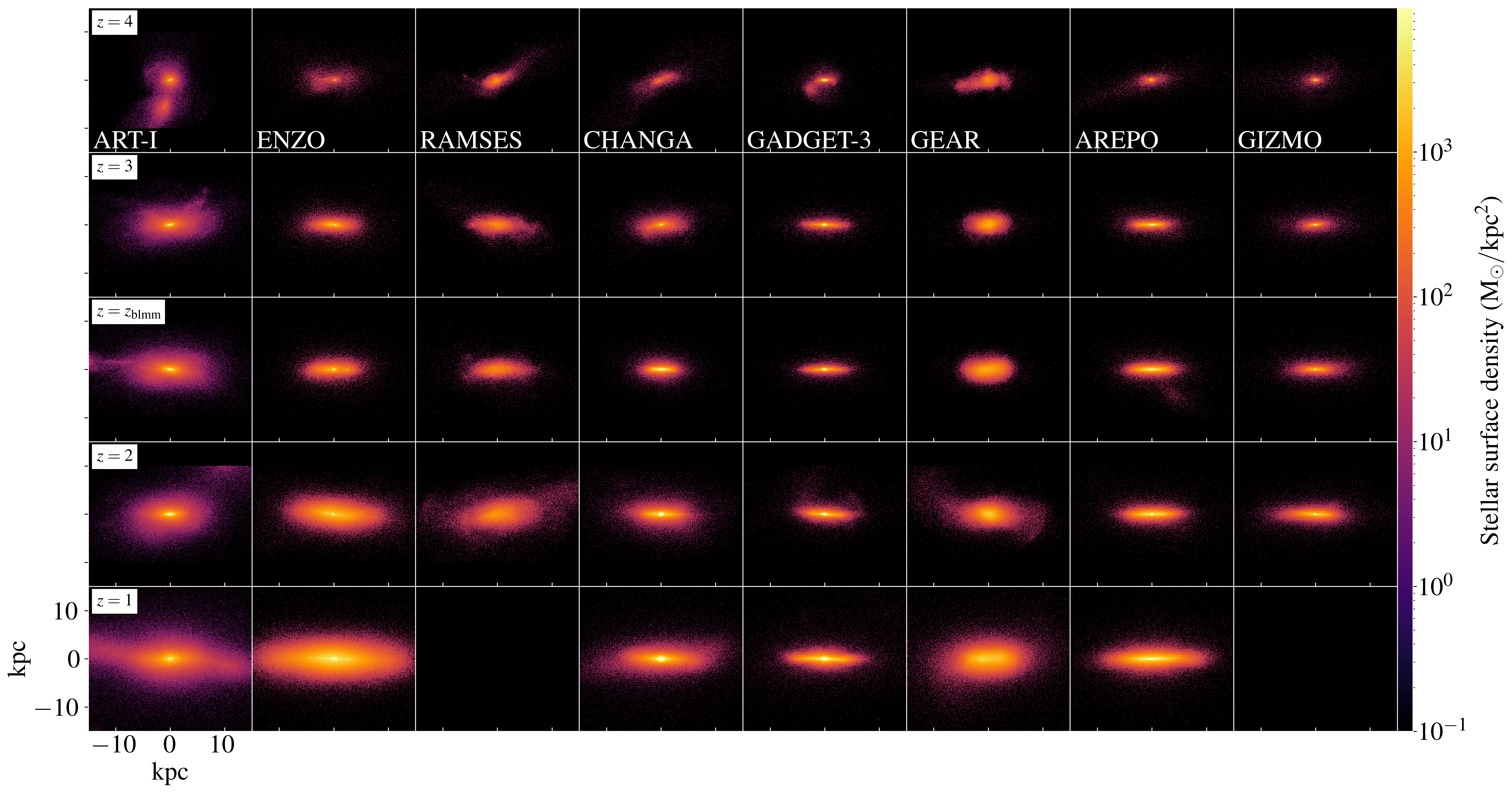}
    \vspace{-5mm}    
    \caption{Edge-on views of the projected stellar density of the {\tt CosmoRun} galaxies from $z=4$ ({\it top}) to $z=1$ ({\it bottom}).
     While some codes (e.g., {\sc Gadget-3}, {\sc Arepo}, and {\sc Gizmo}) form an apparent thin disk and bulge before $z=3$, other codes do not form these structures even at $z=2$.
     See Section \ref{sec:stellar_morph} for more information on {\tt CosmoRun} and this figure.}	     
    \label{fig:star_snap_edgeon}
    \vspace{2mm}    
\end{figure*}

In the bottom left panel of Figure~\ref{fig:Mhz}, we show the evolution of the total stellar mass inside $R_{\rm vir}$. 
Unlike Figure 12 of Paper III where we extract the stellar mass growth history from stellar ages or creation times at a single snapshot, here we show the real stellar mass inside $R_{\rm vir}$ at each epoch, which is the mass of the stellar particles after suffering the effects of stellar evolution (i.e. mass loss through stellar winds and SNe).  
Therefore, we show the real stellar mass growth (SMG), not the stellar mass assembly history (SMAH). 
Even though our stellar mass calibration was made at $z=4$ (see Paper III, Section 5 for details on the calibration process), we confirm that the total stellar mass still converges within 0.5$-$1.0 dex long after the last calibration redshift. 
One noticeable feature of the SMG is that some models are more sensitive to disruptions by major mergers than others. 
In particular, the merger at $z\sim4.5$ induces a strong star formation event in {\sc Ramses}, {\sc Arepo}, and {\sc Gizmo},  while the one at $z\sim2.2$ does that only in {\sc Changa}. 
The other codes show gentle changes in the SMG after the mergers (e.g., in {\sc Art-I} and  {\sc Gadget-3}) or no change at all (e.g., in {\sc Gear}). 
Also, it should be noted that our chosen target halo is somewhat atypical as it grows quickly with only a few mergers at low redshift. 

The early assembly of stellar mass can also be seen in the bottom right panel of Figure~\ref{fig:Mhz}, where we show the ratio of the stellar mass inside the galactic system (i.e., within 0.15 $R_{\rm vir}$) to the halo mass $M_{\rm vir}$. 
As a comparison, we also show in this panel the average stellar mass predicted by a semi-empirical model for the halo with the same total mass as the one in each simulation, based on an abundance matching technique by  \citet{RodriguezPuebla17} (cyan shadows). 
Most codes agree well with predictions from a semi-empirical model at high redshift, but they overproduce stars from $z\sim 2$ to 4. 
Nevertheless, they again match the semi-empirical model at $z\sim$1 after experiencing a period of quenched star formation. 

In Table~\ref{tab:2}, we list the masses of the dark matter, stellar, and gas content in the target halo at $z=4$, 3, $z_{\rm blmm}$, and 2 (see Table~\ref{tab:0} for information on the $z_{\rm blmm}$ epoch). Notice that in this paper, we used $R_{\rm vir}$ as the reference radius to compute these quantities instead of the $R_{200}$ used in Table 2 of Paper III. The total stellar mass, labelled $M_\star$, includes the stellar mass of the central galaxy ($M_{\rm \star, \, gal}$, that is the stellar mass inside $0.15R_{\rm vir}$) and all satellites, streams and halo stars inside $R_{\rm vir}$. 
The first discrepancy we find in this table is that {\sc Art-I} and {\sc Gear} show a slightly smaller virial radius than the other codes at low-z. Consistently, they also have a smaller total virial mass. These discrepancies are a direct consequence of the timing discrepancies presented in the previous section and are well illustrated in Figure~\ref{fig:3}. These two codes have the last major merger below $z=2$ while the others have it at higher-z.
{\sc Enzo} and {\sc Ramses} also show differences with the other codes but in the opposite direction. These codes show relatively small stellar masses at $z_{\rm blmm}$ (that is $= 2.80$ for these two codes) when comparing with all other codes but {\sc Gizmo}, which is in line with the fact that they experienced the major merger at higher-z than the other codes (i.e., the target halo has had less time to produce stars {\it in-situ}).
Differences in the gas content in both the galaxy and the CGM are noticeable at all redshifts, although they are of the same order of magnitude in all simulations at all epochs. 
Earlier, from the SMG in Figure~\ref{fig:Mhz} we showed that some simulations are more responsive to the major mergers than others. 
From this, one would have expected that the codes with stronger star formation bursts would exhibit  larger gas content.  
However, this is not what we find in Table~\ref{tab:2}. 
Although further analysis of the effects of mergers on the host galaxy will be presented in a future paper from the {\it AGORA} Collaboration, here we confirm that the gas content is not a good proxy for how responsive the galactic star formation is to mergers (for example, compare the gas contents of  {\sc Enzo} and {\sc Gear} at $z=z_{\rm blmm}$ and 2 in Table~\ref{tab:2}, and their SMGs in the bottom-left panel of Figure~\ref{fig:Mhz}).  
Nonetheless, gas is the component that is heavily dependent on the stellar feedback implementations. 
Deeper analyses of the gas content and properties (focusing on the CGM) are presented in Paper VI (Strawn et al. 2023 accepted) of the Collaboration.

In the last column of Table~\ref{tab:2}, we present the ratio between the outer stellar halo mass and the total stellar ($M_{\star, \rm hout}/M_{\star}$). Several recent papers studied what is called the ``missing outskirts'' problem which was first presented in \citet{Merritt2020}. 
In this paper, the authors found that the stellar surface density of the outskirts of MW-mass galaxies in the {\sc TNG100} suite was several dex higher than the ones obtained in the observations by Dragonfly Nearby Galaxies Survey (DNGS) \citep{Merritt2016}. More recently, \cite{Keller2022b} using zoom-in cosmological simulations of MW-mass galaxies showed that changes in the stellar feedback can reduce this tension and almost match the observations. The results presented here show that, independently of the code and the stellar feedback used, in all {\tt CosmoRun} simulations, the outer stellar halo masses agree well with the range of values obtained by available models that is log(M$_{*,{\rm hout}}/$M$_*$)~$={\rm-1}-{\rm-3}$ \citep[e.g., see Figure 3 in][]{Wright2023}.
In other words, there is no sign of an ``missing outskirts'' problem. Only the  {\sc Gear} code shows a fraction that is systematically 0.5 dex above the results presented by \citet{Wright2023}. This result may be a consequence of that stellar systems in the {\sc Gear} model are less compact (see Section~\ref{sec:stellar_morph}) and, thus, satellite galaxies can be disrupted faster and incorporated into the external stellar halo. A deeper analysis of the stellar density and luminosity profile of the outer halo in the {\tt CosmoRun} simulations will be presented in a future paper from the Collaboration.

\subsection{Stellar Disk Properties}\label{sec:stellar_dist}

In this section, we analyze the {\tt CosmoRun} galaxies' morphology. 
In particular, we study the stellar density profiles and look for the presence of rotationally-supported disks at $z=4$, 3, z$_{\rm blmm}$, and 2 for all participant codes, and at $z=1$ for the codes that have reached that redshift.

\subsubsection{Stellar Disk Morphology}\label{sec:stellar_morph}

It is well known that the formation of rotationally-supported disks is highly dependent on the mass assembly history of the halo and on the gas available to cool down and fall deep inside the halo potential well where to get denser and form new stars \citep[see, e.g.,][]{Governato07}. 
Here, for the first time, we study the formation of disks in galaxy-scale ``zoom-in'' simulations that started from the same ICs, that share the same merger history, and that have the same amount of intergalactic medium (IGM) gas available for accretion. 
We show that because of the variation in stellar feedback and the intrinsic inter-code differences a wide variety of morphologies are obtained.
 
\begin{figure*}
    \vspace{2mm}
    \centering
    \includegraphics[width = 1.01\linewidth]{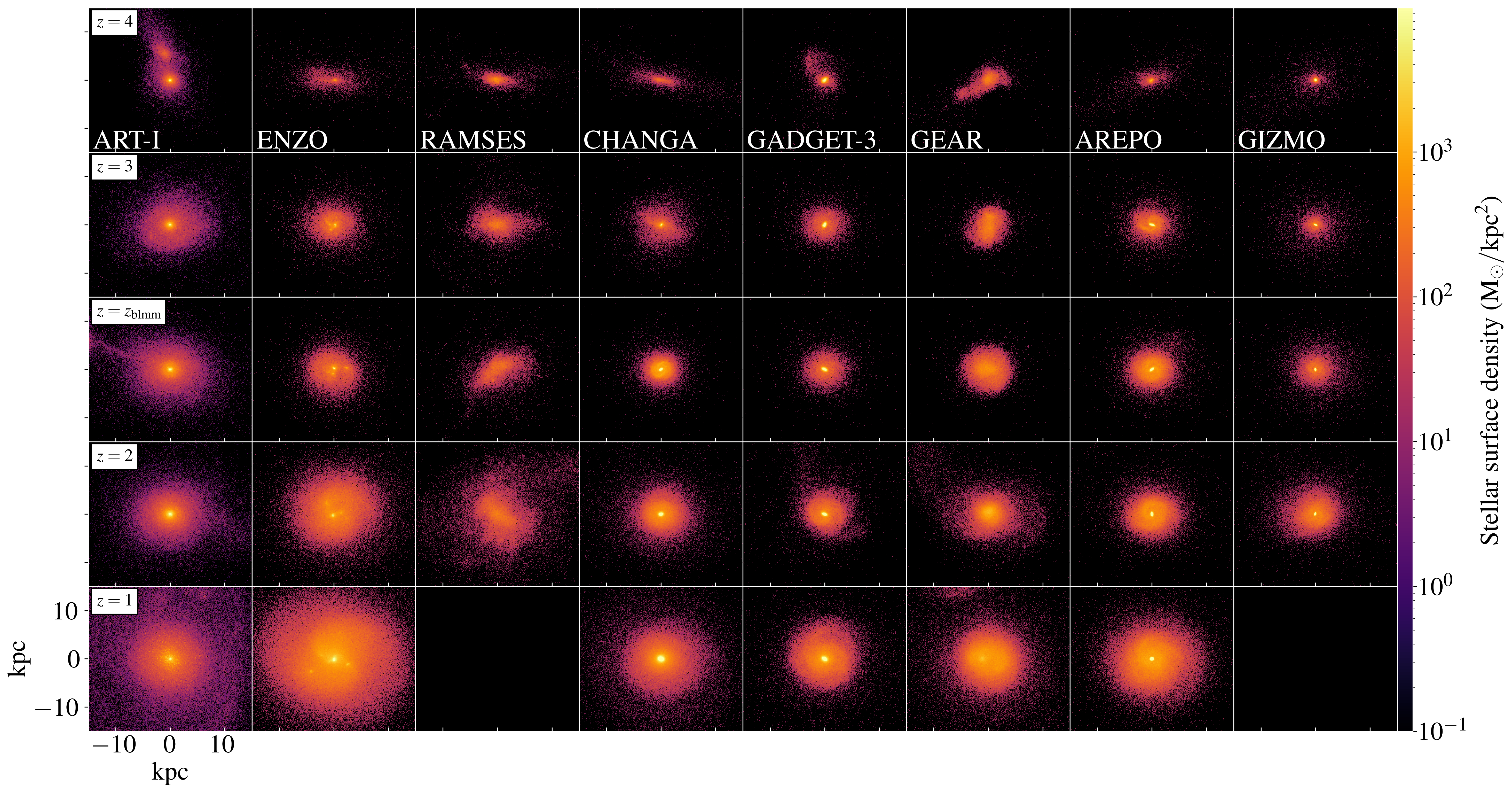}
    \vspace{-6mm}    
    \caption{Similar to Figure \ref{fig:star_snap_edgeon}, but for face-on views.
    See Section \ref{sec:stellar_morph} for more information on {\tt CosmoRun} and this figure.}	     
    \label{fig:star_snap_faceon}
    \vspace{2mm}    
\end{figure*}

In Figures \ref{fig:star_snap_edgeon} and \ref{fig:star_snap_faceon}, we show the stellar distribution of the {\tt CosmoRun} galaxies from the edge-on and face-on perspective, respectively, from $z=4$ (top) to $z=1$ (bottom). 
The face-on viewing angle aligns with the direction of each galaxy's stellar angular momentum vector.\footnote{The centre of each galaxy is determined as follows: {\it (i)} We select a snapshot far from the main merger events; {\it (ii)} We compute the centre of mass of the dark matter$+$stellar particles and we select the particle that is closest to the centre and with less kinetic energy; {\it (iii)} We iteratively ``refine'' the centre by calculating the centre of mass within decreasing radii of [10, 5, 3, 1] kpc; {\it (iv)} We select the particle that is closer to the final centre, and that has the lowest kinetic energy. {\it (v)} In the next snapshot, we use the position of this particle in the next snapshot as the first guess for the centre, and we begin the process at step {\it (iii)}. The face-on orientation aligns with the stellar angular momentum vector, determined by considering all the stellar particles within $0.1 R_{\rm vir}$ from the centre.\label{footnote_1}} 
From these surface stellar density plots, especially from the edge-on view in Figure~\ref{fig:star_snap_edgeon}, we see that as early as $z=4$ one SPH code ({\sc Gadget-3}) and both ``moving-mesh'' codes ({\sc Arepo} and {\sc Gizmo}) already form disk structures. Readers may find it particularly interesting to notice that once formed these disks survive until the last major merger at $z\sim2.2$ and beyond. The presence of these extended thin disks in an SPH code may initially appear surprising as in the past there was a common belief that SPH codes generally yield smaller disk galaxies due to excessive angular momentum transfer \citep[see discussions on the {\it angular momentum catastrophe} in e.g.][]{Steinmetz1999,Navarro2000}. Since then, the SPH community included new formulations\footnote{In recent years, most SPH code groups included new formulations like entropy-conserving, entropy-indenpendent SPH \citep[e.g][]{Hopkins2013,Saitoh2013}, artificial viscosity models \citep[e.g.][]{Cullen2010,Hu2014}, and kernels \citep[e.g.][]{Wendland1995,Dehnen2012} that improve the treatment of shear and viscosity.} that helped to solve this problem. Now, the differences in the disk formation between the AMR and the SPH approaches are minor, if any, as is evident from the analysis of Figure~\ref{fig:star_snap_edgeon}. For instance, while both two AMR codes ({\sc Art-I} and {\sc Enzo}) and an SPH code ({\sc Changa}) produce a well-defined disk-like structure by $z\sim$3, also an AMR and an SPH code ({\sc Ramses} and  {\sc Gear}) do not form a disk by $z\sim 2$. 
These differences between codes on the disk formation and morphologies can have multiple origins, from the effect that different feedback implementations can have on the disk mass fraction and thus on the stability of disks to other more technical ones like the maximum resolution reached by each code in the disk region. In particular, as shown in Figure 1 of Paper VI, codes with a tree-based solver like {\sc Arepo} and {\sc Gizmo} reach higher resolution in the disk region at high-z. As a consequence, these codes may better  resolve unstable modes like spiral arms and bars that transport angular momentum and thus regulate the galaxy size. 
In the face-on view of Figure~\ref{fig:star_snap_faceon}, we see that the well-defined disks in {\sc Enzo}, {\sc Gadget-3}, and {\sc Arepo} have also developed non-axisymmetric structures at low redshift ($z\sim$1) including spiral arms and an incipient stellar bar.  
Such structures in, e.g., {\sc Arepo} are visible as early as $z=3$.

All results presented in the previous paragraphs on the {\tt CosmoRun} disks' morphology are obtained from visual inspection of Figures~\ref{fig:star_snap_edgeon} and \ref{fig:star_snap_faceon}, and thus are qualitative. Readers should be reminded that in a representation where stellar particles are not weighted by age, the rotationally-supported disks, which are made of young stars, are diluted by the older spheroids.  
The {\it AGORA} Collaboration is currently working on generating more realistic images of the {\tt CosmoRun} galaxies including the effects of dust that will allow a better comparison with observations of real galaxies, results will be featured in a future paper. Additionally, in Section~\ref{sec:stellar_kinem} we present a quantitative analysis of the presence of rotationally-supported disks by analyzing the stellar particle kinematics.

\begin{figure*}
        \centering
        \vspace{1mm}
        \includegraphics[width=0.96\linewidth]{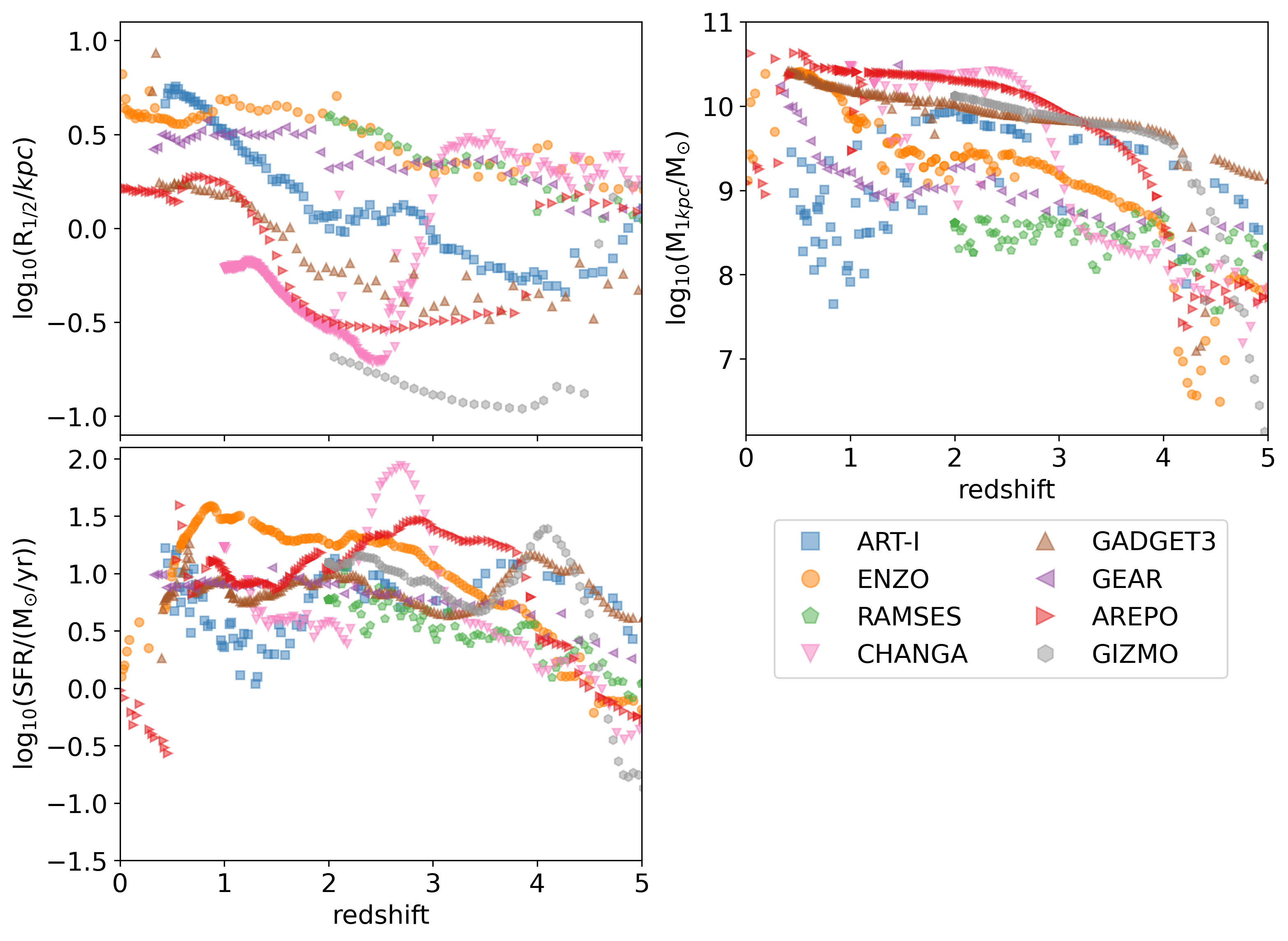}
        \vspace{-3mm}
        \caption{The half-mass radius of the {\tt CosmoRun} galaxies ($R_{1/2}$; top left panel), central stellar mass ($M_{\rm 1\,kpc}$; top right panel), and the star formation rate (SFR; bottom panel) as  functions of redshift for the eight participant codes at $z < 5$. 
        %JP in the $z=1-5$ range.
       See Section \ref{sec:stellar_morph} for more information on {\tt CosmoRun} and this figure.}	         
        \label{fig:R12}
        \vspace{2mm}        
\end{figure*}

Our next step in the analysis of stellar disks is to compute some of the most common morphological parameters used in the community by both simulators and observers to characterize the stellar systems are the half-mass radius ($R_{1/2}$),  
%JP ; or effective/half-light radius), 
the fractional mass within the central 1 kpc, and the Sérsic index \citep[see, e.g.,][]{Zhiyuan23}.
Therefore, in Figure~\ref{fig:R12} we show the $R_{1/2}$ computed as the radius that contains half of the mass of all the stars inside $0.15\,R_{\rm vir}$, the mass within the central 1 kpc, ($M_{\rm 1\,kpc}$), and the disk star formation rate (SFR) as functions of redshift (i.e., as functions of the host's virial mass) for all available snapshots from $z=5$ to $z=1$. 

In the top left panel of Figure~\ref{fig:R12}, we see that although the total half-mass radius is consistent across all {\tt CosmoRun} simulations at high redshift ($z\sim5$),  there is an abrupt change in {\sc Changa}, {\sc Arepo}  and {\sc Gizmo} runs towards a more compact stellar distribution --- in {\sc Changa} after the major merger at  $z\sim2.2$, and in {\sc Arepo}  and {\sc Gizmo} after the one at $z\sim4.5$. 
This change coincides with the time of the star formation bursts seen in the bottom panel of Figure~\ref{fig:R12}.
%JP, and in the bottom left panel of Figure~\ref{fig:Mhz}. 
On the other hand, most other models show a fast increase in the half-mass radius at $z\le2$, which can be related to the acquisition of fresh gas that becomes star forming in the disk. 
It is also noticeable that the {\sc Gadget-3} and {\sc Arepo} (and {\sc Gizmo} until $z=2$) models produce more compact stellar structures than in the other models, at almost all the epochs. 
These two models also show a quick evolution to a more extended structure at low redshift. 
Overall, the half-mass radii obtained from {\sc Art-I}, {\sc Enzo} and {\sc Gear} runs (and {\sc Ramses} until $z=2$) agree within 0.25 dex with the ones obtained by observations of MW-like progenitors at $z<2$  \citep[see, e.g.,][that shows ${\rm log}_{10}(R_{1/2}) = 0.3-0.6$ at $z<2$]{Hasheminia22}, while {\sc Changa}, {\sc Gadget-3}, {\sc Arepo} and {\sc Gizmo} runs show more compact structures. It is also worth mentioning that the {\sc Gadget-3} run shows a larger spread in its redshift evolution than the other codes.  
We suspect that this originates from the presence of non-axisymmetric structures in the disk like the bar and spiral arms. 
The formation and evolution of these structures produce a radial redistribution of material that may be the cause of the observed oscillations. 
The properties and evolution of disk substructures in the {\tt CosmoRun} suite will be studied in the future within the {\it AGORA} Collaboration.

\begin{figure*}
    \centering
    \includegraphics[width = 1.01\linewidth]{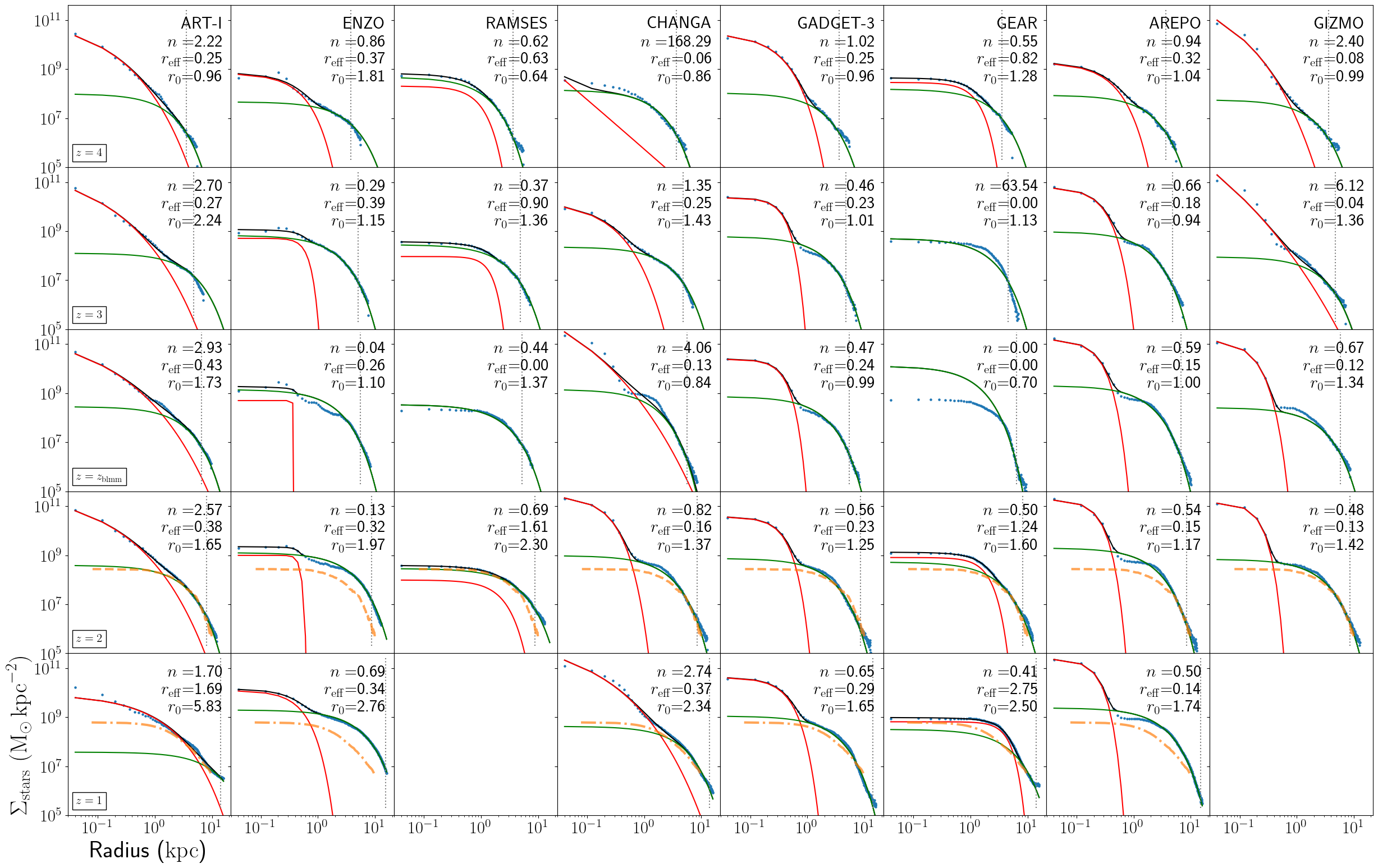}
    \caption{Disk surface density profile (blue dots) of the simulated {\tt CosmoRun} galaxies from $z=4$ ({\it top}) to $z=1$ ({\it bottom}). 
    We fit the profile using the Sérsic profile ({\it red lines}) and an exponential disk profile ({\it green lines}). The final fit (composition of the Sérsic and the exponential disk) is shown as {\it black lines}. 
    The Sérsic index ($n$), the Sérsic effective radius ($r_{\rm eff}$), and the exponential disk's scale length ($r_0$) are presented in each panel. 
    The vertical {\it gray dotted lines} indicate $0.1 R_{\rm vir}$, which is the maximum radial extent used for the fitting. 
    The {\it orange dashed} and {\it dot-dashed curves} in the $z=2$ and $z=1$ panels show the observed median stellar surface density of MW-type progenitors at $z=1.56$ and 1.06, respectively \citep[see Figure 4 in][]{Hasheminia22}.  
    Some codes lack an apparent bulge-like stellar structure, failing the fitting process.
    See Section \ref{sec:stellar_morph} for more information on {\tt CosmoRun} and this figure.}	     
    \label{fig:star_profile}
    \vspace{2mm}    
\end{figure*}

In the top right panel of Figure~\ref{fig:R12}, we show the evolution of $M_{\rm 1\,kpc}$. 
Here we detect three different post-merger behaviors in the {\tt CosmoRun} models. 
First, {\sc Changa}, {\sc Arepo} and {\sc Gizmo} show a star formation burst followed by an increase on the $M_{\rm 1\,kpc}$ and a decrease of the  $R_{1/2}$ as discussed above --- in {\sc Changa} during and after the major merger at $z\sim2.2$, and in {\sc Arepo} and {\sc Gizmo} during and after the merger at $z\sim4.5$.  
This behavior is in line with what is expected in a compaction event.
Second, {\sc Gadget-3} features a systematically high $M_{\rm 1\,kpc}$  and low $R_{1/2}$ across all redshifts, but it converges with the rest of the codes at low redshift. 
Third, the remaining codes show some post-merger variations but promptly move back to the original evolution track soon after the merger (see, e.g.,  {\sc Enzo} and {\sc Gear} during and after the merger at $z\sim4.5$).
Another result that is worth discussing is the behavior of the {\sc Ramses} run which is different from all other codes: the $M_{\rm 1\,kpc}$ in {\sc Ramses} shows a larger spread in its temporal evolution. 
This spread is caused by a poor definition of the center of the stellar structure due to the flatness of the central stellar density profile, which results in a high uncertainty of the total mass enclosed inside 1~kpc (see the next paragraph for a discussion of the density profiles).
From this figure, we argue that the evolution of stellar structures in the {\tt CosmoRun} galaxies is a complex process that deserves more studies including not only the analysis of the disk gas fraction  but also the properties and stability of this gas (e.g., turbulence) and how responsive it is to external perturbations. 
Future papers from the Collaboration will present a detailed study of the effects of mergers on our target galaxy, including processes like compaction \cite[e.g.,][]{Zolotov15,Lapiner23}.

Finally, for our study of the stellar disk morphology, in Figure~\ref{fig:star_profile} we show the disk stellar surface density profiles (blue dots) from $z=4$ (top) to $z=1$ (bottom). 
The profiles have been obtained by following the strategy presented in \citet{Marinacci14}.\footnote{A face-on projection is first created for all the stellar particles in a cylinder centered on the main host's center (for more information on the centering strategy see footnote \ref{footnote_1}), with a height equal to $0.1R_{\rm vir}$ (vertical gray dotted lines in Figure~\ref{fig:star_profile}).  Then, stars are binned in circular annuli in the projection plane; and, the total stellar mass in each bin is divided by its surface area to acquire the surface density. \label{footnote_2}} 
Also following  \citet{Marinacci14} we apply a two-model fit to the disk stellar surface density --- a Sérsic profile in the inner regions (red lines) and an exponential disk profile in the outer regions (green lines), changing the parameters of each function at once \citep[see Eq.(1) in ][]{Scannapieco11}. 
With this method we first find the optimal parameters of the exponential profile, thus, model the residual excess only with the central Sérsic profile \citep[for details, see Section 5.1 of][]{Scannapieco11}.  % check if this is right? (??)
In each panel, we present the Sérsic index ($n$), the Sérsic effective radius ($r_{\rm eff}$), and the exponential disk's scale length ($r_0$) obtained from the fit.  

\begin{figure*}
    \centering
    \includegraphics[width = 1.\linewidth]{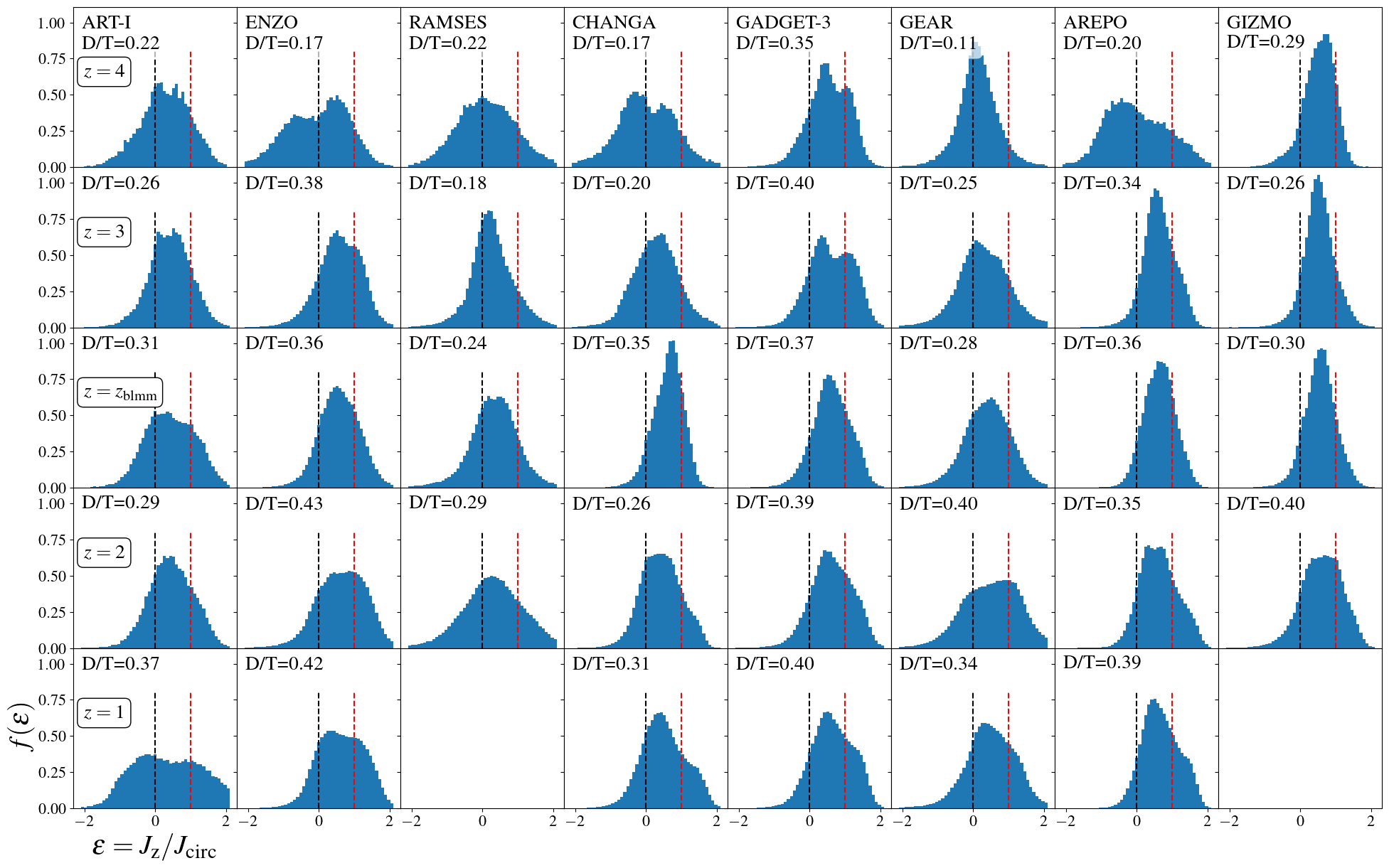}
    \caption{The probability density function of the stellar circularities, $\epsilon = J_z / J_{\rm circ}$,  within a cylinder of radius $0.1 R_{\rm vir}$ for the simulated {\tt CosmoRun} galaxies. 
    The disk-to-total (D/T) ratio is determined by the fraction of stars with $\epsilon > 0.8$.
    The D/T ratio increases with redshift, reaching $\sim 0.4$ at $z=1-2$, which is indicative of the presence of an extended disk.  
    The majority of the panels exhibit a distribution that can be fitted using two overlapped Gaussian: one centered at $\epsilon = 0$ ({\it black vertical dashed line}) indicating the bulge component, the other centered at $\epsilon = 1$ ({\it red vertical dashed line}) representing a co-rotating disk. 
%    Both {\sc Gadget-3} and {\sc Gizmo} tend to have a distinct peak at the pre-merger time. %, whereas the separation becomes less prominent at $z=1$ across all codes.
    See Section \ref{sec:stellar_kinem} for more information on {\tt CosmoRun} and this figure.}	 
    \label{fig:circularity_z2}
    \vspace{2mm}        
\end{figure*}

The fits show that most {\tt CosmoRun} galaxies contain an extended structure at large radii that is well fitted by an exponential disk model, and at the same time contain a bulge-like structure dominating in the central radii ($r_{\rm eff}<1$kpc) in most cases. 
The central concentrations are significant in most particle-based codes (except {\sc Gear}) and ``moving-mesh'' codes. 
Exceptions are the {\sc Gear} simulation that hosts an extended bulge structure all along its evolution with an $r_{\rm eff}$ as large as 2.75 at $z=1$, and the {\sc Ramses} simulation that also features an extended bulge structure of $r_{\rm eff} =1.61$ at $z=2$. 
As an intermediate case, {\sc Art-I}, and also {\sc Enzo} at low redshift, produce an extended exponential disk at large radii and an extended bulge-like structure in the center, but not as concentrated as in, e.g., the ``moving-mesh'' codes. 
The absence of a compact bulge can be related to the efficiency of the stellar feedback in driving outflows, thus removing the gas from the galactic center out to the CGM. 
The properties of the CGM gas and the disk-CGM interplay are analyzed in Paper VI of this Collaboration. 

Regarding the disk-like structure in the outer regions of galaxies, we find that its exponential scale length ($r_0$) gradually increases with time, reaching its maximum between $z=1$ and 2, well after the last major merger.
However, it should be noted that some simulations (e.g., {\sc Arepo} or {\sc Enzo}) already created extended disk-like structures as early as $z=4$.
In the 
%JPbottom two 
$z=2$ and $z=1$ rows of Figure~\ref{fig:star_profile}, we include as orange dashed and dot-dashed lines the median stellar surface density of MW-type progenitors observed at $z=1.56$ and 1.06, respectively \citep[see Figure 4 in][]{Hasheminia22}. 
Excluding the central $\sim$3 kpc, where the bulge structure is located, we see that most models host exponential disks that well match the observations. %with a softer exponential decay than the one observed. (??)
{\sc Gizmo} at $z=2$ and {\sc Gadget-3} at $z=1$ are  models that show an especially good fit agreement with the observations. 
Nonetheless, we warn the reader that the extended exponential structures seen in the stellar surface density profiles may not be good proxies of thin rotationally-supported disks. 
This proxy claim is because for Figure~\ref{fig:star_profile} we used all the stars within a sphere of 0.1$R_{\rm vir}$ in radius to fit the profiles (see footnote \ref{footnote_1}), instead of selecting stars by age or kinematics. 
As a result, these profiles can include both thin disks and oblate spheroids (with an exponential decay; see, e.g., {\sc Art-I} and {\sc Gear} vs. {\sc Arepo} at $z=3$ in Figure~\ref{fig:star_snap_edgeon}). 
In the next section, we use stellar particle kinematics to better understand which simulations host well-defined rotationally-supported thin disks. 
We also remind the reader that in some codes the Sérsic profile fit fails (e.g., for {\sc Ramses} and {\sc Gear} at $z=z_{\rm blmm}$). 
This fit failure suggests that the bulge-like and disk-like structures are less well-defined at high redshift, especially when they experience catastrophic events such as a major merger.

\subsubsection{Kinematic Decomposition}\label{sec:stellar_kinem}

As discussed in Section \ref{sec:stellar_morph}, the analysis of the stellar surface density profiles provides us with a good first impression of the galaxy's morphology, but it is not the best tool to determine whether or not a well-defined rotationally-supported disk exists. 
In this section, we use the stellar particles' kinematic information to better characterize the galactic structures in the {\tt CosmoRun} simulations. 

In Figure \ref{fig:circularity_z2}, we show the probability distribution function of the stellar circularities, $\epsilon = J_z / J_{\rm circ}$, where $J_{\rm circ}(r)$ is the angular momentum of a circular orbit at the radius $r$ of a stellar particle in consideration \citep[for detailed definitions see, e.g.,][]{Scannapieco12}. 
The red vertical dashed line represents the value $\epsilon = 1$ at which particles have the same vertical angular momentum as an object moving in a pure circular motion (for which the circular velocity is computed assuming a spherical potential).\footnote{Notice that with the $\epsilon$ definition used here, it can take values that are greater than 1 and lower than -1. This is not true when using a  $J_{\rm circ}$ that is computed as the angular momentum of a circular orbit that has the same total specific energy as the real orbit which would take values only between -1 and 1 (see \citep{Abadi2003,ElBaldri2018}).} 
Readers can immediately notice that the majority of the panels exhibit a bimodal distribution: one (Gaussian) distribution centered close to $\epsilon = 0$ (black vertical dashed line) indicating a velocity dispersion-supported bulge component in some cases slowly rotating, the other centered at $\epsilon = 1$ representing a co-rotating disk. 
From this bimodal distribution, we can conclude that most of the {\tt CosmoRun} galaxies host rotationally-supported structures, especially at low redshift, though they may not be obvious in the stellar surface density plots in Figures \ref{fig:star_snap_edgeon} and \ref{fig:star_snap_faceon}. 
It should be noted that all the codes that reached $z=1$ show rotationally-supported stellar disks at $z=1$ (the bottom row of Figures \ref{fig:circularity_z2} and \ref{fig:star_snap_edgeon}). 

In Figure~\ref{fig:circularity_z2} we have also 
%JP computed and 
shown the kinematic disk-to-total  (D/T) ratio \citep[][see labels at the top-left corner of each panel]{Scannapieco09}.\footnote{We note that the authors of \citet{Scannapieco09} warn the reader that the kinematic definition therein may underestimate the mass of the geometrically thin disk component.} 
For disk stars we include all the stellar particles with $\epsilon>0.8$, as suggested by \citet{Scannapieco12} and \citet{Marinacci14}.  
Overall, we see an increase in the D/T ratio with redshift, reaching $\sim 0.4$ at $z=1-2$, which is indicative of the presence of an extended disk \citep[Section 3.1 of][see also their Figure 4 to compare the values obtained from our D/T definition and the ones from a more careful kinematic decomposition]{Marinacci14}. 

\begin{figure*}
     \centering
     \includegraphics[width = 0.6\linewidth]{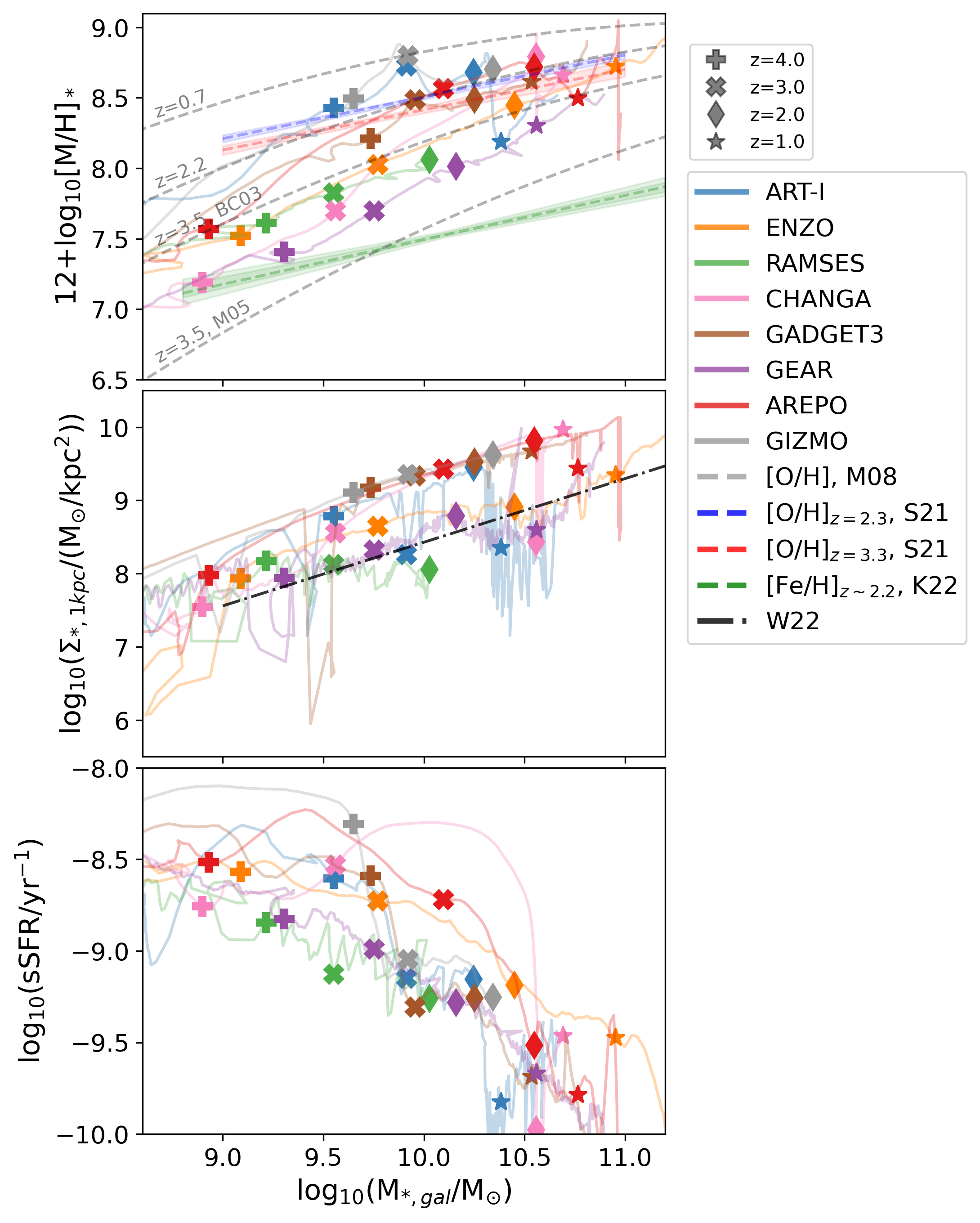}
     \vspace{-1mm}
     \caption{{\it Top panel}: The mass-metallicity relation (MZR) for the eight participant codes from $z=4$ to $z=1$ (when available). 
     The stellar mass has been computed using all the stellar particles within a sphere of radius 0.1$R_{\rm vir}$.\footref{fnlabel1}\ 
     The total metal mass inside stellar particles younger than 200~Myr is used to compute the abundances to mimic results obtained from observations of gas-phase metallicity in HII star-forming regions. 
     The filled symbols in the bottom panel indicate the redshift evolution of the MZR in each code: at $z=4$ (plus sign), 3 (x), 2 (diamond), and 1 (star, when available).  
     {\it Dashed lines} are the fits to observations of high-$z$ galaxies:  four {\it gray dashed lines} for the analytical fits to the $12+{\rm log}[{\rm O/H}]-M_\star$ data by \citet{Miolino08}, the {\it red} and {\it blue dashed lines} for the $12+{\rm log}[{\rm O/H}]-M_\star$ data by \citet{Sanders21}, and the {\it green dashed line} for the $12+{\rm log}[{\rm Fe/H}]-M_\star$ data by \citet{Kashino22}. 
     {\it Middle panel}: Central stellar surface density ($\Sigma_{\star, \rm 1\,kpc}$) as a function of the total stellar mass for the same models and redshift shown in the top panel. 
     {\it The dot-dashed line} is the $\Sigma_{\star, \rm 2\,kpc}$ least-squares fit to the star-forming galaxies in TNG at $z=0$ \citep[Figure 16 of][]{Walters2022}. 
     {\it Bottom panel}: Specific star formation rate (sSFR) as a function of the total stellar mass. 
    See Section \ref{sec:stellar_metal} for more information on {\tt CosmoRun} and this figure.}	     
    \label{fig:MstarMet}
    \vspace{2mm}    
\end{figure*}

Detecting the rotationally-supported structures in these diagrams is not straightforward due to the presence of bulges in most simulations. 
This is especially true with the {\sc Arepo} run which shows well-defined thin disks in the stellar surface density plots of Figure \ref{fig:star_snap_edgeon}, but only a skewed distribution in Figure~\ref {fig:circularity_z2} across all redshifts. 
We can also see that out-of-equilibrium galaxies may host complex structures, seen as multiple peaks in this diagram (e.g., {\sc Art-I} at $z=4$).

\subsection{Metallicity and Young Stellar Populations}\label{sec:stellar_metal}

Galaxy formation and evolution processes leave strong imprints on the metallicity distribution in galaxies, both in gas and stars. 
For instance, stellar feedback boosts the metallicity of the ISM, while intergalactic primordial gas inflows dilute it. 
Understanding the evolution of gas and stellar metallicities in time and their distribution inside the galaxy allows us to unveil the details of the galaxy formation process and how efficient the stellar feedback processes are in removing metal-rich gas via outflows and/or stopping fresh gas inflows. 
The mass-metallicity relation is an empirical scaling relation that helps us to undertake a comparison between simulations and observations of galaxies at low and intermediate redshift. 
This comparison enables us to know which feedback model better fits the real observed galaxies. 
Many recent observational efforts presented a detailed analysis of the MZR using thousands of external galaxies \citep[e.g.,][]{Sanders21,Kashino22}. 
The authors showed that there exists a clear redshift dependence in the MZR. 
They also discussed the differences in the MZR when different tracers are used for the metallicity, e.g., [O/H] vs [Fe/H], owing to the redshift variation of the [O/Fe] from 0 at $z=0$ to $\sim0.42$ at $z=2.2$ \citep[see Section 4.3 of][for a detailed description of the observed differences]{Kashino22}. 
In the {\tt CosmoRun} models each code group were asked to use their own prescription to simulate the production of metals in SNe, close to the most widely-used practice in each code community (see Table \ref{tab:1} and Section \ref{sec:intro}).  
For this reason, here we do not analyze the iron or the $\alpha$ element abundances separately but only study the total metallicity and  compare this total value with observations of both [Fe/H] and [O/H], assuming the former to be a lower limit of the total metallicity.  

In the top panel of Figure~\ref{fig:MstarMet}, we show the MZR of the {\tt CosmoRun} galaxies at various redshifts, where the stellar mass has been computed using all the stellar particles within a sphere of radius 0.1$R_{\rm vir}$.\footnote{Notice that here we use $0.1R_{\rm vir}$ instead of $0.15R_{\rm vir}$ used in previous sections to be consistent with the models and observations shown in the same panel.\label{fnlabel1}}
We use the direct summation of all the mass in metals inside stellar particles that are younger than 200~Myr, regardless of the metal species, to represent the gas-phase metallicity, to mimic the observed metallicity of the newly formed stars. 
With filled symbols, we indicate the position of the MZR of each code at $z=4$ (plus symbol), 3 (cross), 2 (diamond), and 1 (star, when available). 
We also include some observational results\footnote{Notice that the ${\rm [O/H]}$ data shown here should be taken as a lower limit according to recent results presented by \citep{Cameron2023}.}: 
{\it (i)} the analytical fit to the MZR relations obtained by the AMAZE program using the near-IR data to obtain stellar masses and ${\rm [O/H]}$ abundances, at $z=0.7$, 2.2, and 3.5, including two different techniques to compute the stellar mass at $z=3.5$  \citep[four gray dashed lines; for details, see][M08, hereafter]{Miolino08}; 
{\it (ii)} the ${\rm [O/H]}$ values obtained by the MOSDEF survey for $\sim 300$ star-forming galaxies at $z\sim2.3$ and $\sim150$ galaxies at $z\sim3.3$ \citep[blue and red dashed lines, respectively, including a 1$\sigma$ dispersion as a shade;][S21, hereafter]{Sanders21}; 
{\it (iii)} the ${\rm [Fe/H]}$ values of 1336 star-forming galaxies at $1.6 \le z \le 3.0$ obtained by the zCOSMOS-deep survey \citep[green dashed line, including 1$\sigma$ and 2$\sigma$ dispersions as shades;][K22, hereafter]{Kashino22}. 
In this figure, we see that at $z=3$ (crosses) most MZRs from the {\tt CosmoRun} suite, excluding {\sc Art-I} and {\sc Gizmo}, are slightly below the observations. This is specially true if we assume that the ${\rm [Fe/H]}$ values by K22 are a lower limit for the $z\sim2.2$ MZR\footnote{The total metallicity is usually computed as the summation of Fe and O abundances, being Fe the less abundant metal specie} and that at high stellar masses the ${\rm [O/H]}$ values from S21 and M08 agree well with each other. At $z=2$ (diamonds), we observe that the MZRs from the {\sc Art-I},  {\sc Arepo} and {\sc Gizmo} agree relatively well with S21 and M08. Otherwise, {\sc Ramses} (green data points) and {\sc Gear} (purple data points) are almost always $\sim 0.5$ dex below the observed values, although they are still above the limit given by the [Fe/H] observations by K22. 
By $z\sim1$ many codes show a decline in metallicity, or a slow-down in the increase of metallicity, thus diverging from the low-redshift observed values in star-forming galaxies. 
This may be because some of the {\tt CosmoRun} galaxies are no longer star-forming at $z\sim1$ as shown in Figure~\ref{fig:R12}. The accretion of fresh low-metallicity gas combined with a slowdown in the production of metal-rich gas by newly formed stars and  the absence of mergers can result in a global decline in metallicity of the star-forming regions. The cause of this decrease in the metallicity will be studied in a future paper. 
Overall, the MZR redshift evolution in the {\tt CosmoRun} models is similar to the ones presented in \citet{Garcia2024} for the TNG, Illustris and EAGLE models. For instance, from $z=4$ to $z=1$ (i.e. from log$_{10}$(M$_*$/M$_{\odot}$)$=$8 to 11)) the metallicity increses $\sim$1 dex in most of the models. However, because here we trace the evolution of a single galactic system, with a stellar mass of $\sim$10$^{12}$M$_{\odot}$ at $z=0$, we can not break the redshift-M$_*$ degeneracy in the same way they do it in their figure 2. Within the uncertainties, our results fit well with the MZR evolution shown by the lower envelope of all curves shown in their figure 2.

\begin{figure*}
     \centering
     \includegraphics[width = 0.8\linewidth]{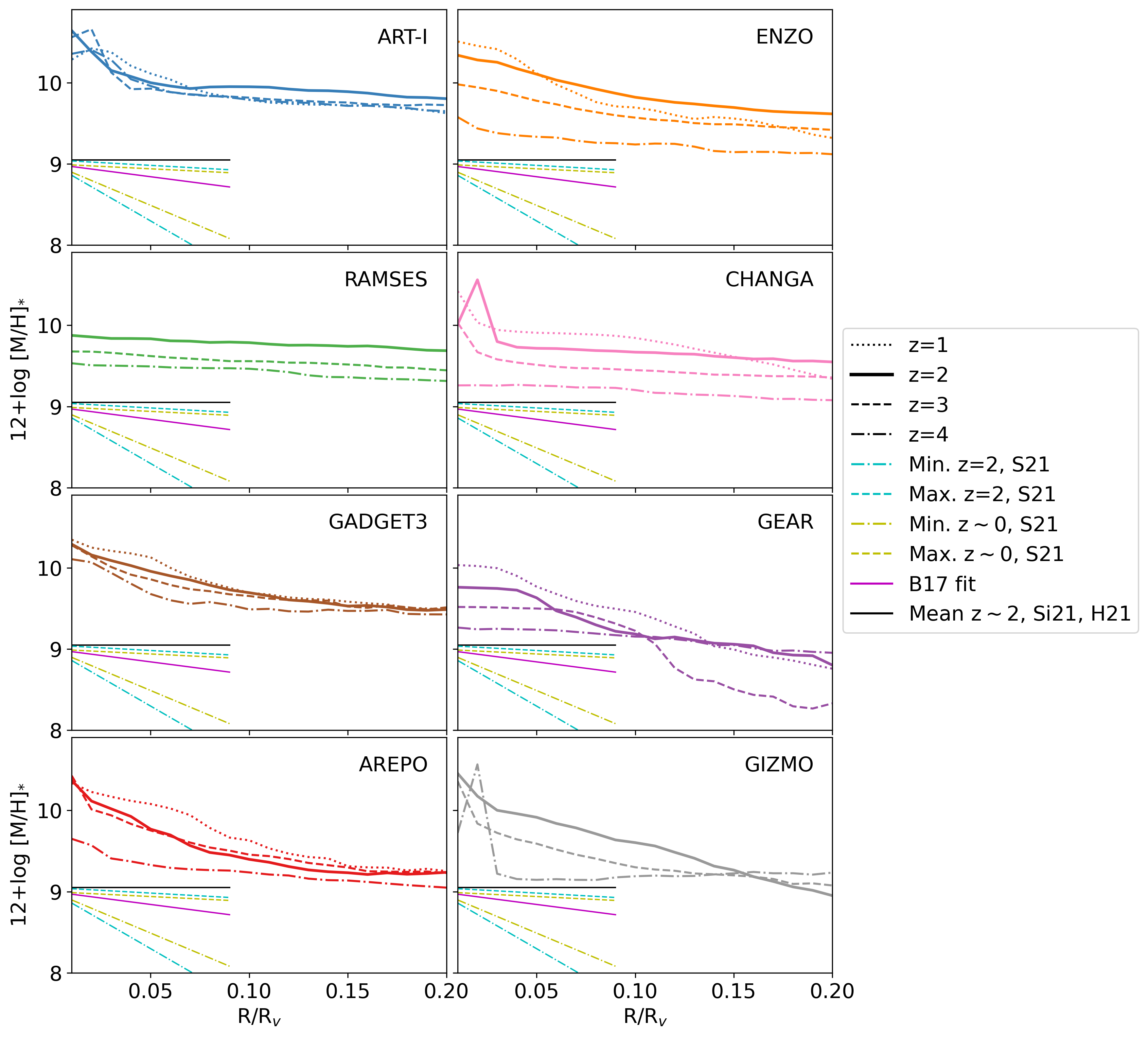}
     \vspace{-1mm}
     \caption{The galaxy radial metallicity profiles obtained from the young stellar population, i.e., younger than 500~Myr, which mimics observations of gas in star-forming regions, for the eight participant codes from $z=4$ to $z=1$ (when available).  
     In the bottom-left corner of each panel, we show the highest and lowest predicted gradient of the gas metallicity ({\it dashed} and {\it dot-dashed lines}, respectively) for high-redshift galaxies and local star-forming spirals  ({\it cyan} and {\it yellow}, respectively) in \citet{Sharda2021}. 
     In a {\it magenta solid line}, we also show the best-fit slope to the average metallicity profiles of local spirals in the MaNGA survey \citep{Belfiore2017}. Finally, in a {\it black solid line}, we show a zero slope line that reflects the results obtained by \citet{Hemler2021} and \citet{Simons2021}, indicating a flat or slightly positive gradient in observed galaxies at $z=0.6-2.6$.
     Note that these lines have been shifted downward for visualization purposes.
    See Section \ref{sec:stellar_metal} for more information on {\tt CosmoRun} and this figure.}	          
    \label{fig:MgasMetRad}
    \vspace{2mm}    
\end{figure*}

The middle panel of Figure~\ref{fig:MstarMet} illustrates the evolution of the central surface density ($<1$ kpc), $\Sigma_{\star, \rm 1\,kpc}$, with stellar mass and redshift. 
Here, we also include the least-square fit to the star-forming galaxies in IllustrisTNG100 at $z=0$ \citep{{Walters2022}}, shown as a dot-dashed line. In their work, the authors show the $\Sigma_{\star, \rm 2\,kpc}$ profile instead of the commonly used $\Sigma_{\star, \rm 1\,kpc}$ due to limitations on the spatial resolution \citep[see Figure 16 in][]{Walters2022}. We also computed $\Sigma_{\star, \rm 2\,kpc}$ in the redshift range from $z=4$ to $z\sim1$ (not shown in the figure), and we have found that, as expected, these are lower than the $\Sigma_{\star, \rm 1\,kpc}$, thus agreeing well with the results presented in \citet{{Walters2022}} for the MANGA and the TNG100 at $z\sim0$.
Notice also that during major mergers it is difficult to find the center of a simulated galaxy since the center of mass, the location of the highest density, and the location of the deepest potential well do not coincide. 
Our approach here is to compute the center as the center of mass of both the dark matter and stellar particles in a shrinking sphere (readers can find the details about the centering process in footnote~\ref{footnote_1}). As a result of using the center of mass to determine the center of the galaxy, during mergers, there is a mismatch between it and the peak of the stellar density, and thus some of the codes show unphysical excursions down to lower stellar surface densities in the central surface density plots (see, e.g., {\sc Gadget-3}, {\sc Gear} and {\sc Arepo}).
Excluding those unphysical excursions, in most {\tt CosmoRun} galaxies, we see the  common trend in the literature, where galaxies show higher central surface densities with increasing stellar mass.
It is particularly interesting that, for {\sc Ramses}, $\Sigma_{\star, \rm 1\,kpc}$ is constant in time and independent on the stellar mass. The {\sc Ramses} code group is currently performing new calculations with different stellar feedback to better understand this behavior. Another interesting information we extract from this figure is that, as it happens for $R_{1/2}$ (see Figure~\ref{fig:R12}, top-left panel), the codes with a central concentration, i.e. higher $\Sigma_{\star, \rm 1\,kpc}$, are the ones that show the formation of thinnest disks (e.g. {\sc Gadget-3}, {\sc Changa} and {\sc Arepo} at $z=1$). This is in agreement with our previous discussion that these models which reach a higher resolution in the disk region may be able to resolve the formation of unstable modes that bring material to the central parts of the galaxy thus keeping the stellar disk thinner and smaller.
%Another feature that is worth mentioning is that most of the codes have similar $\Sigma_{\star, \rm 1\,kpc}$ from $z=3$ to 1 (with some exceptions like {\sc Enzo},  {\sc Ramses} and {\sc Gear}).  %??

In the bottom panel of Figure~\ref{fig:MstarMet}, we show the evolution of the specific star formation rate (sSFR), a ratio of the stellar mass born in the last 20 Myr to the total stellar mass inside $0.15 R_{\rm vir}$. 
If we compare our results at $z=4$ to $z\sim1$ with the $z\sim0$ observations presented  in \citet[][in their Figures 4 to 8]{Salim2023}, we first see that the sSFR in our models is higher at all stellar mass ranges. 
This result is expected because our models are of galaxies at higher redshift. 
More interestingly, we can see that the tracks followed by all our models  start at high sSFR at high redshift and move towards lower sSFR when the stellar mass increases and the redshift decreases. 
All simulations move from being actively star-forming at high redshift (low stellar mass) to quenching at low redshift (high stellar mass). 
This result is independent of the adopted stellar feedback scheme and the numerical code, implying that the process of quenching is likely dictated by the mass of the galaxy and its mass assembly history, not by the particular stellar feedback physics.  
 
Finally, in Figure~\ref{fig:MgasMetRad}, we show the radial metallicity profiles inside the galaxy obtained from the young stellar population, i.e., younger than 500~Myr which mimics observations of gas in star-forming HII regions for the eight participant codes at $z=4$, 3, 2, and 1 (when available). 
In the bottom-left corner of each panel, we show the highest and lowest predicted gradient of the HII gas metallicity (dashed and dot-dashed lines, respectively) for high-redshift galaxies and local star-forming spirals  (cyan and yellow, respectively) compiled in \citet{Sharda2021} assuming a low yield reduction factor (i.e., most supernova-produced metals are lost immediately in metal-enhanced galactic winds, instead of mixing with the ISM).\footnote{Note that these are only guiding lines showing the predicted/observed slope of the radial metallicity distribution.  They do not represent real data from observed metallicity distribution.}
In a magenta solid line, we also show the best-fit slope to the average metallicity profiles of local spirals in the MaNGA survey \citep[][using the calibration described in \citealt{Pettini2004}]{Belfiore2017}. Finally, in a black solid line, we show a zero slope line that reflects the results obtained by \citet{Hemler2021} and \citet{Simons2021}, indicating a flat or slightly positive gradient in observed galaxies at $z=0.6-2.6$.
These lines have been arbitrarily shifted downward so that they start at an abundance of 9.0, simply for visualization purposes. 
In many of the {\tt CosmoRun} simulations, the overall young stellar metallicity increases with redshift while maintaining a nearly constant radial slope (e.g., {\sc Art-I},  {\sc Ramses},  {\sc Enzo}, {\sc Changa}, and {\sc Gadget-3}). 
In the remaining codes, we see some variations in the profile's slope, but without a clear trend (e.g., {\sc Gear}, {\sc Arepo}, and {\sc Gizmo}). None of the codes shows a gradient that steepens over time, differently from what was suggested by the observations presented in \citet{Sharda2021}.  
Regarding the slope of the radial metallicity profile, \citet{Hemler2021} and \citet{Simons2021} showed in their Figures 11 and 5-6, respectively, that this number is either zero or even slightly positive in observed galaxies at $z\sim2$. \citet{Simons2021} also showed in their Figure 7 that many cosmological simulations fail to produce the flat metallicity profiles, but instead show steep negative gradients (e.g., TNG50, FIRE, MUGS and MaGICC). This negative slope is also present in some of the {\tt CosmoRun} simulations, but not for {\sc Art-I}, {\sc Ramses}, and {\sc Changa} that show the gradients compatible with the observations.  They are close to being flat if the very central region is excluded.

\vspace{2mm}

\section{Summary and conclusions} \label{sec:conc}

In this fourth paper from the {\it AGORA} Collaboration, we have analyzed the evolution of the {\tt CosmoRun} suite of simulations presented in Paper III down to lower redshifts, while presenting a new {\tt CosmoRun} fiducial model by the {\sc Arepo} code group, and revised calibrated versions of some of the previous runs.
This paper is the second {\it AGORA} paper that is devoted to analyzing and comparing the {\tt CosmoRun} suite of ``zoom-in'' cosmological simulations. 
Many new sub-projects have been initiated to analyze different aspects of these simulations and those that will be added to the {\tt CosmoRun} suite shortly.  
These efforts include the analysis of the properties of the satellite galaxies (Paper V; Jung et al. 2024 accepted), the CGM (Paper VI; Strawn et al. 2023 accepted), the internal structures of disks, and the particularities of the major mergers in each model.   

Since the publication of Paper III, the community has shown interest in creating new {\tt CosmoRun} simulations by using new codes \citep[e.g., {\sc Swift}][]{Schaller2023} or by using the already participating codes with different stellar feedback strategies. 
Several code groups  completed these new runs or are running new models \citep[e.g., {\sc Ramses-Vintergatan} model][]{Agertz2021}; some of these runs are presented in the Appendix of this paper, and others will be presented in future papers from the Collaboration. 
We encourage the community to produce new {\tt CosmoRun} simulations with their favorite stellar feedback strategy, enhancing this library of calibrated models, and, thus, helping to gain insight into the effects of different simulation codes and/or different stellar feedback strategies on the processes of galaxy formation and evolution.

In the present paper, we have introduced the halo merger trees obtained by the {\sc Rockstar} halo finder and studied the timing discrepancies in the merger timings that were first reported in Paper I. 
After careful analysis, we have concluded that timing discrepancies come mostly from differences in the choice of parameters for the orbit integration and in the variable length of the minimum timestep (Section \ref{sec:MergerTree} and 
Appendix~\ref{subsec:timdisc}).\footnote{After the publication of this paper, the merger trees, together with the centers, virial radii, and other relevant properties of the target halo as functions of redshift will be made public at \url{https://flathub.flatironinstitute.org/agora}.}   

In the second part of this paper, we have studied the main properties of the host halos at redshifts $z < 4$. 
Most models show a good convergence towards the galaxy stellar mass values predicted by abundance matching when far from merger events and close to lower redshift (Section \ref{sec:massevolution}). 
We have also analyzed the morphology of the stellar disks,
including fitting stellar surface density profiles and the evolution of the D/T ratios.
%and focused on the formation of rotationally-supported disks. 
Most codes produce rotationally-supported structures at low redshift but the density profiles are surprisingly diverse for systems that share identical initial conditions, common cooling and other astrophysics, and nearly identical merger history and mass evolution (Section \ref{sec:stellar_dist}). 
%Also interesting is that some codes generate thin disks at redshift as high as 3, including a bar and a spiral structure in the case of the {\sc Arepo} model. 
Some simulations show clear signs of a compaction event caused by a major merger, followed by a starburst and quenching. 
Why only some models show these processes after a major merger while others keep their rotationally-supported structures will be a matter of further studies in the Collaboration.

Finally, we have presented the properties of the young stellar population and the gas contained inside the galaxy. 
We have found that most codes agree well with observations of the MZR down to intermediate redshifts ($z\sim2$) when the main host galaxy is still star-forming, but they diverge at lower redshifts when the main host starts to quench (Section \ref{sec:stellar_metal}). 
All simulated galaxies follow trends in central surface density $\Sigma_{\star, \rm 1\,kpc}$ and sSFR that are similar to the ones in observed galaxies, nearly independently of the stellar feedback prescription adopted by each code group. 
The slopes of the radial gas metallicity gradients in three of the codes are similar to the one predicted by analytical models (that assume a low degree of metal mixing with the ISM) and the one observed at low to intermediate redshift. Nonetheless, many other codes show a steep negative gradient that is inconsistent with observations, a result however similar to the one found in several other state-of-the-art cosmological simulations. All models show a small redshift dependence. 

Cosmological dark-matter-only simulations using different numerical codes now show very good convergence \cite[e.g.][and papers cited there]{Kim2014,AnguloHahn2022}.  In our {\tt CosmoRun} simulation suite we find that this convergence is not yet true of galaxy simulations including baryons.  While the eight simulations analyzed here that started from the same ICs produce galaxies that resemble each other in some respects, they also differ in important respects.  Because of the complexity of the astrophysical processes involved and the different sub-grid prescriptions used for feedback, it may be some time before hydrodynamic simulations achieve reproducibility approaching that achieved by DMO simulations. 

%All material will be made public and we hope it will become part of a larger database of models obtained using different codes and feedback prescriptions \al{resulting} from a common playground.

\section{The {\it AGORA} Project: Past, Present, and Future} \label{sec:agora}

This is the first of the three new papers (Papers IV, V and VI) that present results from the {\tt CosmoRun} simulations of the Assembling Galaxies of Resolved Anatomy ({\it AGORA}) international collaboration, which has simulated a Milky Way mass galaxy using eight of the main state-of-the-art simulation codes to evolve the same initial conditions with the same astrophysical assumptions concerning the cosmic ultraviolet background radiation, gas cooling and heating, and star formation.  All the simulations achieved a resolution better than 100 physical parsecs.  Each simulation used sub-grid stellar feedback prescriptions commonly used with each code.  They were calibrated to give nearly the same star formation histories down to redshift $z=4$, which continued to be true down to $z=2$.  More than 200 snapshots were saved from each simulation between $z=15$ and $z=2$, and all were analyzed using the same methods.  The analysis focused especially on redshifts $z=4$ and 3 and before the last major merger at $z\sim2$.  Timing discrepancies in the halo assembly history caused the major mergers at $z\sim4$ and $z\sim2$ to occur at slightly different redshifts in the different simulations.  Many of the simulations have now reached lower redshifts, but since these simulations did not include the effects of active galactic nuclei (AGNs) it is not clear how relevant they are to observed Milky Way mass galaxies at these lower redshifts. 

The $\Lambda$CDM missing satellites problem was that the number of satellite dark matter halos is much larger than the number of observed satellite galaxies. This so-called missing satellites problem was subsequently explained when baryonic effects that greatly reduced the number of predicted satellite galaxies were taken into account. In Paper V (Jung et al. 2024, accepted), we show that this is true across all eight {\tt CosmoRun} galaxy simulations.  After identifying the matched pairs of halos between the {\tt CosmoRun} simulations and the DMO simulations, each {\tt CosmoRun} halo also tends to be less massive than its DMO counterpart. There was reasonable inter-code agreement in other properties of satellite galaxies such as the stellar mass-halo mass and the mass-metallicity relation.

The circumgalactic medium (CGM) is the gas surrounding galaxies out to their virial radii.  In Paper VI (Strawn et al. 2023, accepted), we show that all the codes achieved similar resolutions as a function of radius.  Nevertheless, the CGM varied significantly in the different {\tt CosmoRun} simulations, with some simulated galaxies retaining most of their metals with other simulations ejecting most of the metals into the CGM and beyond.  These differences were mostly caused by the different sub-grid stellar feedback prescriptions used in the different simulations.  All the simulated galaxies have cool inflowing streams and some amount of hot metal-rich biconical outflows. The outflows are significantly faster in grid codes and slower in particle codes, with ``moving-mesh'' codes in between.  Some of the simulations predicted surface densities and covering fractions for commonly observed ions such as Si~{\sc iv}, C~{\sc iv}, O~{\sc vi}, and Ne~{\sc viii} that were consistent with observations, while predictions of the simulations that did not eject most metals into the CGM were significantly lower than observations.

The {\it AGORA} Collaboration represents a new collaboration paradigm bringing competing code groups into close cooperation.  It has helped improve the codes, both finding and correcting errors and understanding better how parameters in each code control astrophysical processes including star formation.  All {\tt CosmoRun} outputs and codes will be made public.  Several simulations with new codes starting from the same {\tt CosmoRun} initial conditions are now running, and other code groups are welcome to join.  New {\it AGORA} projects are underway including comparisons of the {\tt CosmoRun} major mergers and comparisons of various ways of including the effects of AGNs. 

The Collaboration is open to the participation of community members interested in any of the current projects, or in new ones they want to host that will make use of the currently available data (or future data). We also invite new code groups and groups that are using different stellar feedback strategies than the ones already in the {\tt CosmoRun} project to test their code’s compatibility on their own, by adopting the common initial conditions, the common physics package, and the proposed calibration steps, and then comparing their results with the simulations presented by the Collaboration.\footnote{To locate our increasing collection of publicly available datasets, see footnote \ref{dataset_url}.}   
The larger the {\tt CosmoRun} set of simulations becomes, the better we will be able to understand the intricate process of galaxy formation. 

\vspace{4mm}
%\acknowledgments

We thank all of our colleagues participating in the {\it AGORA} Project for their collaborative spirit which has allowed the {\it AGORA} Collaboration to remain strong as a platform to foster and launch multiple science-oriented comparison efforts.  We also thank Volker Springel for providing the original versions of {\sc Gadget-3} to be used in the {\it AGORA} Project. We thank the UCSC Foundation Board Opportunity Fund for supporting the {\it AGORA}  Project papers as well as the {\it AGORA}  annual meetings.
This research used resources of the National Energy Research Scientific Computing Center, a DOE Office of Science User Facility supported by the Office of Science of the U.S. Department of Energy under Contract No. DE-AC02-05CH11231 using NERSC award HEP-ERCAP0024062.
S.R.-F. and O.A. acknowledge support from the Knut and Alice Wallenberg Foundation, the Swedish Research Council (grant 2019-04659), and the Swedish National Space Agency (SNSA Dnr 2023-00164). S.R.-F. also acknowledges financial support from the Spanish Ministry of Science and Innovation through projects PID2020-114581GB-C22, PID2022-138896NB-C55 and PID2021-123417ob-i00. %??
J.K. acknowledges support from the Samsung Science and Technology Foundation under Project Number SSTF-BA1802-04.  
His work was also supported by the National Research Foundation of Korea (NRF) grant funded by the Korean government (MSIT) (No. 2022M3K3A1093827 and 2023R1A2C1003244). 
His work was also supported by the National Institute of Supercomputing and Network/Korea Institute of Science and Technology Information with supercomputing resources including technical support, grants KSC-2020-CRE-0219, KSC-2021-CRE-0442 and KSC-2022-CRE-0355.
A.G. would like to thank Ruediger Pakmor, Volker Springel, Matthew Smith and Benjamin Keller for help with {\sc Arepo} and {\sc Grackle}.
{\sc Art-I} simulations were performed on the {\sc Brigit/Eolo} cluster at the Centro de Proceso de Datos, Universidad Complutense de Madrid, and on the {\sc At\'ocatl} supercomputer at the Instituto de Astronom\'ia de la UNAM. 
{\sc Ramses} simulations were performed on the {\sc Miztli} supercomputer at the LANACAD, Universidad Nacional Aut\'onoma de M\'exico, within the research project LANCAD-UNAM-DGTIC-151 and on the Laboratorio Nacional de Superc\'mputo del Sureste-Conacyt.
{\sc Changa} simulations were performed on the {\sc At\'ocatl} supercomputer at the Instituto de Astronom\'ia de la UNAM.
{\sc Gadget3-Osaka} simulations and analyzes were performed on the XC50 systems at the Center for Computational Astrophysics (CfCA) of the National Astronomical Observatory of Japan (NAOJ), {\sc Octopus} at the Cybermedia Center, Osaka University, and {\small Oakforest-PACS} at the University of Tokyo as part of the HPCI system Research Project (hp190050, hp200041).
{\sc Arepo} simulations were performed on the High-Performance Computing resources of the {\sc Freya} cluster at the Max Planck Computing and Data Facility (MPCDF, https://www.mpcdf.mpg.de) in Garching operated by the Max Planck Society (MPG).
The publicly available {\sc Enzo} and {\tt yt} codes used in this work are the products of collaborative efforts by many independent scientists from numerous institutions around the world.  
Their commitment to open science has helped make this work possible.   

\bibliographystyle{aasjournal}
\bibliography{sample63}

\clearpage

\begin{table*}
\vspace*{1mm}
\footnotesize
\caption{\footnotesize Stellar feedback implementation adopted by the {\sc Arepo} code group, complementing Table 1 of {\it AGORA} Paper III\tablenotemark{\textdagger}}
\centering
\vspace{-2mm}
\begin{tabular}{c || c | c | c | c }
\hline\hline 
Code & Stellar feedback & SN \& metal production model & Effective metal yield & Runtime parameters\\ 
\hline
{\sc Arepo} & Thermal  & SN Type II & 0.034& $E_{\rm thermal}= 2\times 10^{52}\,{\rm ergs/SN}$ released 3 Myr after star formation \\ 
\hline 
\end{tabular}
\tablenotetext{$\textdagger$}{\scriptsize While the total returned mass via feedback is constrained across the code platforms, the exact feedback scheme and the metal yield are left to each code group's discretion to be as close to the most widely-used practice in its community as possible.  
For more information on the items listed here, see Table \ref{tab:1} of this paper, and Section 3.2 of Paper III.  
For more information on the ``effective'' metal yield by stellar feedback measured in the entire simulation box at $z=4$ for the {\tt CosmoRun} suite of simulations (fourth column), see  Section 6.2.2 and Table 1 of Paper III.}
\label{tab:arepogeneral}
%\vspace*{2mm}
\vspace{-5mm}
\end{table*}

\begin{appendix}\label{sec:Appendix}

In Paper III from the {\it AGORA} Collaboration, we presented a suite of cosmological ``zoom-in'' simulations obtained after a careful calibration process. 
Since then, some participant code groups found and corrected small errors in the original {\tt CosmoRun} models (e.g., {\sc Art-I}, {\sc Changa}). As a consequence, these groups decided to get the new models presented here and that will become the fiducial {\tt CosmoRun} models. 
In addition, some groups decided to enlarge the original {\tt CosmoRun} suite by adding simulations that adopt a different stellar feedback strategy.  
Furthermore, a new code group has joined the {\it AGORA} Collaboration ({\sc Arepo}), and other ones are currently working to do so (e.g., {\sc Swift}). 
In Appendix \ref{sec:code-arepo} and \ref{sec:ArtChanga}, we present the new {\tt CosmoRun} suite  including the new and revised code entries, to be used as a fiducial set for accompanying and future comparisons, including Papers V and VI. %JP
In particular, we discuss the results of the calibration steps with these new simulations, as we did in Paper III. 
This procedure will become a standard as we include more and more code groups in our growing library of simulations in the {\it AGORA} Project. 

\vspace{5mm}

\section{Calibration of the \texorpdfstring{\sc A\MakeLowercase{repo}}{Arepo} code}\label{sec:code-arepo}

In this section, we show the results of the calibration process by the {\sc Arepo} code group, to produce their fiducial {\tt CosmoRun} simulation entry.  
We present the same figures as in our Paper III, now for a total of eight codes.
Our Figures 12-19 and 22-24 correspond to Figures 2-4, 6, 9, 11-12, 14, 19-20, and 22 of Paper III, in that order.   
Our Figure 20 corresponds to the $z=4$ panels across Figures 15-16, 18, and 21 in Paper III.   
We do not provide the full details and discussion about observed differences between the codes; for more information on the origin of such differences, see Sections 5 and 6 of Paper III.  
Additionally, as we did for all other participant codes in Section 3.2 of Paper III, here we briefly explain the physics included in the {\sc Arepo} fiducial run, focusing only on the part that is unique in this code.  
Interested readers are encouraged to see our previous work to grasp the full picture of how each code in {\tt CosmoRun} works --- Paper I for gravitational dynamics, Paper-II for hydrodynamics, and Paper III for the cosmological simulation techniques and the calibration steps we followed.

\begin{figure*}
        \centering
        \vspace{2mm}
        \includegraphics[scale=0.64]{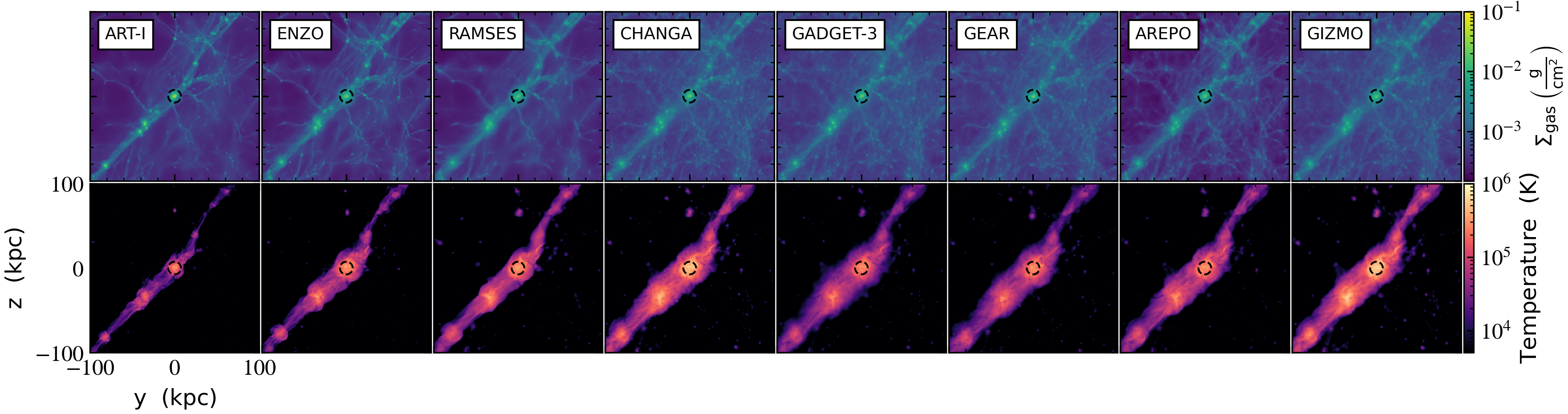}  
        \vspace{-5mm}      
        \caption{Gas density projection ({\it top}) and density-weighted temperature projection ({\it bottom}; each projected through a slab of thickness 200 kpc) at $z=7$ from the first calibration step, {\tt Cal-1} (adiabatic evolution test). 
        We indicate the mean $R_{\rm vir}$ among the codes ($\sim 7.5$ kpc at $z=7$) with a black dashed circle. 
        Units are proper kpc. 
        See Appendix \ref{sec:calstep1} for more information on {\tt Cal-1} and this figure.
	}
        \label{Ap:Cal1_0}
        \vspace{3mm}
\end{figure*}

\subsection{Code-dependent Physics In {\sc Arepo}} \label{sec:AREPOphysics}

{\sc Arepo} is a moving-mesh code that solves coupled equations of gravity and hydrodynamics \citep{Springel2010}. 
Gravitational forces are computed using the TreePM scheme. 
The moving-mesh in {\sc Arepo} is unstructured and represents a Voronoi tesselation of a set of mesh-generating points, which evolves in time based on the local fluid velocity. 
{\sc Arepo} takes advantage of the strengths of both Lagrangian and Eulerian approaches since the resolution is automatically adapted to the local tracer distribution as in SPH approaches, while the finite-volume discretization of the Euler equations is employed to solve hydrodynamics as in mesh-based methods. 
The fluxes of conserved quantities are computed across the interfaces of Voronoi cells using a second-order unsplit Godunov scheme with an exact Riemann solver.\footnote{Although not used in the {\tt CosmoRun} model, {\sc Arepo} has been modified to solve magnetohydrodynamics, where instead an HLLD (Harten-Lax-van Leer with contact and Alfven mode) approximate Riemann solver is utilized, as in the {\sc IllustrisTNG} simulations \citep{Pakmor2011,Miyoshi2005,Pillepich2018}}
Hierarchical, adaptive time-stepping is used to evolve the gas cells and collisionless particles, while time integration is carried out using a second-order Runge-Kutta scheme, which takes into account the time-evolution of the mesh cell geometry \citep{Pakmor2016}.  
The gas cell softening in many {\sc Arepo} implementations is typically adaptive, though in this work we fix the gas softening to the standard value adopted by the participating codes.
Radiative cooling is computed with the {\sc Grackle-v3.1.1} library, previously implemented in {\sc Arepo} by \citet{Smith2021} and \citet{Keller2022}. 

In the {\sc Arepo} simulation presented in this paper, star formation is treated stochastically and stellar yields are computed using the tables of \citet{Portinari1998}.  
To model Type II supernovae, the {\sc Arepo} group employs a thermal feedback model with the energy of $2\times10^{52}$ ergs/SN, assuming a Chabrier Initial Mass Function (IMF). 
The energy, together with mass and metals, from the SNII is distributed among $50\pm1$ weighted neighboring gas cells using a cubic spline SPH kernel. 
Table~\ref{tab:arepogeneral} summarizes the key stellar feedback parameters and the effective metal yield in {\sc Arepo}, which can be compared with Table 1 of Paper III.
Note that each  code group is given a freedom to choose its own stellar feedback scheme for energy and metals, close to the most widely-used practice in its code community.

Although with a much simpler approach for the stellar feedback than the one used in e.g., {\sc IllustrisTNG} or {\sc Auriga} simulations, the {\sc Arepo} entry for {\it AGORA} {\tt CosmoRun} fulfills all the {\tt Cal-4} standards, as we discussed in the main text and will discuss in the following sections.\footnote{In the accompanying {\it AGORA} Papers V (Jung et al. 2024 accepted) and VI (Strawn et al. 2023 accepted), we label this 
%JP entry 
simulation as ``{\sc Arepo-t}'' (rather than the simple notation ``{\sc Arepo}'' in this paper) to emphasize that it is different from the typical {\sc Arepo} runs seen in the literature (e.g., {\sc IllustrisTNG}).}
The {\sc Arepo} group has also provided the Collaboration with an alternative run with the same physics as in {\sc IllustrisTNG}  \citep{Pillepich2018}, except without the implementation of massive black holes.  
While this model is not used anywhere else, in Figure~\ref{fig:4} of Appendix \ref{subsec:timdisc} we use this particular run to test if different feedback implementations may cause changes in the mass assembly history.

\subsection{Physics Calibration Steps For {\sc Arepo}} \label{sec:calibration}

In Section 5 of Paper III we presented a calibration process designed to reduce the number of free parameters to be accounted for in the comparison across different code platforms. 
In this process all code groups need to achieve a set of four steps: checking the convergence in adiabatic gas evolution ({\tt Cal-1}), the implementation of the {\sc Grackle} cooling-heating library ({\tt Cal-2}), the convergence in the created stellar mass at $z=7$ when no stellar feedback is included ({\tt Cal-3}), and finally the convergence to the stellar mass predicted by the semi-empirical models at $z=4$ when stellar feedback is included ({\tt Cal-4}). 
In the next subsections, we present the results of each one of these calibration steps for the {\sc Arepo} code. 
Notice also that in the figures for {\tt Cal-4} we include the new {\sc Art-I}  and {\sc Changa} entries (see Appendix~\ref{sec:ArtChanga} for more information on these new simulations).

\subsubsection{Calibration Step One ({\tt Cal-1}): Adiabatic Evolution of Gas} \label{sec:calstep1}

\begin{figure}
        \centering
        \vspace{2mm}
        \includegraphics[scale=0.25]{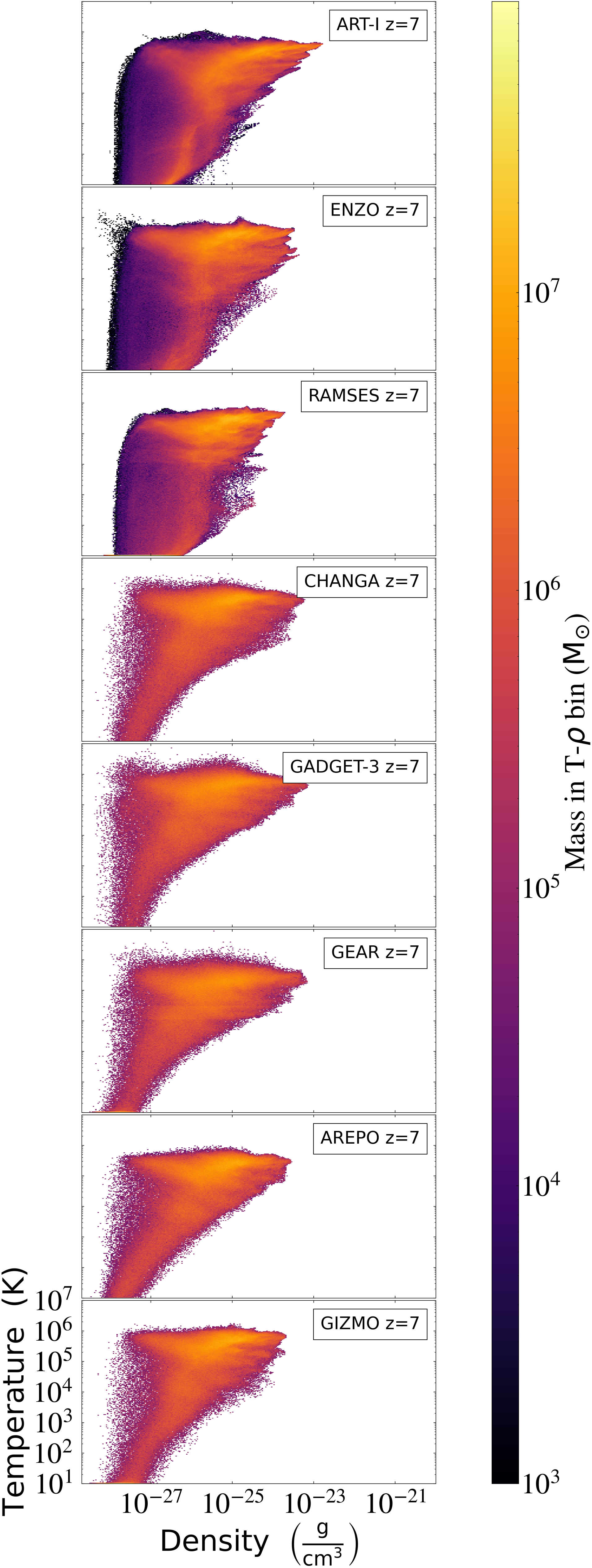}        
        \caption{The $z=7$ composite of probability distribution function (PDF) of density and temperature for the gas within 100 kpc from the center of the main galactic system in the {\tt Cal-1} runs.  
        The 100 kpc-radius sphere encloses the main galaxy, the CGM, and the nearby IGM.  
        Colors represent the total gas mass in each 2-dimensional bin.  
        See Appendix \ref{sec:calstep1} for more information on {\tt Cal-1} and this figure.
        }
        \label{Ap:Cal1_1}
\end{figure}

In Figure~\ref{Ap:Cal1_0} we show the projected density (top row) and temperature (bottom row) from the {\tt Cal-1} runs at $z=7$.
We also present the 2-dimensional density-temperature probability distribution function (PDF) in Figure~\ref{Ap:Cal1_1}.  
The overall large-scale density structures in Figure~\ref{Ap:Cal1_0} and the multiphase density-temperature structure in Figure~\ref{Ap:Cal1_1} are remarkably similar across all eight codes, including the {\sc Arepo} model. 
Readers can find the details on the origin of the minute differences in Section 5.1 of Paper III. 

\begin{figure}
        \centering
        \vspace{2mm}
        \includegraphics[scale=0.25]{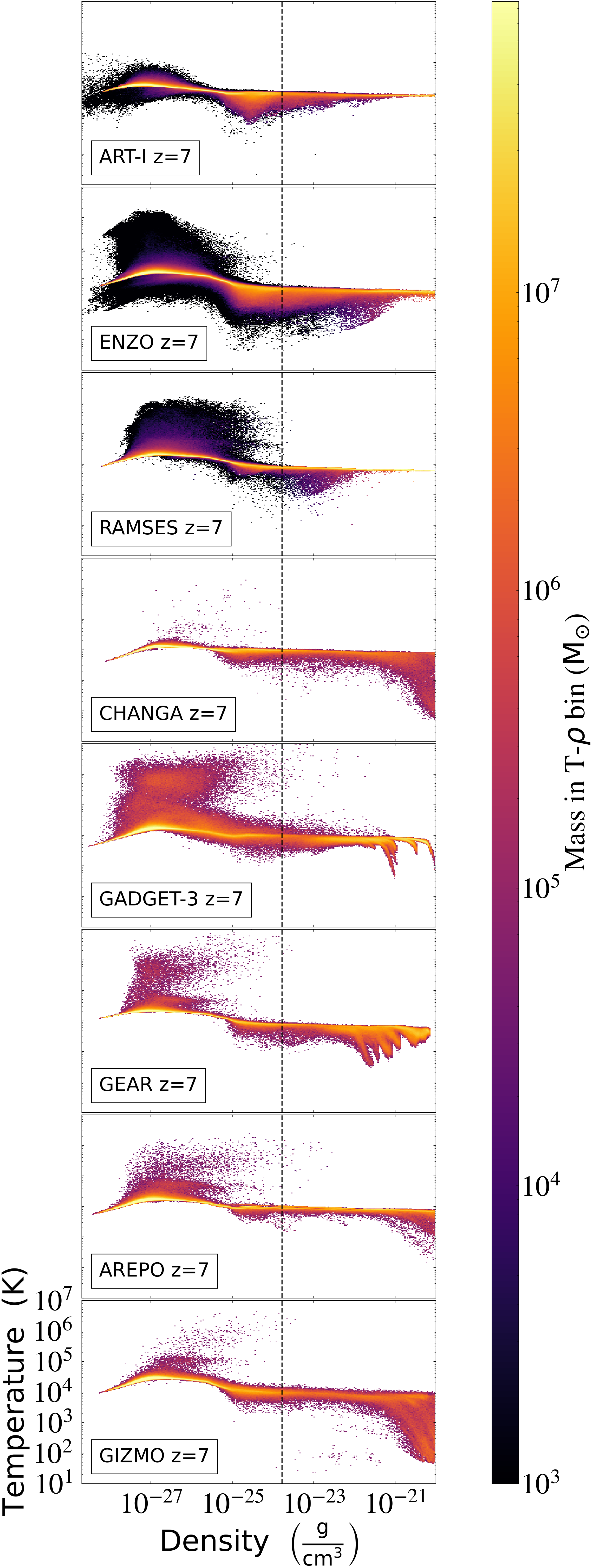}
        \caption{The $z=7$ composite of 2-dimensional PDF of density and temperature  for the gas within 100 kpc from the center of the main galactic system in the {\tt Cal-2} runs (cooling and heating test).  
        The 100 kpc-radius sphere encloses the main galaxy, the CGM, and the nearby IGM.  
        A black dashed vertical line is placed at the value of the star formation density threshold (see Section 3.1 of Paper III).
        See Appendix \ref{sec:calstep2} for more information on {\tt Cal-2} and this figure.
        }
        \label{Ap:Cal2_0}
\end{figure}

\begin{figure}
        \centering
        \vspace{2mm}
        \includegraphics[scale=0.25]{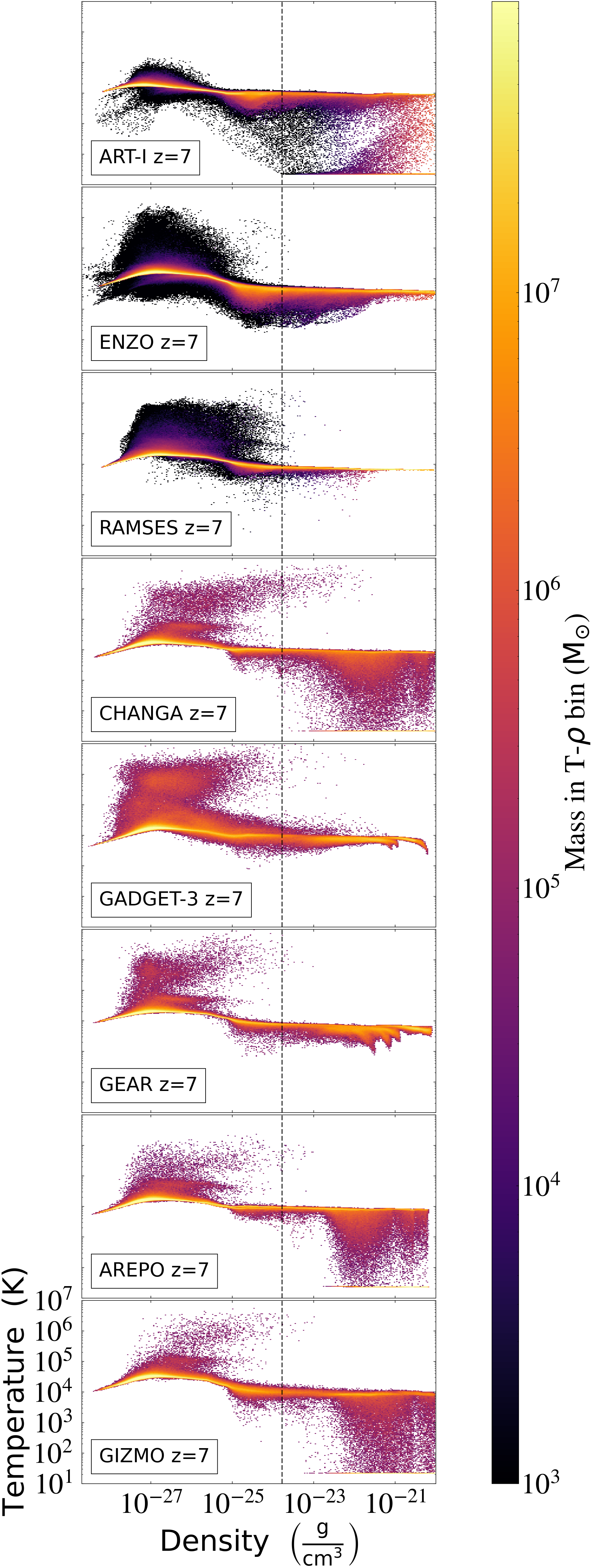}
        \caption{The $z=7$ composite of 2-dimensional PDF of density and temperature for the gas within 100 kpc from the center of the main galactic system in the {\tt Cal-3} runs (star formation test).  
        The 100 kpc-radius sphere encloses the main galaxy, the CGM, and the nearby IGM.  
        A black dashed vertical line marks the density threshold for star formation.
        See Appendix \ref{sec:calstep3} for more information on {\tt Cal-3} and this figure.        
        }
        \label{Ap:Cal3_0}
\end{figure}

\subsubsection{Calibration Step Two ({\tt Cal-2}): Cooling and Heating of Gas By Common Physics Package} \label{sec:calstep2}

In Figure~\ref{Ap:Cal2_0} we show the 2-dimensional PDF of density and temperature for the eight participant codes, at $z=7$, from the {\tt Cal-2} runs (cooling and heating test). 
Just as in Figure 4 from Paper III, we see some differences in both the low-density high-temperature gas, and the high-density low-temperature gas. 
A detailed discussion of these differences is provided in Section 5.2 of Paper III. 
The {\sc Arepo} {\tt Cal-2} run shows similar features in the high-density low-temperature gas as in the {\sc Changa} and {\sc Gizmo} runs, which are the result of using tabulated {\sc Cloudy} tables. 
All other features observed in the {\sc Arepo} are well within the variations seen in other codes. 
It is worth mentioning that during the {\sc Arepo} {\tt Cal-2} calibration process we found a problem in the \citet[][HM12]{HaardtMadau12} {\sc Cloudy} table with self-shielding that was used with {\sc Grackle-v3.1.1}.  
It causes an overheating of the low-density low-temperature ($T<2\times10^4\,{\rm K}$) gas, mostly in the IGM, and at redshift lower than 10$-$12.\footnote{Details on this problem with the HM12 {\sc Cloudy} table with self-shielding can be found in the {\sc Grackle} Project GitHub webpage: \url{https://github.com/grackle-project/grackle\_data\_files/issues/7}.}
This could have had a minor effect on the cooling of the IGM gas at lower redshift, but would have been in the same direction for all {\tt CosmoRun} simulations since all the codes used the same table. 
For Papers IV to VI we continue to compare simulations that used the old {\sc Cloudy} table for the consistency of  comparison.  
In the future projects within the Collaboration we will adopt a revised {\sc Cloudy} table, and the real effect of this fix will be analyzed by running one of the current {\tt CosmoRun} simulations with the corrected table.

\begin{figure}
        \centering
        \vspace{2mm} 
        \includegraphics[scale=0.475]{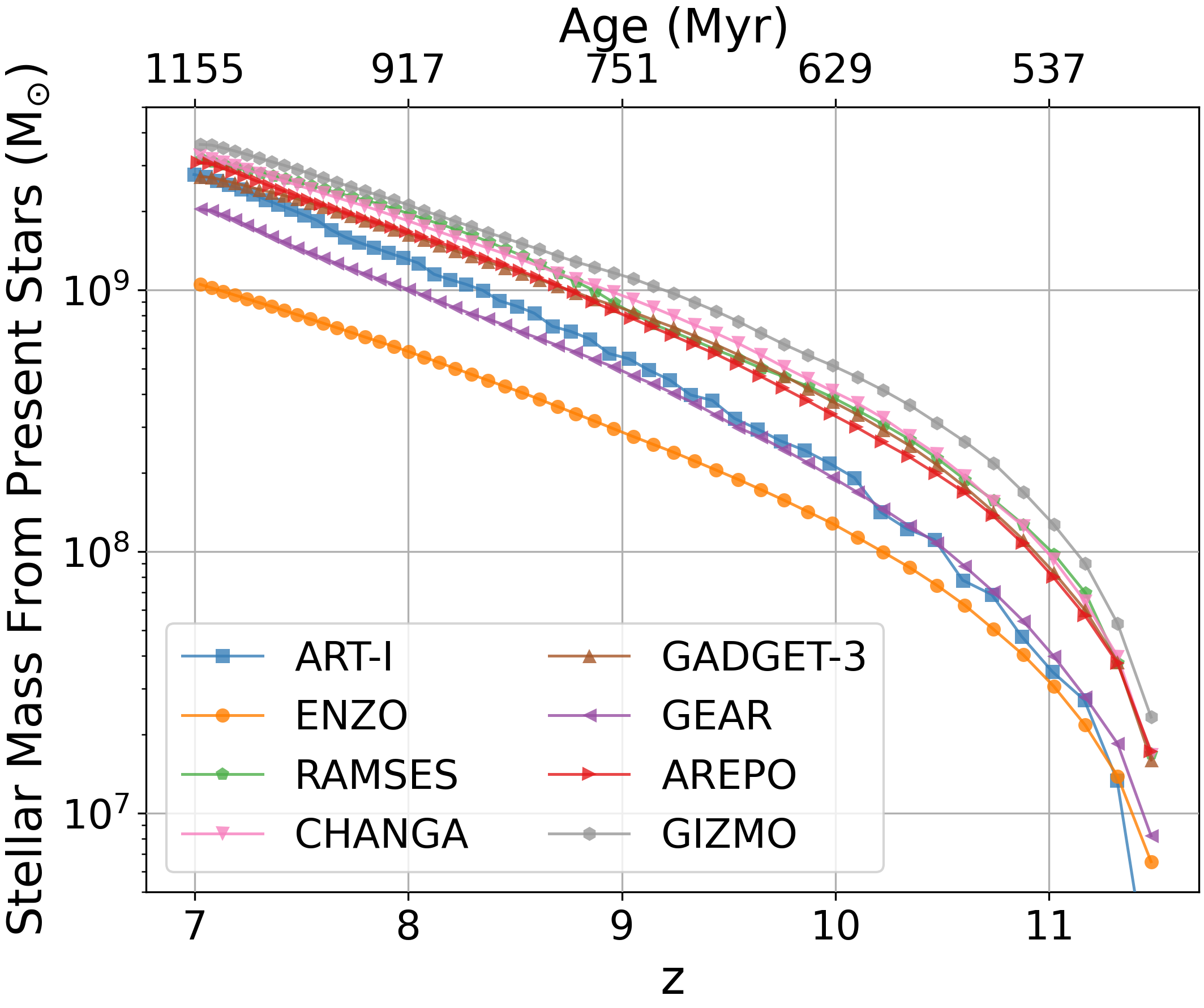}        
        \caption{Stellar mass growth histories for the {\tt Cal-3} runs in a 100 kpc sphere centered at the target progenitor. 
        The curve is computed using the ages or creation times recorded in star particles at $z=7$.
        See Appendix \ref{sec:calstep3} for more information on {\tt Cal-3} and this figure.           
        }
        \label{Ap:Cal3_1}
        \vspace{2mm}
\end{figure}

\subsubsection{Calibration Step Three ({\tt Cal-3}): Common Star Formation Physics} \label{sec:calstep3}

\begin{figure*}
        \centering
        \vspace{0mm}
        \includegraphics[scale=0.64]{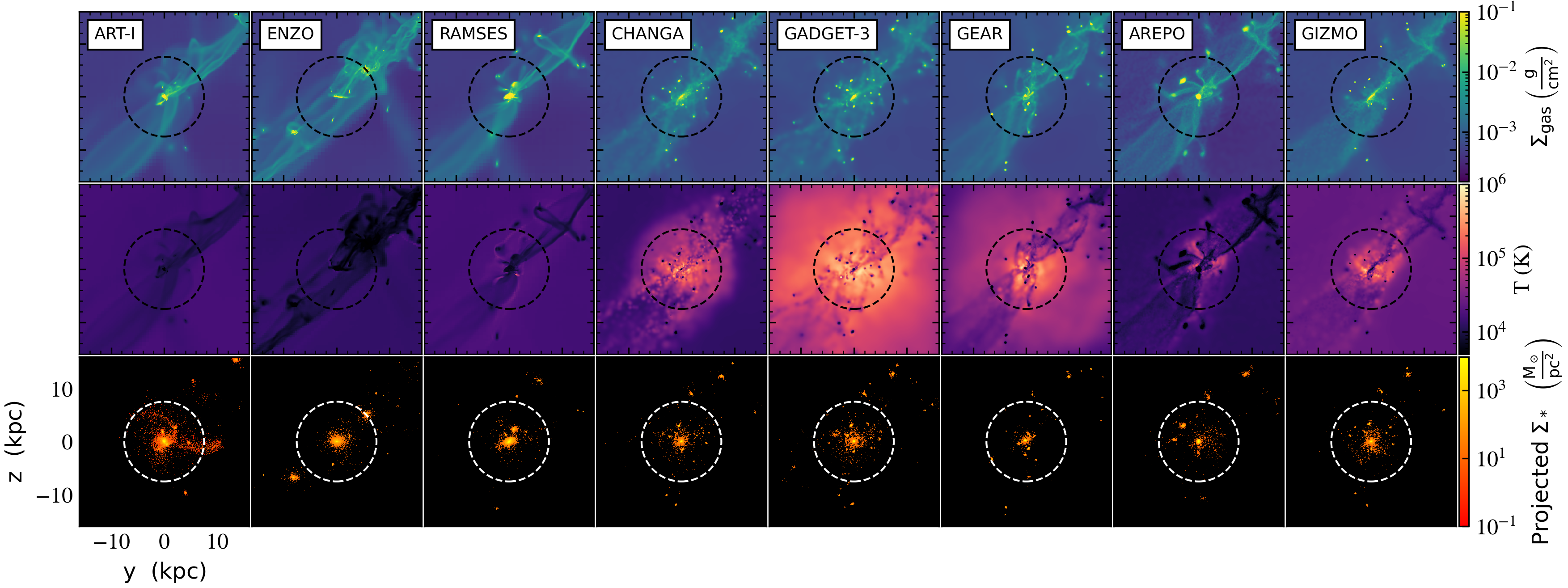}
        \caption{Gas density projection ({\it top}), density-weighted temperature projection ({\it middle}), and stellar surface density ({\it bottom}) at $z=7$ from the third calibration step, {\tt Cal-3}. 
        The width of each panel is $4R_{\rm vir} = 30 \,{\rm kpc}$. 
        The mean $R_{\rm vir}$ among the codes ($\sim 7.5$ kpc) is indicated with a black/white dashed circle.
        See Appendix \ref{sec:calstep3} for more information on {\tt Cal-3} and this figure.            
	}
        \label{Ap:Cal3_2}
        \vspace{5mm}
\end{figure*}

Figure~\ref{Ap:Cal3_0} shows the $z=7$ composite of 2-dimensional PDF of gas density and temperature in the {\tt Cal-3} runs (star formation test). 
As discussed in Section 5.3 of Paper III we see an overall agreement on the temperature-density features, now including the {\sc Arepo} run. 
In particular, results by {\sc Arepo} are remarkably similar to the ones from the finite-volume code  {\sc Gizmo}. 
The {\tt Cal-3} success is demonstrated in Figure~\ref{Ap:Cal3_1} where we show how all the code groups reach a final stellar mass at $z=7$ that converges to $\sim 2\times10^9\, {\rm M}_{\odot}$ within 0.35 dex. 
In Section 5.3.1 of Paper III, readers can find a discussion of the differences in the stellar mass growth. 
Lastly, in Figure~\ref{Ap:Cal3_2}, we show the projected gas density (top row), temperature (middle row), and stellar surface density (bottom row), of the eight {\tt Cal-3} runs at $z=7$. 
The convergence of the {\sc Arepo} {\tt Cal-3} run with the other codes in the gas density and temperature is good, with the minor differences in the IGM gas temperature already discussed in Section 5.3.2 of Paper III. 

\subsubsection{Calibration Step Four ({\tt Cal-4}): ``Favorite'' Stellar Feedback Prescription By Each Code} \label{sec:calstep4}

\begin{figure}
        \centering
        \vspace{2mm}
        \includegraphics[scale=0.47]{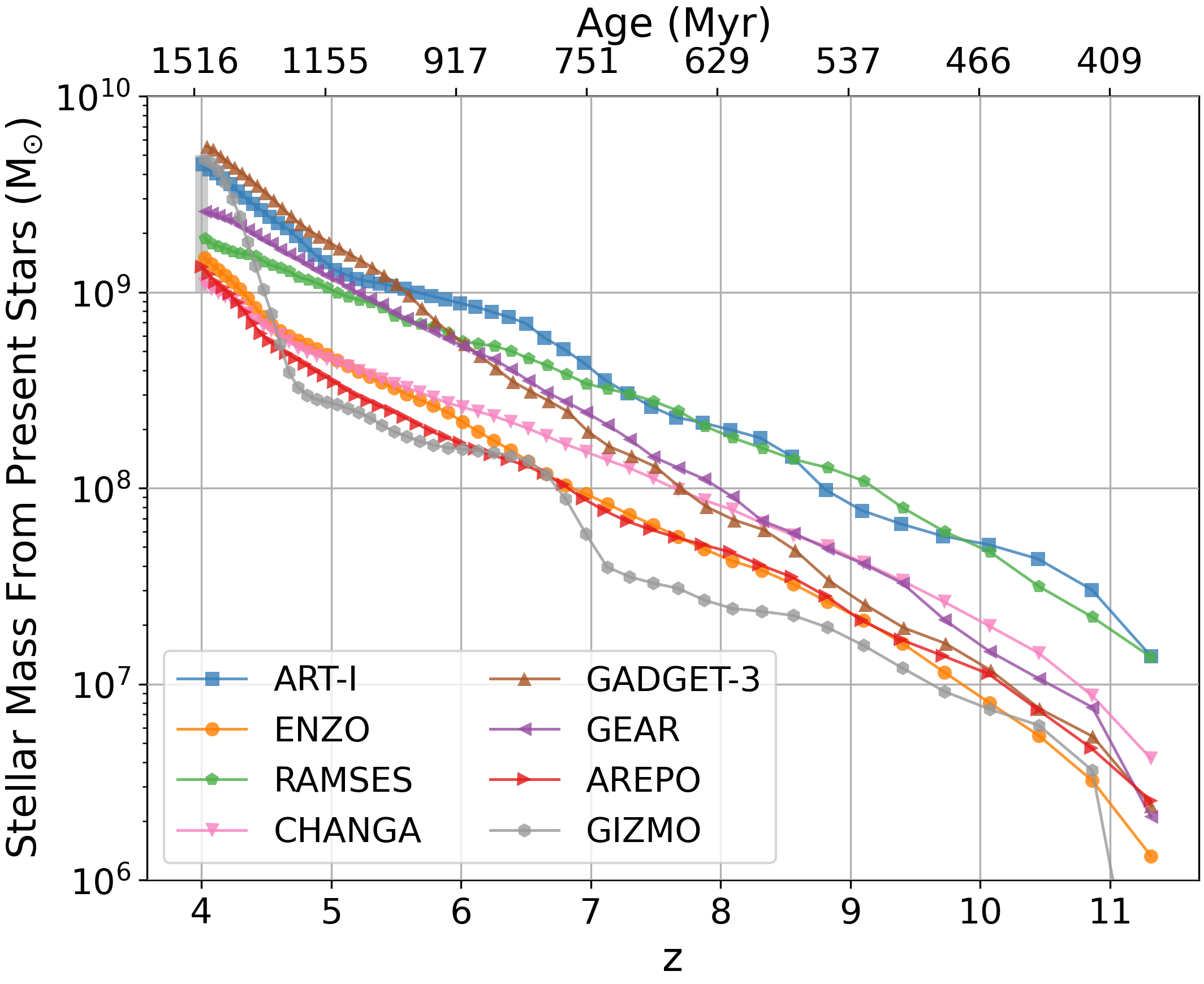}
        \caption{Stellar mass growth histories for the {\tt Cal-4} runs inside a $R_{\rm vir}$ sphere centered at the target progenitor. 
        The curve is computed using the ages or creation times recorded in star particles at $z=4$.
        Therefore, what we show here is an upper limit for the total $M_\star$ formed inside $R_{\rm vir}$.        
        The stellar mass range at $z=4$ targeted in our calibration is $M_\star \sim 1-5\times 10^9\, {\rm M}_{\odot}$, as motivated by semi-empirical models. 
        See Appendix \ref{sec:calstep4} for more information on {\tt Cal-4} and this figure.          
        Notice that in the subsequent figures for {\tt Cal-4} and {\tt CosmoRun} we also include the new {\sc Art-I}  and {\sc Changa} entries (changes on the SNe feedback strategy or on the minimum timestep have no, or little, effect in the previous calibration steps but have a strong impact on the {\tt Cal-4} results); see Appendix~\ref{sec:ArtChanga} for more information on these runs.}
        \label{Ap:Cal4_0}
        \vspace{3mm}        
\end{figure}

In this final calibration step, {\tt Cal-4}, we compare the cosmological simulations obtained using each group's ``favorite'' stellar feedback, now including the new {\sc Art-I} and {\sc Changa} simulations, and the {\sc Arepo} simulation.
Figure~\ref{Ap:Cal4_0} shows the result of this final calibration step which is a convergence in the stellar mass at $z=4$ to the predictions by semi-empirical models. 
The convergence is good for all eight codes: the {\sc Art-I} run (blue) falls well inside the range defined by the semi-empirical models (gray shadowed region at $z=4$), unlike in Figure 12 of Paper III where {\sc Art-I} produced the lowest stellar mass. 
The stellar mass growth seen in the {\sc Arepo} run is also very similar to the other codes, including {\sc Enzo}.

With this {\tt Cal-4} result, we conclude the calibration procedure for {\sc Arepo} (and for the new {\sc Art-I} and {\sc Changa} runs). 
The eight simulations shown in Figure~\ref{Ap:Cal4_0} constitute the final {\tt CosmoRun} suite. 
In the next section, we present and analyze the properties of these final simulation entries focusing on the {\sc Arepo} result at $z = 4$.

\begin{figure}
        \centering
        \vspace{2mm}        
        \includegraphics[scale=0.25]{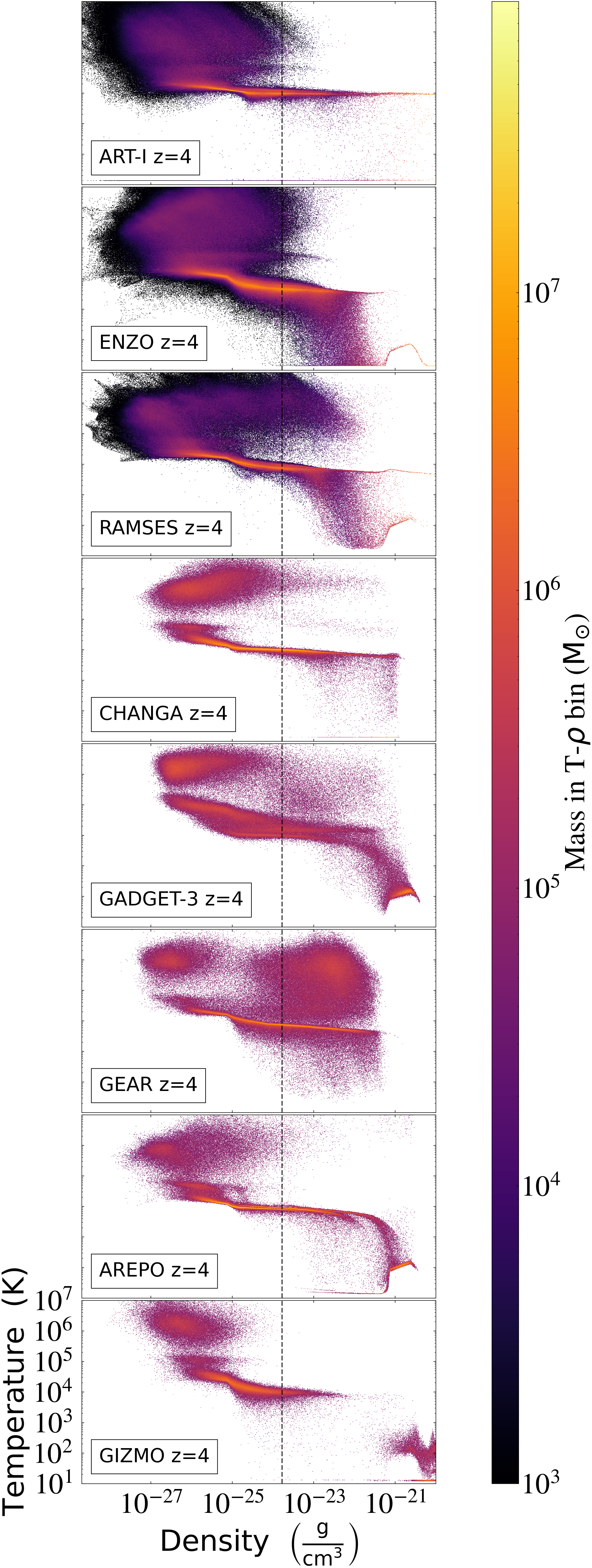}
        \caption{The $z=4$ composite of 2-dimensional PDF of density and temperature for the gas within the mean $R_{\rm vir}$ among the codes  ($\sim 25.4$ kpc)  from the target galaxy's center in the {\tt CosmoRun} simulations.
        A black dashed vertical line marks the density threshold for star formation. 
        See Appendix \ref{sec:propertiesz4} for more information on {\tt CosmoRun} and this figure.                 
        }
        \label{Ap:Cal4_1}
\end{figure}

\subsubsection{Global Properties of The Target Galaxy Progenitor In The {\sc Arepo} {\tt CosmoRun} At $z=4$ } \label{sec:propertiesz4}

In this section, we present a set of comparison plots similar to the ones in Section 6 of Paper III. 
As we have found in the previous papers by the Collaboration, these plots show that clear differences in the gas distribution and properties exist when using different stellar feedback strategies, although many overall properties converge very well.

\begin{figure*}
        \centering
        \vspace{2mm}          
        \includegraphics[scale=0.66]{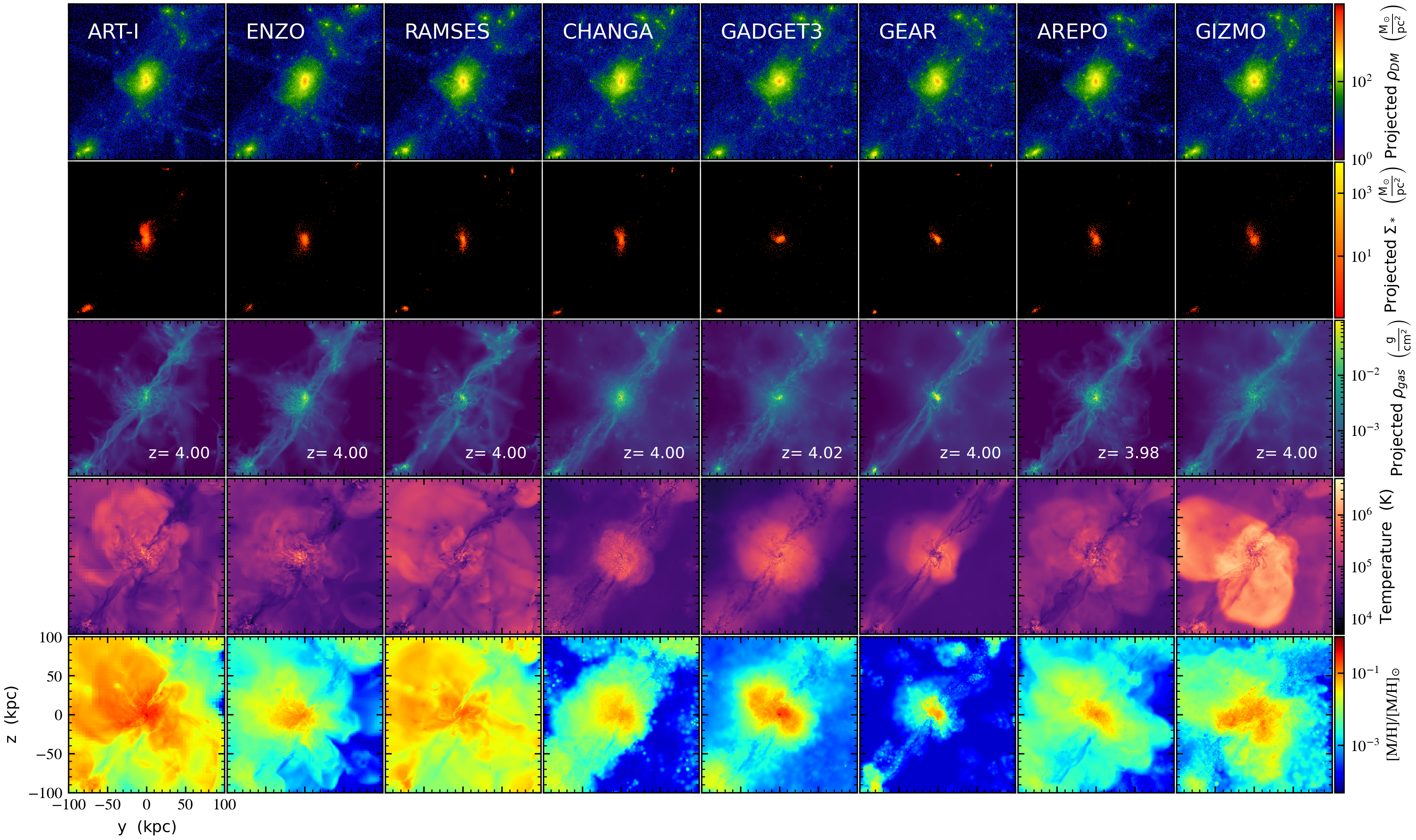}
        \caption{Dark matter surface density ({\it 1st row}), stellar surface density ({\it 2nd row}), gas surface density ({\it 3rd row}), density-weighted projection of gas temperature ({\it 4th row}), and density-weighted projection of gas metallicity ({\it 5th row}) through a slab of thickness 200 kpc at $z=4$ in our {\tt CosmoRun} simulation suite. 
	The full-color version of this figure is available in the electronic edition.  
	The high-resolution versions of this figure and article are available at the Project website, \url{http://www.AGORAsimulations.org/}.
        See Section \ref{sec:overview} for more information on {\tt CosmoRun}, and Appendix \ref{sec:propertiesz4} in particular on this figure.
        	Simulations performed by:  Santi Roca-F\`{a}brega ({\sc Art-I}, {\sc Ramses}), Ji-hoon Kim ({\sc Enzo}), Johnny Powell and H\'ector Vel\'azquez ({\sc Changa}), Kentaro Nagamine and Ikkoh Shimizu ({\sc Gadget-3}), Loic Hausammann and Yves Revaz ({\sc Gear}), Anna Genina ({\sc Arepo}), and Alessandro Lupi and Bili Dong ({\sc Gizmo}).\footref{fnlabel2}}
        \label{Ap:Cal4_2}
        \vspace{2mm}
\end{figure*}

In Figure~\ref{Ap:Cal4_1} we present the 2-dimensional PDF of density and temperature for the gas within the mean $R_{\rm vir}$ among the codes  ($\sim 25.4$ kpc) from the galactic center at $z=4$. 
The {\sc Arepo} run is in good overall agreement with other codes, especially with the particle-based codes and {\sc Gizmo}. 
The new {\sc Art-I} and {\sc Changa} runs show a density-temperature PDF that agrees well with all other codes and is almost identical to the one presented in Paper III, only the new {\sc Changa} run shows a difference that is the production of less high-density low-temperature gas.
Then, Figure~\ref{Ap:Cal4_2} shows the dark matter, star and gas distribution and the gas temperature and metallicity (from top to bottom) for the eight participant codes at $z=4$.\footnote{A movie with the temporal evolution from $z=10$ to the lowest $z$ reached by each model is available in \url{https://youtu.be/egKBqW-mOdc}.\label{fnlabel2}}
In this figure, while the overall agreement is evident in the large-scale structure around the target galaxy, we see the differences in galactic morphology, gas temperature and gas metallicity.   
These differences are the consequence of different stellar feedback strategies adopted. 
We also see  already at $z=4$ the effect of the timing discrepancies that are discussed in detail in Section \ref{sec:MergerTree} and Appendix~\ref{subsec:timdisc}. 
The {\sc Arepo} run is in overall agreement with all other codes, being an intermediate case between the particle-based and the mesh-based codes, both in the dark matter evolution and in the gas temperature and metallicity.

\begin{figure}
        \centering
        \vspace{2mm}
        \includegraphics[scale=0.46]{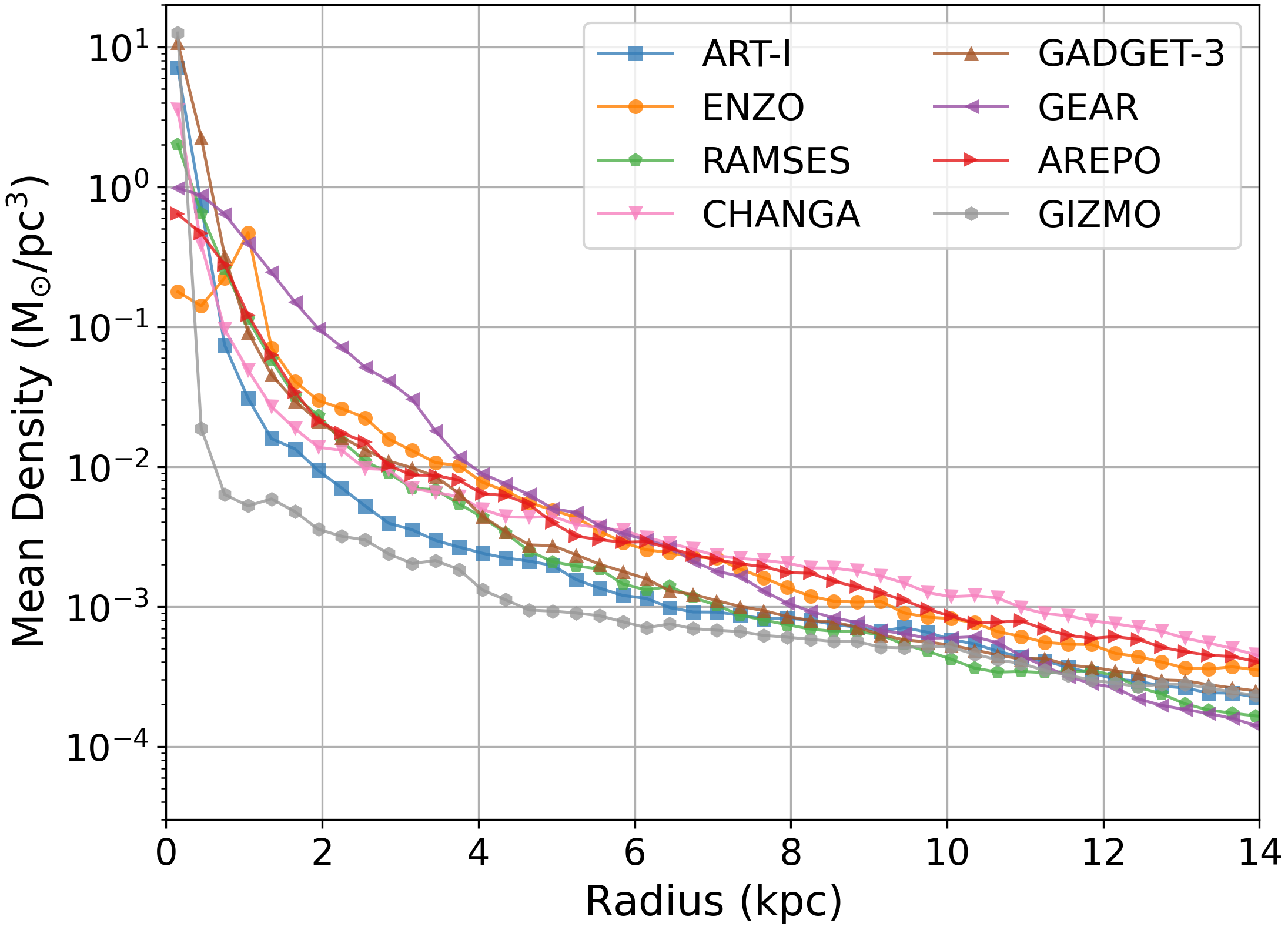}
        \caption{Spherically-averaged gas density profile as a function of distance from the target progenitor's center at $z=4$ in our {\tt CosmoRun} simulation suite.         
        See Appendix \ref{sec:propertiesz4} for more information on {\tt CosmoRun} and this figure.        
        }
        \label{Ap:Cal4_3}
\end{figure}

\begin{figure}
        \centering
        \vspace{2mm}        
        \includegraphics[scale=0.465]{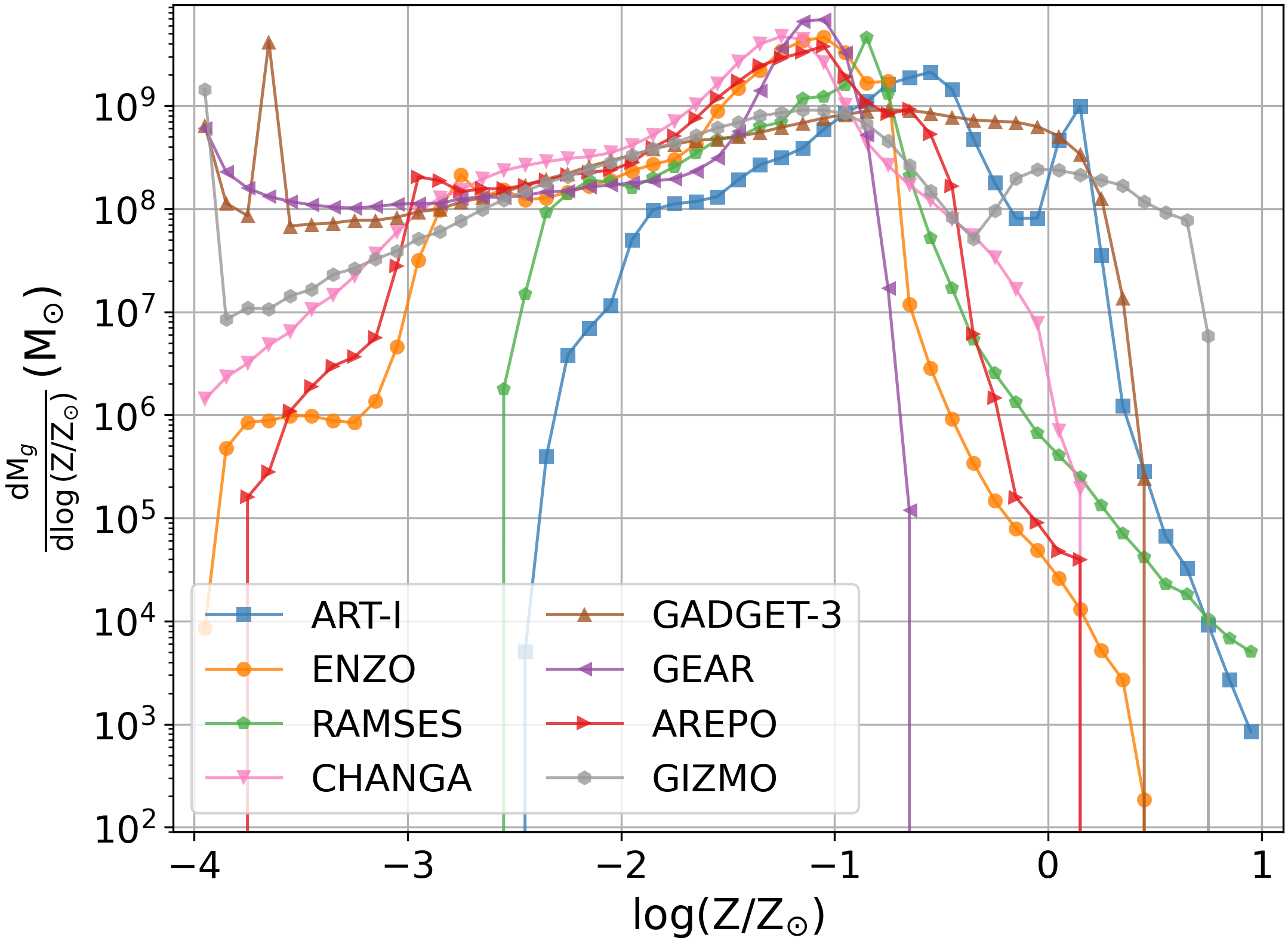}
        \caption{Distribution of gas mass as a function of gas metallicity at $z=4$ for all the gas inside the target progenitor's $R_{\rm vir}$ in our {\tt CosmoRun} simulation suite.
        See Appendix \ref{sec:propertiesz4} for more information on {\tt CosmoRun} and this figure. 	
        }
        \label{Ap:Cal4_4}
        \vspace{0mm}         
\end{figure}

\begin{figure}
        \centering
        \vspace{2mm}         
        \includegraphics[scale=0.25]{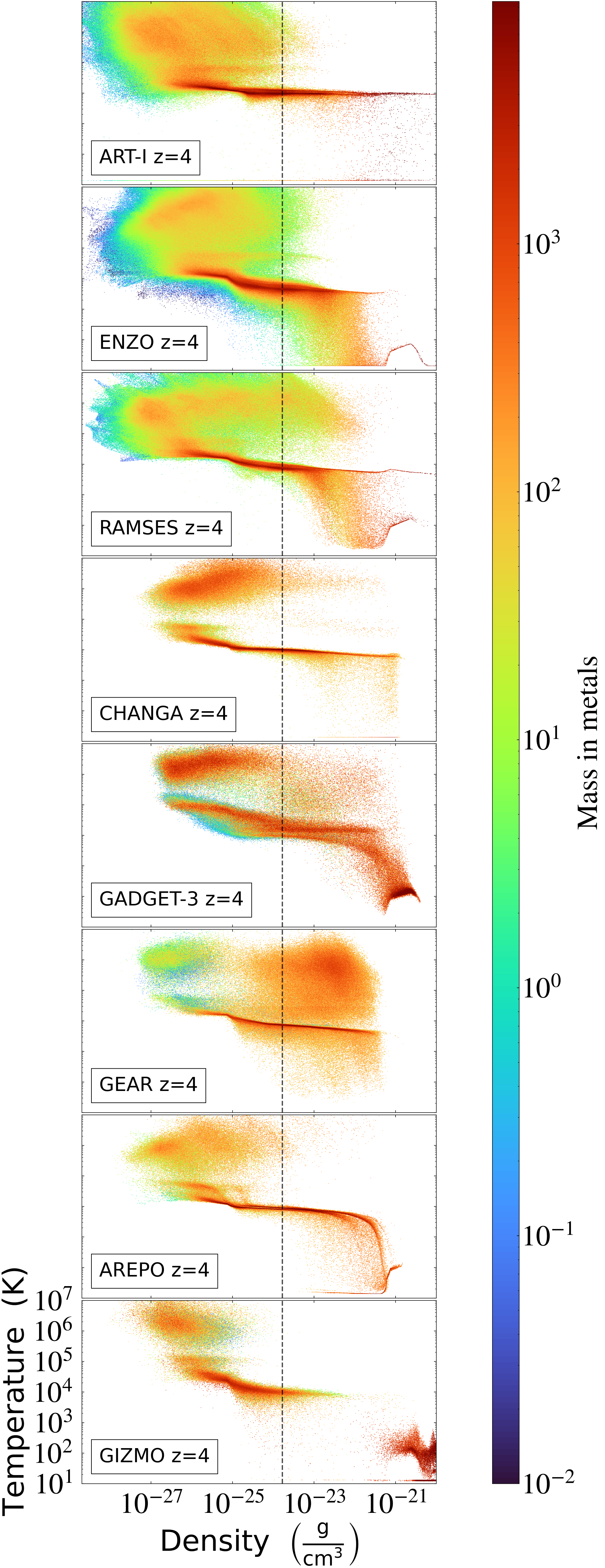}        
        \caption{Similar to Figure~\ref{Ap:Cal4_1}, but now with colors representing the total metal mass in each 2-dimensional bin in our {\tt CosmoRun} simulation suite.   
        Note that the PDF is for the gas within $R_{\rm vir}$ from the center of the target galaxy in the {\tt CosmoRun} simulations.  
        A sphere of radius $R_{\rm vir}$ encloses the main galaxy and CGM, but not the IGM.  
        See Appendix \ref{sec:propertiesz4} for more information on  {\tt CosmoRun} and this figure.
        }
        \label{Ap:Cal4_5}
        \vspace{0mm}          
\end{figure}

\begin{figure}
        \centering
        \vspace{2mm}
        \includegraphics[scale=0.455]{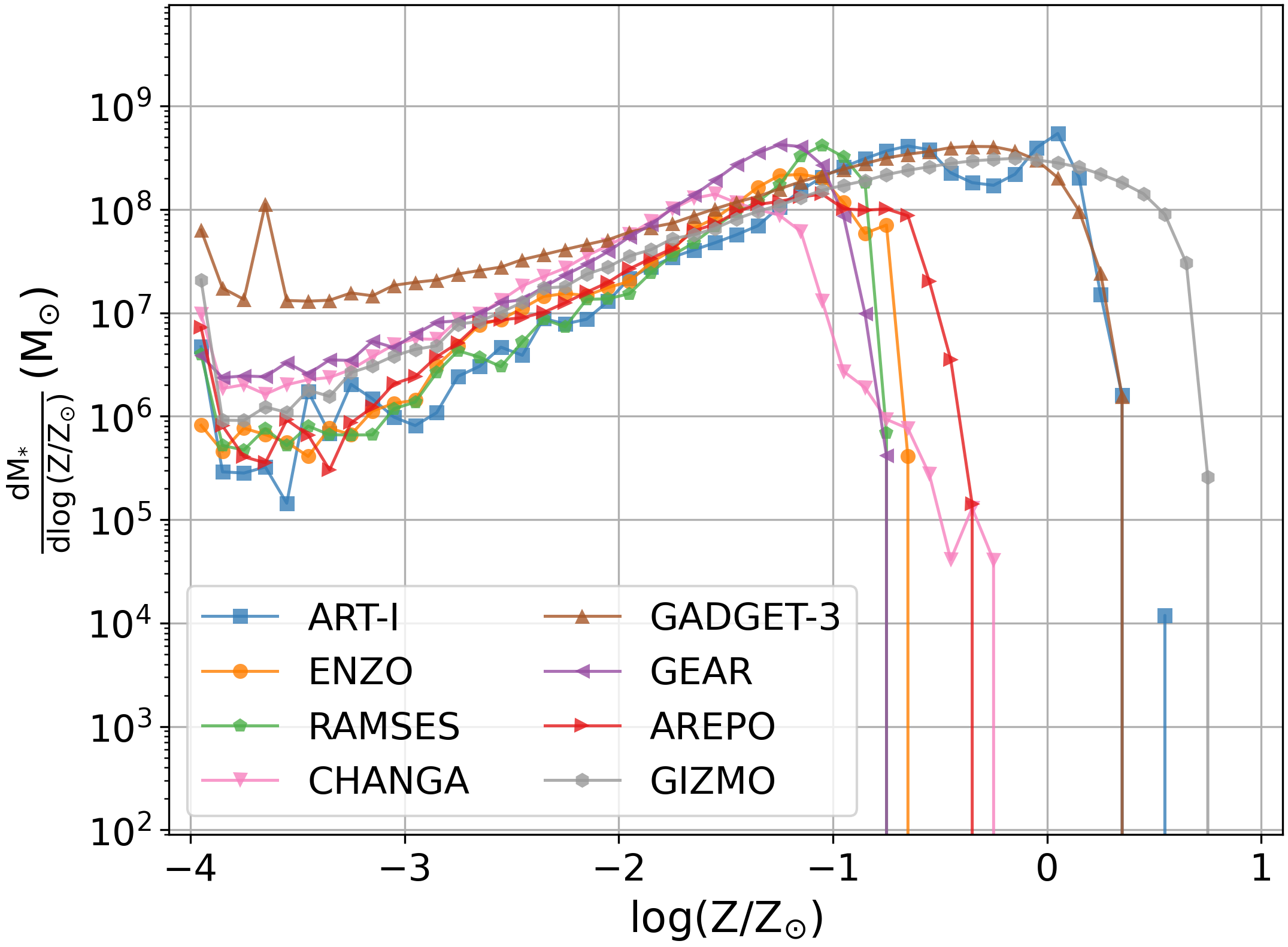} 
        \caption{Distribution of stellar mass as a function of stellar metallicity at $z=4$ for all the stars inside the target progenitor's $R_{\rm vir}$ in our {\tt CosmoRun} simulation suite.
        See Appendix \ref{sec:propertiesz4} for more information on {\tt CosmoRun} and this figure.}  
        \label{Ap:Cal4_6}
        \vspace{0mm}        
\end{figure}

In Figures~\ref{Ap:Cal4_3} to \ref{Ap:Cal4_6} we present the detailed gas properties. 
Figure~\ref{Ap:Cal4_3} demonstrates that all the codes show good convergence of the mean gas density as a function of the distance from the target galaxy. 
In Figure~\ref{Ap:Cal4_4} and \ref{Ap:Cal4_5} readers can see that the differences in the metal distribution are largely in the highest and the lowest metallicity bins. 
These differences are direc consequences of the following: differences in the code-dependent physics, namely, the stellar feedback strategy, the diffusion scheme, and the metal yield. 
Most of the high-metallicity gas resides above the hydrogen cooling threshold in all the codes. 
This result is reflected in Figure~\ref{Ap:Cal4_6}, which shows the distribution of stellar metallicity. 
Since stars inherit the metallicity of the cold star-forming gas, it is expected that we see the same trend and inter-code differences at high metallicity.  
From Figures~\ref{Ap:Cal4_1} to \ref{Ap:Cal4_6}, it is notable that the {\sc Arepo} run and the {\sc Enzo} run are remarkably similar e.g., in the metallicity distribution of Figures \ref{Ap:Cal4_4} and \ref{Ap:Cal4_6}, and in the bottom row of Figure~\ref{Ap:Cal4_2}. 
This resemblance is likely due to the same stellar feedback strategy adopted by the two code groups --- i.e., pure thermal SNe feedback of $\,\gtrsim 10^{52}\,{\rm ergs/SN}$. 

From these thorough calibration processes and analyses, we conclude that the results from the newly included code group {\sc Arepo} are compatible with those presented in Paper III, now making the eight-code {\tt CosmoRun} simulation suite.

\section{The New  \texorpdfstring{\sc A\MakeLowercase{rt-}I}{Art-I} and  \texorpdfstring{\sc C\MakeLowercase{hanga}}{Changa} Entries To The {\it AGORA} \texorpdfstring{\tt C\MakeLowercase{osmo}R\MakeLowercase{un}}{CosmoRun} }\label{sec:ArtChanga}

\vspace{1mm}

\begin{figure}
        \centering
        \vspace{2mm}
        \includegraphics[scale=0.47]{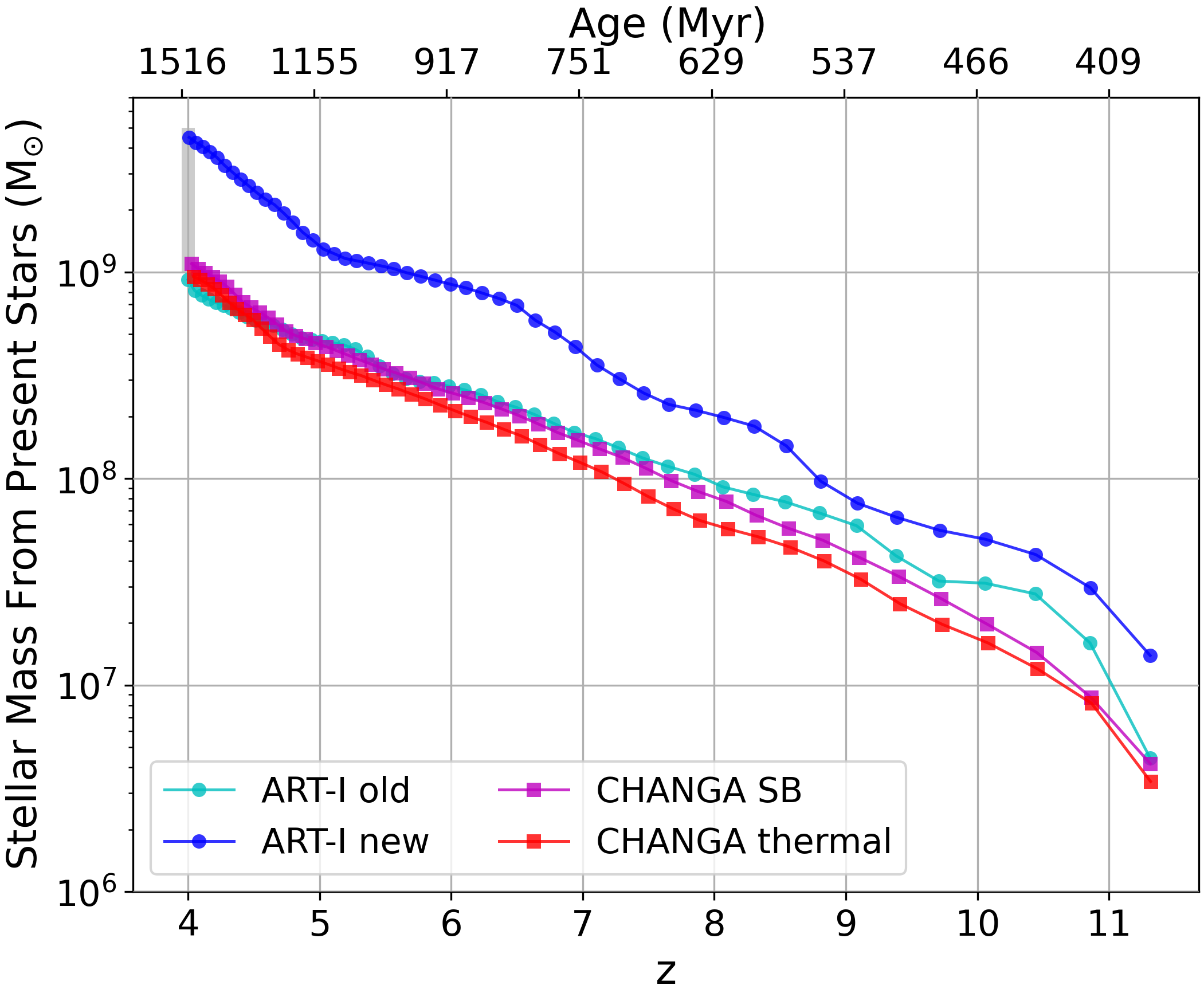}
        \vspace{0mm}
        \caption{Similar to Figure~\ref{Ap:Cal4_0}, but for the old and new {\sc Art-I} runs ({\it cyan} and {\it blue}, respectively), and for the two new {\sc Changa} runs ({\it magenta} and {\it red}; with thermal$+$``superbubble'' feedback, and with only thermal feedback, respectively).
        See Appendix \ref{sec:ArtChanga} for more information on these four runs and this figure.}
        \label{fig:NewArtCanga}
        \vspace{0mm}                
\end{figure}

\begin{figure}
        \centering
        \vspace{2mm}
        \includegraphics[width=1.09\linewidth]{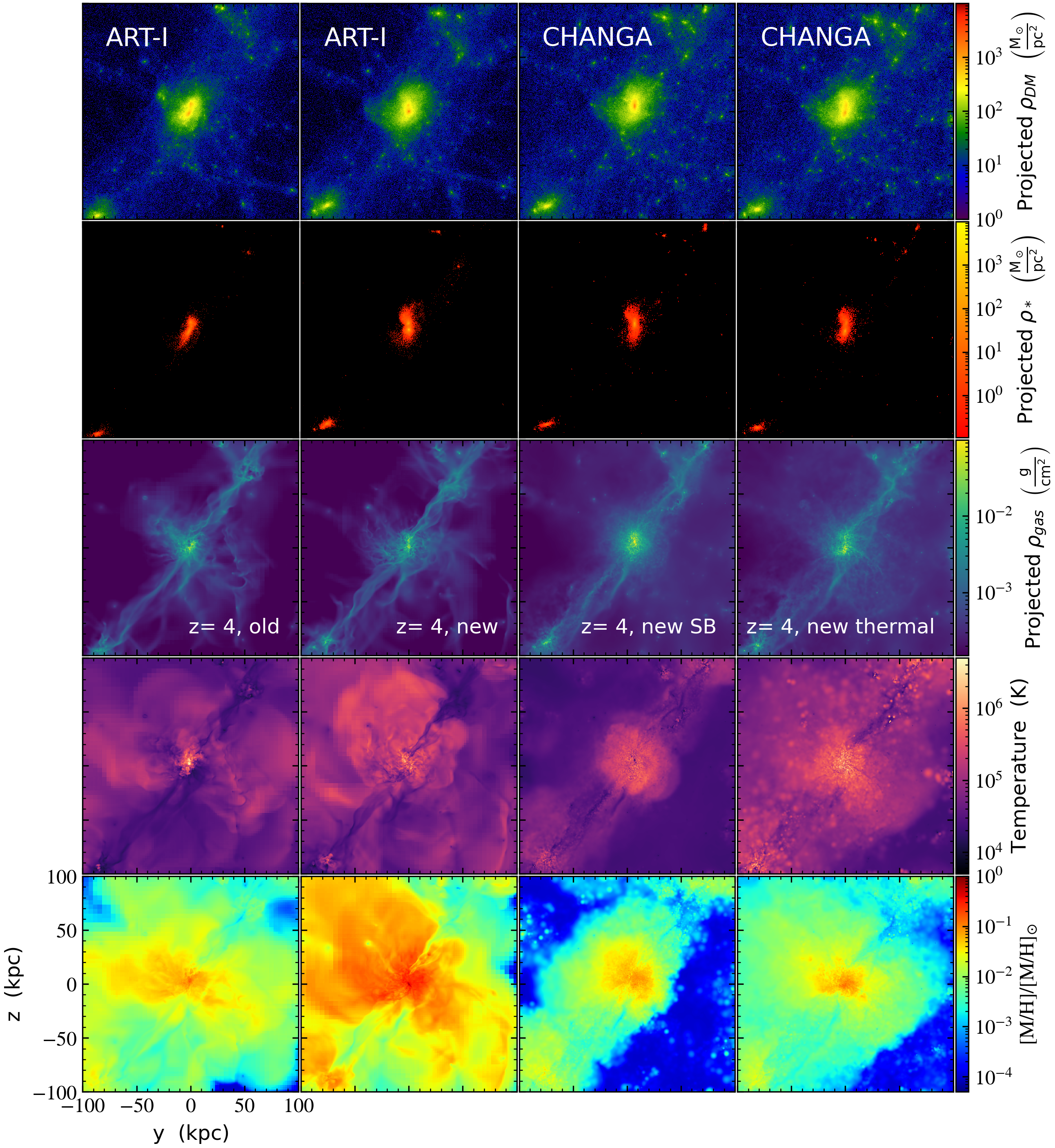}
        \vspace{-3mm}
        \caption{Similar to Figure~\ref{Ap:Cal4_2}, but for the {\sc Art-I}  and {\sc Changa} simulations comparing different stellar feedback prescriptions.  
        The old and new {\sc Art-I} runs ({\it 1st} and {\it 2nd column}, respectively), and the two new {\sc Changa} runs ({\it 3rd} and {\it 4th column}; with thermal$+$``superbubble'' feedback, and with only thermal feedback, respectively).
        See Appendix \ref{sec:ArtChanga} and the caption of Figure \ref{fig:NewArtCanga} for more information on these four runs.}
        \label{fig:DMGasStarsz4}
        \vspace{1mm}        
\end{figure}

The {\sc Art-I} {\tt CosmoRun} simulation presented in Paper III exhibited a stellar mass at $z=4$ slightly below the {\tt Cal-4}'s accepted range (see Figure 12 of Paper III). 
The  {\sc Art-I} code group decided to reduce the stellar ``kinetic'' feedback from $p = 3.6\times10^6\,{\rm M}_\odot \,{\rm km\,s}^{-1}$/SN to $p = 2.5\times10^6\,{\rm M}_\odot \,{\rm km\,s}^{-1}$/SN to acquire a better result in {\tt Cal-4}. 
The {\sc Art-I} code group also introduced a fix to an issue in their choice of a minimum timestep in the force calculations which previously produced a significant variation on the time of major mergers with respect to the other codes, specially at low redshift with minor effects at $z\ge7$.   
In Figure~\ref{fig:NewArtCanga} we show, in cyan and blue lines with circular markers, the stellar mass evolution for the old and new  {\sc Art-I} simulations, respectively. 
In the left two columns of Figure~\ref{fig:DMGasStarsz4} we also show the dark matter, stellar, and gas surface density, and the gas temperature and metallicity for the two {\sc Art-I} runs (see labels in the figure).
We see that the new {\sc Art-I} model, a dark blue line in Fig~\ref{fig:NewArtCanga}, acquires a stellar mass at $z=4$ that falls within the accepted range of {\tt Cal-4} (a gray shadowed region). 
In Figure~\ref{fig:DMGasStarsz4}, the effect of a wrong minimum timestep in the old {\sc Art-I} run is visible here, where the major merger occurs at slightly different redshift (timing discrepancies; see Section \ref{sec:MergerTree} and Appendix~\ref{subsec:timdisc}). 
In the two leftmost bottom panels of Figure~\ref{fig:DMGasStarsz4} we also see that, as expected, the weaker kinetic feedback in the the new {\sc Art-I} run keeps metals in the target galaxy's inner CGM (see red colors) instead of pushing them out to the IGM and thus leaving a the CGM less metal-rich (green colors). 

Due to a technical problem with storing snapshots, the {\sc Changa} code group decided to rerun their {\tt CosmoRun} simulation while also varying their stellar feedback strategy. 
They chose to run two new simulations, one with and one without the ``superbubble'' approximation \citep{Keller2014}. 
The simulation with the ``superbubble'' strategy is the one close to the original {\tt CosmoRun}, but now with $E_{\rm thermal}= 3.25 \times10^{51}\,$ergs/SN instead of  $5 \times10^{51}\,$ergs/SN in Paper III. 
This is the {\sc Changa} entry we discuss in the present paper.
The model without the ``superbubble'' approach uses only thermal feedback of energy $E_{\rm thermal}= 8 \times10^{51}\,$ergs/SN.\footnote{In the accompanying {\it AGORA} Paper VI (Strawn et al. 2023 accepted), this  
%JP particular entry 
{\sc Changa} simulation with just thermal SN feedback
is used for the CGM analyses.  We label it as ``{\sc Changa-t}'' to distinguish it from the {\sc Changa} run in Papers IV and V, and to emphasize that it is different from the typical {\sc Changa} runs seen in the literature that used ``supperbubble'' feedback \citep[e.g.,][]{Keller2014}. \label{footnote-changa-t}}
In Figure~\ref{fig:NewArtCanga} we show, in magenta and red lines with square markers, the stellar mass evolution for both {\sc Changa} runs, with and without ``superbubble'' feedback, respectively. 
In the right two columns of Figure~\ref{fig:DMGasStarsz4} we also present the $z=4$ morphologies of the two {\sc Changa} runs.
In both figures, we see that varying stellar feedback strategy in {\sc Changa}  does not cause a significant change in the overall properties or the merger timings of the target galaxy. 
The main difference between these two runs is in the presence of hot clumps embedded in a more extended warm-hot corona in the run without superbubbles than in the run with superbubbles (see the two rightmost bottom panels in Figure~\ref{fig:DMGasStarsz4}). 
%Also, it is noticeable the presence of many metal-rich clumps embedded and scattered in the cold CGM of the model with superbubbles.

As is shown in this section, the field-tested {\it AGORA} infrastructure enables rapid comparisons between new simulation entries with different stellar feedback implementations.  
This exercise also illustrates how much insight the community can gain by expanding the {\tt CosmoRun} suite of models. 
Several research groups and code groups have expressed their interest in generating newly calibrated simulations of the formation of the {\it AGORA} {\tt CosmoRun} target galaxy, using their own stellar feedback implementations. 
We encourage the members of the community to join us in this task of generating a large library of calibrated models using different feedback prescriptions, to better understand their effects on  galaxy formation and evolution.

\begin{figure*}
    \centering
    \vspace{2mm}
    \includegraphics[width = 0.71\linewidth]{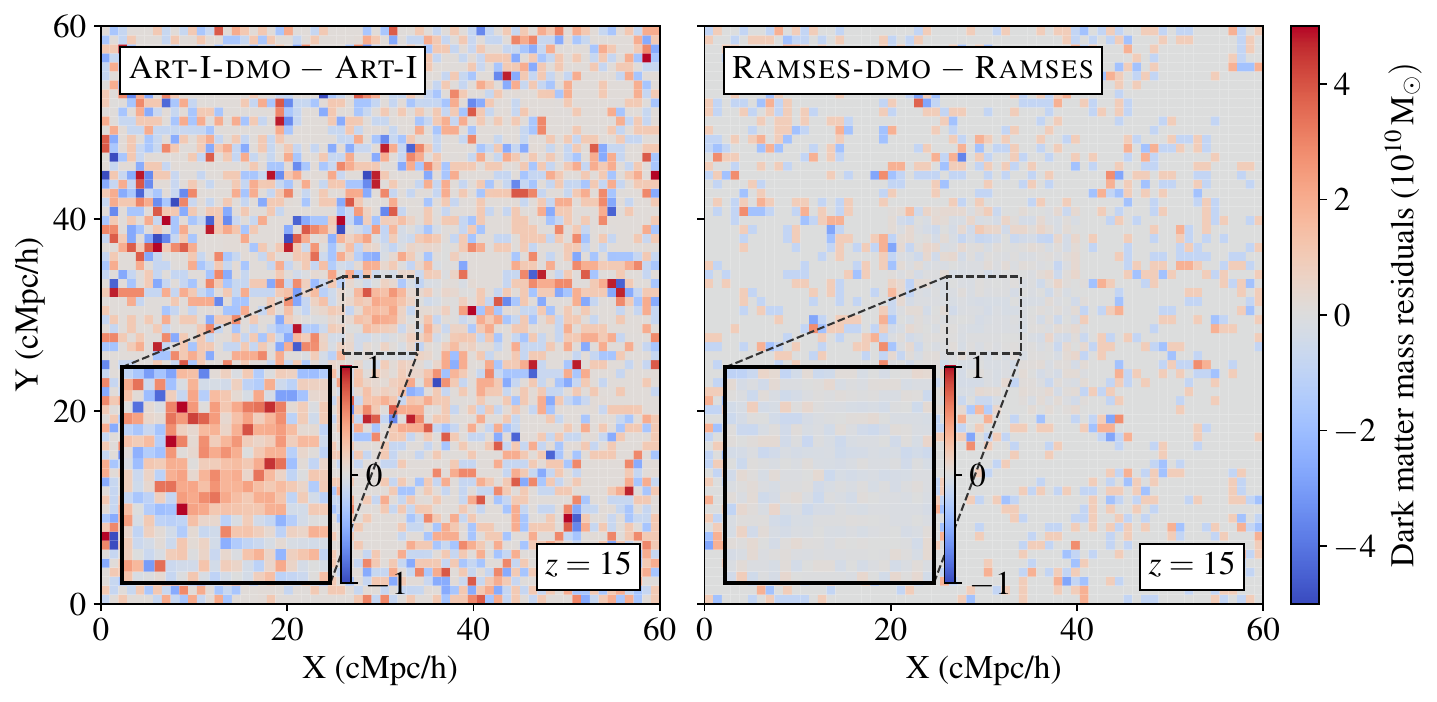}
    \caption{Residuals of dark matter distribution at $z=15$ between DMO run and the {\tt CosmoRun}.  
    The {\it left panel} displays the results from comparing the {\sc Art-I} and {\sc Art-I-dmo} run, while the {\it right panel} shows those from {\sc Ramses} and {\sc Ramses-dmo}. 
    Unlike {\sc Ramses}, {\sc Art-I} exhibits a positive residual in the innermost region, indicating that the enclosed dark matter mass around the target progenitor is higher in the DMO run at $z=15$.
    See Appendix \ref{sec:Arttiming} for more information on this figure.}
    \label{fig:timing_discrepancy_art} 
    \vspace{3mm}    
\end{figure*}

\vspace{1mm}

\section{Possible Origins of the Observed Timing Discrepancies}\label{subsec:timdisc}

In Figures~\ref{fig:1}  to \ref{fig:3} in Section \ref{sec:MergerTree}, we have shown that there exist ``timing discrepancies'' in the halo assembling histories among the {\tt CosmoRun} simulations, among the DMO simulations, and between the {\tt CosmoRun} and the DMO runs.  
These discrepancies are particularly pronounced in the {\sc Art-I} and {\sc Gear} runs, so here we focus on those runs.
Theoretically, one way to explain the timing discrepancy is to attribute it to small differences in the position and velocity of subhalos at high redshift, which are magnified at later times. 
Recent studies have demonstrated that these small differences at high-$z$ can cause changes in the impact parameters of infalling subhalos, thus driving differences in the halo merger histories at low-$z$ \citep[e.g.,][]{Corentin2021, Corentin2022}.\footnote{
%It is important to note that the discrepancies observed near redshift $z=2$ could stem from differences in dark matter distribution in the much earlier universe. 
For instance, at $z=7$, a sphere with a radius of 500 comoving kpc/h encompassing the target halo exhibits $<10\%$ difference in dark matter mass. 
As the majority of dark matter within the sphere falls on to target halo, this high-$z$ discrepancy could cause to differences in target halo's mass at low-$z$.}
%\footnote{Changes on the impact parameter of the infalling subhalos can be identified in the movie presented by the {\it AGORA} Collaboration in this paper, see \textcolor{red}{XXX}}
The origin of such small variations at high-$z$ can be due to the redistribution of energy and angular momentum by the first SNe at $z\gtrsim12$. 
If true, this process would only affect simulations that include baryonic physics, e.g., the {\tt CosmoRun} suite. 
Another process that can generate these discrepancies is the accumulation of small numerical errors in the integration of the equations of motion, or the different runtime configurations (e.g., time-stepping strategy, or in how the minimum timestep is set) that lead to variations in the precision of the orbit integration \citep[see e.g.][]{Power2003}. 
This process would affect both the {\tt CosmoRun} and the DMO runs. 

In Paper I we showed that a relatively small timing mismatch in the numerical integration of the equations of motion can prompt a non-negligible disparity when the runs are compared after a long integration. 
We found that this timing discrepancy precipitates the mismatch in the relative positions of small substructures and the timing of their infall. 
Indeed, the discrepancies in the effective timing of the simulations were found to be an important factor in many comparison studies, including the Santa Barbara Cluster Comparison Project \citep[see the discussion in][]{Frenk1999}. 
In addition, if the accumulation of small differences in the orbit integration dominates over the angular momentum redistribution by the SNe winds, the merger histories in a simulation where its minimum timestep is linked to hydrodynamics (e.g. {\sc Art-I}; see Section~\ref{sec:Arttiming}) will diverge from those in simulations where it is not linked to hydrodynamics. 

In the following sections, we examine in detail the timing discrepancies in the {\sc Art-I} and {\sc Gear} codes (discussed already in Section \ref{sec:MergerTree} and Figures \ref{fig:1}  to \ref{fig:3}) focusing on the differences between their {\tt CosmoRun} and DMO runs. 

\begin{figure}
    \centering
    \vspace{2mm}
    \includegraphics[width = 1.0\linewidth]{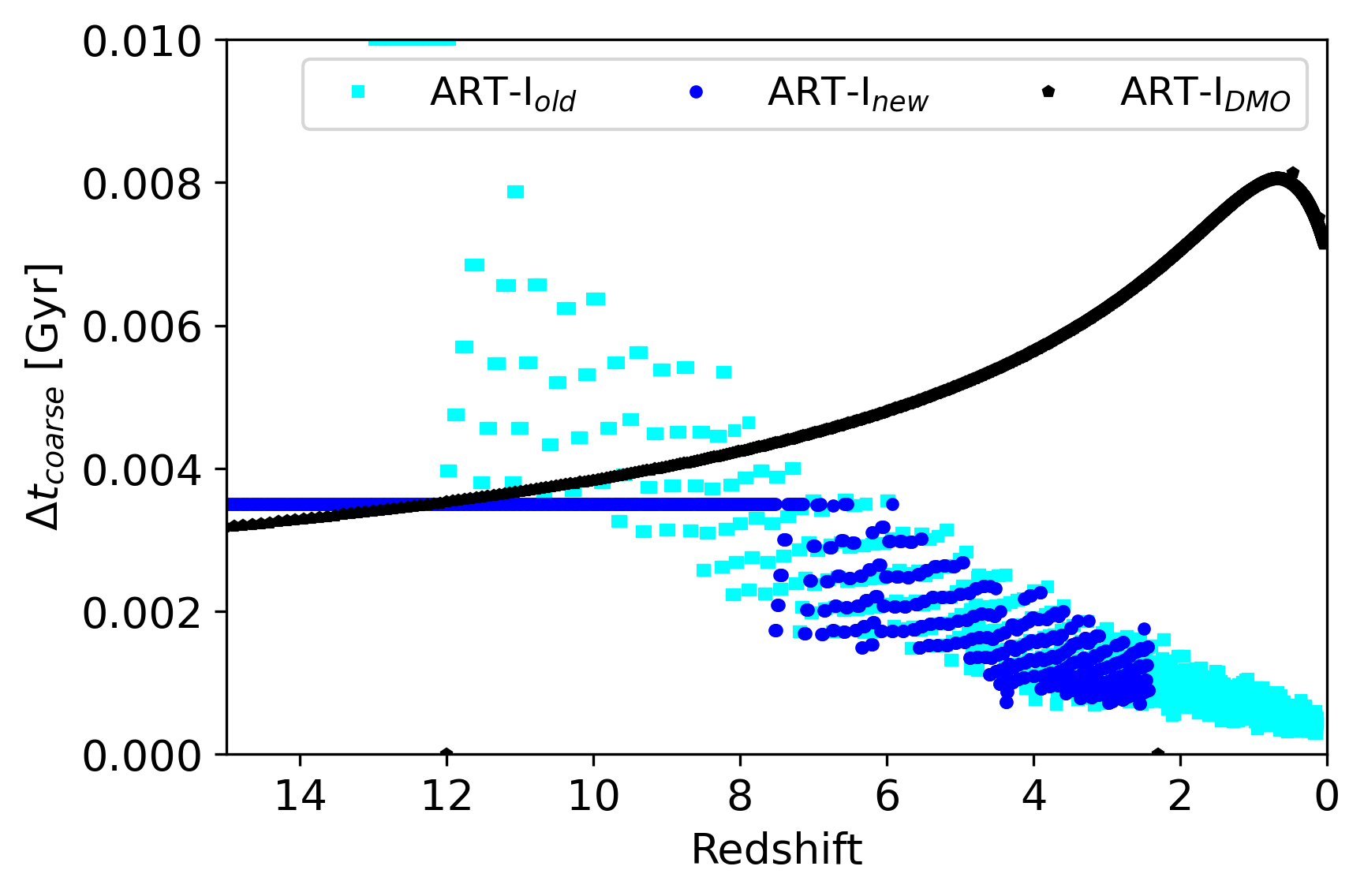}
    \vspace{-5mm}    
    \caption{Timesteps of the coarsest refinement level in the {\sc Art-I} and {\sc Art-I-dmo} run, as a function of redshift.  
    The {\it cyan dots} show the old {\sc Art-I} run, {\it blue dots} the new  {\sc Art-I} run (see  Appendix~\ref{sec:ArtChanga} and Figure \ref{fig:NewArtCanga}), and {\it black dots} the DMO run.
    See Appendix \ref{sec:Arttiming} for more information on this figure.}
    \label{fig:dt_ART}
    \vspace{2mm}    
\end{figure}

\subsection{Timing Discrepancy In The {\sc Art-I} Simulations }\label{sec:Arttiming}

\begin{figure}
        \centering
        \vspace{2mm}
        \includegraphics[width=1.03\linewidth]{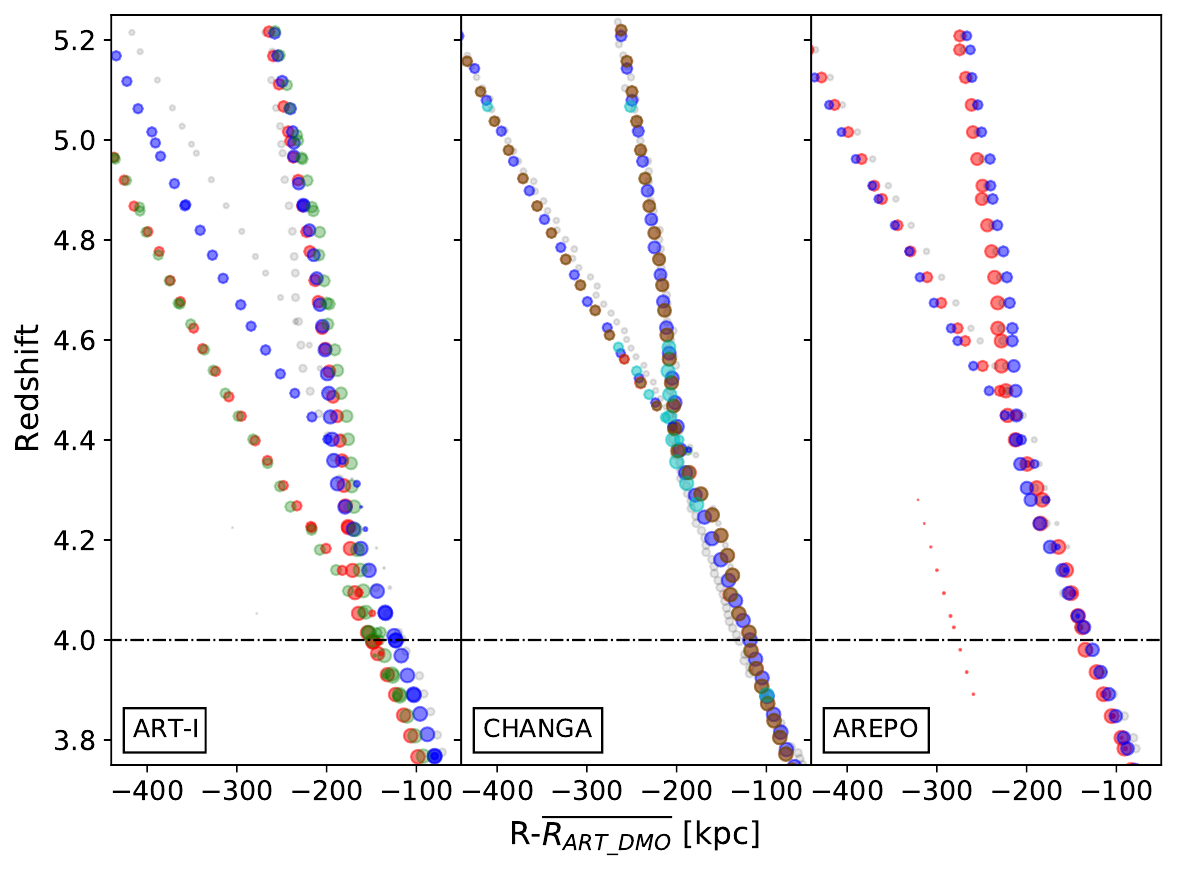}
        \vspace{-4mm}
        \caption{Similar to Figure \ref{fig:2} (around the $z\sim4.5$ major merger), but for various {\sc Art-I}, {\sc Changa}, and {\sc Arepo} runs using different stellar feedback prescriptions and minimum timesteps for the force computation. 
        In {\it blue circles} we show the fiducial run, {\sc Art-I} (the ``new'' model; see Appendix \ref{sec:ArtChanga}, \ref{sec:Arttiming} and \ref{sec:feedbactiming}), {\sc Changa} (with ``superbubble'' feedback; see Appendix \ref{sec:ArtChanga} and \ref{sec:feedbactiming}), and {\sc Arepo} (with only thermal feedback; see Appendix \ref{sec:code-arepo} and \ref{sec:feedbactiming}).  
        In {\it gray circles}, we include each DMO counterpart run.
        Other {\it colored circles} represent alternative feedback prescriptions in each code (see Appendix  \ref{sec:feedbactiming}).         
%        The sizes of the circles are proportional to the virial mass of each halo at each epoch. 
%        Horizontal {\it dot-dashed lines} indicate $z=4$. 
      The $x$-axis shows the spherical radial position ($R$) of the target halo and the subhalos residing inside the zoom-in region in each run with respect to the median $R$ in the redshift interval from $z=15$ to 0 of the target halo in the {\sc Art-I-dmo}  (this is a reference point arbitrarily selected).}
        \label{fig:4}
        \vspace{2mm}        
\end{figure}

While the {\sc Art-I-dmo} run shows similar merger timings to those from other codes, the timing discrepancy arises solely in the hydrodynamic {\sc Art-I} run (see the {\sc Art-I} panels in Figures  \ref{fig:2}, \ref{fig:3} and  \ref{fig:4}). 
To determine when the discrepancy originated, we compare the distribution of dark matter at different epochs. 
Figure \ref{fig:timing_discrepancy_art} displays the residuals of dark matter distribution at $z=15$ between DMO run and the {\tt CosmoRun} (i.e., $M_{\,\rm DM, \,DMO}- M_{\,{\rm DM}, \,{\tt CosmoRun}}$), with the contribution of gas in the {\tt CosmoRun} factored out. 
In contrast to the results from the {\sc Ramses} code, {\sc Art-I} exhibits a positive residual in the innermost region around the target progenitor (indicated by a zoom-in box). 
This positive residual suggests that, in {\sc Art-I}, hydrodynamic simulations have different accretion timings than their DMO counterparts before $z=15$.

In Figure~\ref{fig:dt_ART} we compare the timesteps in the coarsest refinement level as a function of time in the {\sc Art-I-dmo} run and the old and new {\sc Art-I} runs presented earlier (see Appendix~\ref{sec:ArtChanga} for details). 
The timesteps are clearly different if it is linked to hydrodynamics or not (cyan/blue dots vs. black dots). 
The plot also illustrates that the new {\sc Art-I} model (blue dots) imposes a minimum timestep that is similar to the one in {\sc Art-I-dmo} (black dots). 
The different time-stepping strategy is likely the reason why we see changes in the merger timings in the left panel of Figure~\ref{fig:4} --- the new {\sc Art-I} run shows less discrepancy from the {\sc Art-I-dmo} run (blue circles) than the old {\sc Art-I} run does (green circles).  

It should be clearly noted that the error in the minimum time step selection that the old {\sc Art-I} model suffered was only present in the {\tt CosmoRun} in Paper III.  
Other {\sc Art-I} users in the community might have used a time-stepping approach that better converges with the other codes. 
Nonetheless, the new {\sc Art-I} run presented in this paper agrees better with all other codes throughout the calibration process, as discussed in Appendix \ref{sec:ArtChanga}. 

\subsection{Timing Discrepancy In The {\sc Gear} Simulations }\label{sec:Changatiming}

The timing discrepancy of the {\sc Gear} code is present in both hydrodynamic and DMO simulations (see the {\sc Gear} panels in Figures  \ref{fig:2} and \ref{fig:3}). 
The fact that {\sc Gear} inherited {\sc Gadget-2}'s $N$-body algorithm implies that the discrepancy does not arise from the code itself but rather from the specific parameter choice made in the {\sc Gear} group. 
We have conducted a suite of DMO simulations using {\sc Gadget-2} and found that different parameter choices could reproduce the merger timing seen in the {\sc Gear} run. 
Specifically, higher user-defined values for the parameters {\tt ErrTolIntAccuracy} and {\tt MaxRMSDisplacementFac} can lead to a delayed merger timing in the {\sc Gadget-2} simulations.
Therefore, we conclude that the origin of the timing discrepancies in {\sc Gear}  and {\sc Gear-dmo}  is the user parameters, not the code itself.

\subsection{Timing Discrepancy Caused By Changing Stellar Feedback Implementations} \label{sec:feedbactiming}

In Figure~\ref{fig:4} we show the merger trees around the $z\sim4.5$ major merger for {\sc Art-I}, {\sc Changa}, and {\sc Arepo} obtained from multiple runs that use different stellar feedback prescriptions. 
We show the final fiducial {\tt CosmoRun} suite of models in blue circles (note that the {\sc Art-I} and {\sc Changa} group changed their fiducial models from Paper III to Paper IV; see Appendix \ref{sec:ArtChanga}), the DMO runs in gray circles, and the runs with alternative feedback prescriptions in other colored circles.  
For {\sc Art-I}, two additional models are shown ---  in red circles, the old fiducial {\sc Art-I} model with an injection of momentum $p = 3.6\times10^6\,{\rm M}_\odot \,{\rm km\,s}^{-1}$/SN and without a forced minimum timestep (see Figure \ref{fig:dt_ART} and Appendix \ref{sec:Arttiming}); and in green circles, a model with the new momentum value $p = 2.5\times10^6\,{\rm M}_\odot \,{\rm km\,s}^{-1}$/SN  (see Figure \ref{fig:NewArtCanga} and Appendix \ref{sec:ArtChanga})  but without the forced minimum timestep. 
For {\sc Changa}, we also include three additional models --- in cyan circles, the {\sc Changa-t} model with no ``superbubble'' feedback and $E_{\rm thermal}= 8 \times10^{51}\,$ergs/SN  (see Appendix~\ref{sec:ArtChanga} and footnote \ref{footnote-changa-t}); in red circles, the old fiducial {\sc Changa} model with ``superbubble'' feedback and $E_{\rm thermal}= 5 \times10^{51}\,$ergs/SN; and, in green circles, one with no ``superbubble'' feedback and $E_{\rm thermal}= 6 \times10^{51}\,$ergs/SN.
Lastly, for {\sc Arepo}, we include one additional model --- in red circles, a run with the baryonic physics used in the {\sc IllustrisTNG} simulations (see Appendix \ref{sec:AREPOphysics}).
In this figure, we see that changes in the stellar feedback can induce some variations in the major merger timings (e.g., red vs. green in the {\sc Art-I} panel), but not as large as the ones produced by changes in the minimum timestep (e.g., green vs. blue in the {\sc Art-I} panel). 
%For instance, the major merger in {\sc Art-I} varies from $z\sim$4.0 (green) to $z\sim$4.15 (red) due to changes in the SNe feedback. 
%Although less evident, in the {\sc Changa} and  {\sc Arepo}  we also see that the position of the target halo varies when changing the feedback strategy. 
%This result is consistent with the nature of the used SNe feedback: while in the {\sc Art-I} runs we change the injection of momentum, in {\sc Changa} and  {\sc Arepo} we change the thermal energy, and thus only in the former feedback has a direct impact on the kinematics of the surrounding material, i.e. in the mergers at low redshift.

In this section, we have shown that small variations in how the equations of motion are integrated, together with the variations in stellar feedback could result in significant 
%JP discrepancies 
differences in the mass assembly history. 
This result will need to be accounted for in future studies when studying systems like the Local Group. 
This is why, for analyses and papers from the {\it AGORA} Collaboration, we define the redshift {\it before the last major merger (blmm)} ($z_{\rm blmm}$) for each of the participant code in the {\tt CosmoRun} suite (see Section \ref{sec:lowz_cosmorun} and Table~\ref{tab:0}). 
This definition ensures a fair comparison for some of the properties of the target halo, e.g., the number of satellites around the target galaxy, or the virial radius, $R_{\rm vir}$.

\end{appendix}

\end{document}